\documentclass[11pt, a4paper]{book}		
\usepackage[latin1]{inputenc}
\usepackage[english]{babel}
\usepackage{graphicx}				
\usepackage{epsf}
\usepackage{amsfonts}
\usepackage{slashed}
\usepackage{amssymb}
\usepackage{amsmath}
\usepackage{mathtools}
\usepackage{cancel}
\usepackage{amssymb}
\usepackage{titlesec}
\usepackage{titletoc}
\usepackage[titletoc]{appendix}
\usepackage[grey]{quotchap}
 \usepackage{pdfpages}
\begin{document}
\pagestyle{empty}

%\input cover1.tex

%\newpage

%\vspace*{-40mm}

\centerline{DEPARTMENT OF PHYSICS, UNIVERSITY OF JYV\"ASKYL\"A}
\centerline{RESEARCH REPORT No. 8/2014}

\vspace{25mm} 

\centerline{\bf MINIJET INITIAL STATE OF HEAVY-ION COLLISIONS  }
\centerline{\bf FROM NEXT-TO-LEADING ORDER }
\centerline{\bf   PERTURBATIVE QCD}

\vspace{13mm}

\centerline{\bf BY}
\centerline{\bf RISTO PAATELAINEN}

\vspace{13mm}

\centerline{Academic Dissertation}
\centerline{for the Degree of}
\centerline{Doctor of Philosophy}

\vspace{13mm}

\centerline{To be presented, by permission of the}
\centerline{Faculty of Mathematics and Natural Sciences}
\centerline{of the University of Jyv\"askyl\"a,}
\centerline{for public examination in Auditorium FYS 1 of the}
\centerline{University of Jyv\"askyl\"a on August 29th, 2014}
\centerline{at 12 o'clock noon}

\vspace{20mm}

\begin{figure}[!h]
\center
\includegraphics[scale=0.12]{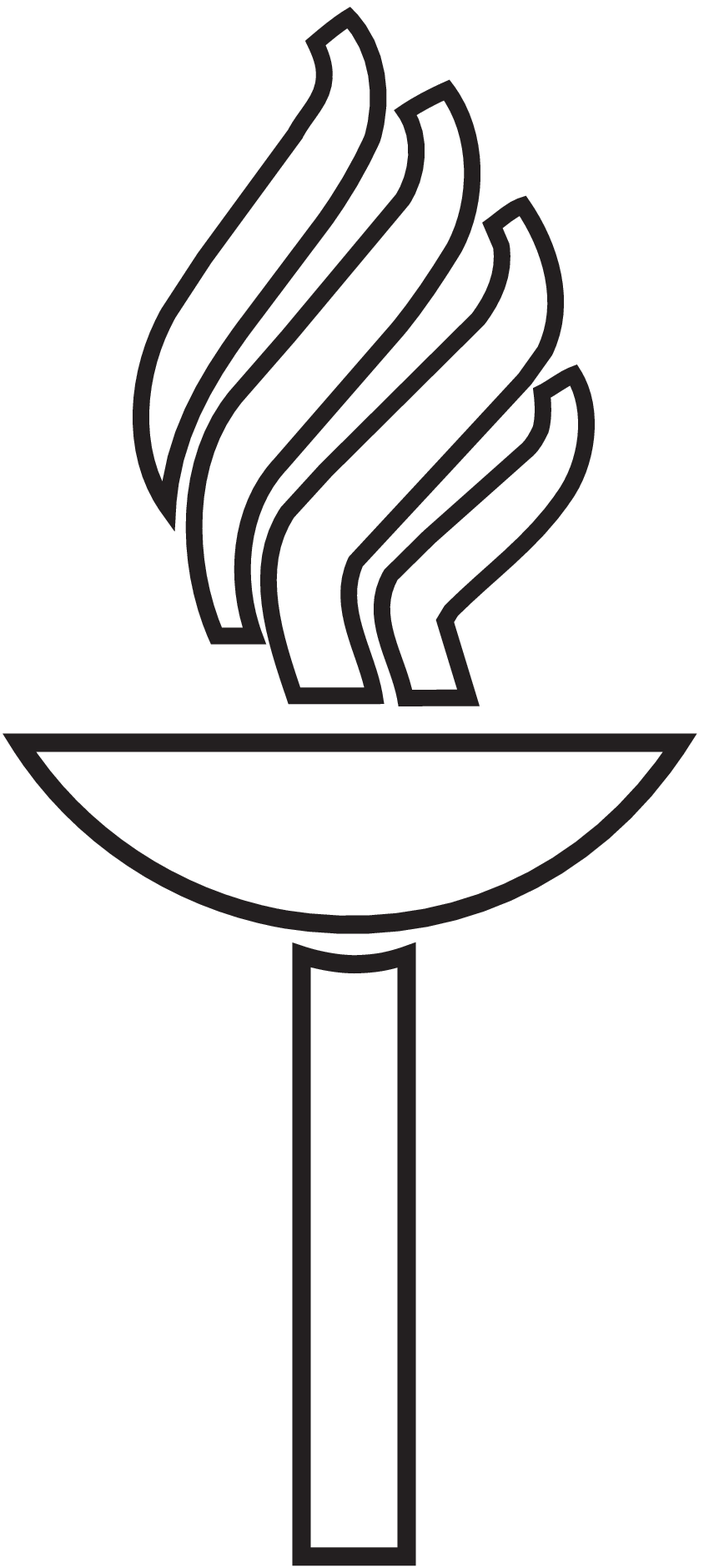}
\end{figure}

%\centerline{\picture{19mm}{50mm}{bwteksti.EPS}}

\centerline{Jyv\" askyl\" a, Finland}
\centerline{July 2014}

\pagebreak

%\newpage
%\phantom{Tyhjaa}
%\newpage

\pagenumbering{roman}

%\newpage

%\input preface.tex

%\newpage

%
%\newpage
%\phantom{Tyhjaa}
%\newpage

\chapter*{\bf{Abstract}}
%This paper is a slightly modified version of the introductory part of a doctoral dissertation. 
The aim of this thesis is to calculate field-theoretically as rigorously as possible the initial state of partonic matter produced in ultrarelativistic heavy-ion collisions at CERN-LHC and BNL-RHIC colliders. The computed minijet initial conditions are then used in the initialization of the relativistic hydrodynamical modeling of these collisions.

\vspace{0.3cm}

In the theoretical introduction part the computation of parton production cross section at next-to-leading order (NLO) perturbative QCD (pQCD) is discussed. Furthermore, the full analytical calculation for the squared quark-quark scattering matrix element  including the systematic ultraviolet renormalization is presented. Finally, the subtraction method allowing for the cancellation of the infrared and collinear singularities in the partonic QCD cross section at NLO is discussed.

\vspace{0.3cm}

In the more phenomenological part of the thesis the original EKRT model, which combines collinearly factorized  leading-order pQCD minijet production with gluon saturation, is introduced. Next, the minijet production is generalized rigorously to NLO. In particular, a new set of measurement functions is introduced to define the produced infrared- and collinear-safe minijet transverse energy, in terms of which the saturation is now formulated. Finally, the framework is updated with the latest knowledge of nuclear parton distribution functions.

\vspace{0.3cm}

Using the NLO-improved EKRT model with hydrodynamics we obtained a good agreement with the measured centrality dependence of the low-transverse-momentum bulk observables, simultaneously at the LHC and RHIC. In particular, aiming at a determination of the QCD matter properties from these measurements, we were able to constrain the temperature dependence of the QCD matter shear viscosity, which is an important result.

\chapter*{List of Publications}

This thesis consists of an introductory part and the following
publications:

\begin{itemize}
\item[{\bf I}]
\textbf{Systematics of the charged-hadron $p_T$ spectrum and the nuclear suppression factor in heavy-ion collisions from $\sqrt{s}=200$ GeV to $\sqrt{s}=2.76$ TeV}, \\
T.~Renk, H.~Holopainen, R.~Paatelainen K.~J.~Eskola,\\
Phys. Rev. C84 (2011) 014906, [arXiv:1103.5308 [hep-ph]].

\item[{\bf II}]
\textbf{Multiplicities and $p_T$ spectra in ultrarelativistic heavy ion collisions from a next-to-leading order improved perturbative QCD + saturation + hydrodynamics model},\\
R.~Paatelainen, K.~J.~Eskola, H.~Holopainen, K.~Tuominen,\\
Phys. Rev. C87 (2013) 044904, [arXiv:1211.0461 [hep-ph]].

\item[{\bf III}]
\textbf{Fluid dynamics with saturated minijet initial conditions in ultrarelativistic heavy-ion collisions},\\
R.~Paatelainen, K.~J.~Eskola, H.~N.~Niemi, K.~Tuominen, 
Phys. Lett. B731 (2014) 126-130, [arXiv:1310.3105 [hep-ph]].

\end{itemize}

The author has written from scratch the necessary numerical programs to calculate the leading-order minijet initial conditions for hydrodynamical evolution in [I]. He participated also in the planning and writing of the first paper.

\vspace{0.3cm}

The author has been in a significant role in developing the new theoretical features of the improved next-to-leading order minijet calculation for the second [II] and third [III] publication. As documented in this thesis, the author clarified the field-theoretical background of the NLO minijet cross sections used in \cite{PHD2} and \cite{PHD3}. The author also implemented all the NLO improvements into the group's original NLO minijet program, developed the code essentially further, and performed all the numerical work for these minijet calculations. The author also wrote the original draft versions for both of these publications.

\pagenumbering{arabic}
\tableofcontents
\pagestyle{plain}
\chapter{Introduction}
\label{intro}

Quantum Chromodynamics (QCD) is a renormalizable field theory that describes the strong interactions between quarks and gluons (partons), and in particular how they bind together to form hadrons. QCD predicts \cite{QCDPRED} that at high temperature and high energy density there will be a transition from hadronic matter to a plasma of deconfined quarks and gluons called quark-gluon plasma (QGP). 

\vspace{0.25cm}

One of the main goals of ultrarelativistic heavy-ion collisions (URHIC) at the Large Hadron Collider (LHC) and the Relativistic Heavy-Ion Collider (RHIC) is to study the thermodynamic and kinetic properties of strongly interacting matter under extreme conditions of high energy density.  When the two nuclei collide a system of particles which are mainly partons is produced. The partonic system then starts to expand and thermalize via reinteractions. If this thermalization or at least near-thermalization takes place quickly enough, a thermodynamically describable QGP is formed \cite{QGP1,QGP2}. Such a collective strongly interacting system then evolves, expanding and cooling down, going through the QCD phase transition back to hadronic matter which then eventually decouples to observable final-state particles.

\vspace{0.25cm}

Unfortunately, it is impossible to observe the QGP directly. Thus, we have to reconstruct its properties from final-state observables, like transverse momentum spectra of the produced hadrons. Consequently, it is then extremely important to have a good control over the initial conditions of the produced system. If this goal is achieved, one can use for example relativistic hydrodynamics to describe the further evolution of the produced system \cite{HYDROQGP1,HYDROQGP2,HYDROQGP3} and compute the final-state observables to be compared with experimental data.

\vspace{0.25cm}

Thanks to the dominance of partonic processes in the initial particle production at collider energies, perturbative QCD (pQCD) makes it possible to compute the properties of the initial state of partonic matter, which can be used as initial condition for the further hydrodynamical evolution \cite{EKRT,LAPPI}. How to compute these initial conditions is the question I will discuss in this thesis. 

\vspace{0.25cm}

This thesis	consists of two parts, the separate introduction part and the published three articles \cite{PHD1,PHD2,PHD3}. The computation of differential $(2\rightarrow 2)$ and $(2\rightarrow 3)$ partonic cross sections at next-to-leading order (NLO) pQCD is discussed in chapter 2. The full calculation of dimensionally regularized and ultraviolet renormalized virtual corrections for the $(2\rightarrow 2)$ quark-quark parton scattering process at NLO is shown in chapter 3. Related to this, in Appedix C constituting an important part of this thesis, I have presented the calculational tools for performing such a tedious NLO calculation.
In chapter 4, I describe how to compute the infrared and collinear safe physical cross section from partonic scattering processes at NLO. In the more phenomenological part of this thesis, the initial state calculations using the original EKRT model \cite{EKRT} and the NLO-improved EKRT model \cite{PHD2,PHD3} are briefly summarized in chapters 5 and 6, respectively. The hydrodynamical equations are presented in chapter 7 and the main results of this thesis are discussed in chapter 8 . Finally, conclusions and outlook are given in chapter 9.

\chapter{Calculation of partonic cross sections at NLO}
\label{partonmodel}

In this chapter I will set up the stage required for the computation of differential parton production cross sections at hadron level in the framework of collinear factorization and NLO perturbative QCD. In particular, I will concentrate on the $(2\rightarrow 2)$ and $(2\rightarrow 3)$ hard scattering sub-processes:
\begin{equation}
\begin{split}
(2\rightarrow 2):\quad a_A + a_B & \rightarrow a_1 + a_2,\\
(2\rightarrow 3):\quad a_A + a_B & \rightarrow a_1 + a_2 +  a_3,
\end{split}
\end{equation}
where a parton of type $a_A$ from hadron $A$ scatters of a parton of type $a_B$ from hadron $B$, yielding partons $a_1, a_2$ and $a_3$.

\section{Partonic cross section}

The calculation of the parton production cross section in high-energy hadron collisions relies on the collinear factorization theorem in QCD \cite{COLFAC}. In this approach a generic hadron-level cross section is given in terms of perturbatively computable pieces (sub-cross sections in leading order (LO)) associated with scattering of gluons $(g)$, quarks $(q)$ and anti-quarks $(\bar{q})$, which are convoluted with parton distribution functions (PDFs) that describe the parton content of the hadrons. Thus,
\begin{equation}
\label{eq:pmodel}
\begin{split}
\sigma(P_A,P_B) = \sum_{\substack{a_A,a_B  \\ \in~g,q,\bar{q}}}\int {\rm d}x_A{\rm d}x_B &f_{A}(a_A,x_A,\mu_F)f_{B}(a_B,x_B,\mu_F)\\
& \times \hat{\sigma}_{a_Aa_B}(p_A,p_B,\mu_F,\mu_R,\alpha_s(\mu_R),Q),
\end{split}
\end{equation}
where $P_A(P_B)$ are the momenta of the hadrons $A(B)$, the quantities $x_{A(B)}$ are the longitudinal momentum fractions of the incoming partons $a_{A(B)}$, and the momenta of the partons which participate in the hard interaction are $p_A = x_AP_A$ and $p_B=x_BP_B$. The functions $f_{A(B)}$ are the PDFs which (in the lowest-order approximation at least) correspond to the probability density to find a parton of a flavor $a_{A(B)}$ in the hadron $A(B)$ with a momentum fraction $x_{A(B)}$. These inherently non-perturbative functions can be determined indirectly from experiments measuring hard processes such as deeply inelastic scattering \cite{DIS} or the Drell-Yan \cite{DY} process. The quantities $ \hat{\sigma}_{a_Aa_B}$ are the perturbative pieces which can be expressed as a fixed-order series expansion in the strong QCD coupling constant $\alpha_s$ as
\begin{equation}
\hat{\sigma}_{a_Aa_B} = \alpha_s^2 \hat{\sigma}^{(0)}_{a_Aa_B} + \alpha_s^3\hat{\sigma}^{(1)}_{a_Aa_B} + \mathcal{O}(\alpha_s^4),
\end{equation}
where the superscripts (0) and (1) denote the leading-order and next-to-leading order contributions, respectively. The characteristic scale of the hard scattering is denoted by $Q$. In field theoretical calculations, one often sets the renormalization $\mu_R$ and factorization $\mu_F$ scales to be equal, $\mu_F = \mu_R=\mu$, with $\mu$ of the order of $Q$. In practice, the more terms are included in the perturbative expansion, the weaker the dependence of the cross section on $\mu$ is.

\vspace{0.3cm}

In the following sections \ref{sec22} - \ref{sec25} I discuss the computation of parton production $(2\rightarrow 2)$ and $(2\rightarrow 3)$ cross sections in some detail up to NLO. In practice, I will explain how to formulate the differential 2- and 3-parton production cross sections at hadron level by using the standard dimensional regularization approach \cite{DIMREG} in $d=4-2\epsilon$ dimensions. This discussion closely follows \cite{KUSO}, keeping the same notation as in the original paper.

\section{Kinematics and phase space}
\label{sec22}

%In this section the two-parton kinematics as a function of the jet variables are introduced.

Natural variables for the analysis of two-parton interactions in hadron-hadron collisions are $y$, $p_{\perp}$ and $\phi$, where the transverse momentum $\mathbf{p}_{\perp}=(p_x,p_y) = (p_{\perp}\cos\phi,p_{\perp}\sin\phi)$ with $p_{\perp} \equiv \vert \mathbf{p}_{\perp}\vert$, azimuthal angle $\phi$ and the rapidity
\begin{equation}
\label{eq:rapidity}
y = \frac{1}{2}\ln\left (\frac{E+p_z}{E-p_z}\right ) = \tanh^{-1}\left (\frac{p_z}{E}\right ).
\end{equation}
Here $E$ and $p_z$ are, respectively, the energy and longitudinal momentum in the hadron-hadron center-of-mass (CMS) frame. Assuming that the partons are massless the energy can be written as $E = \sqrt{p_{\perp}^2 + p_z^2}$. Using Eq.\ \eqref{eq:rapidity}, the particle energy and longitudinal momentum can be rewritten in terms of $y$ and $p_{\perp}$ as
\begin{equation}
\label{eq:EandpT}
\begin{split}
E  = p_{\perp}\cosh(y), \quad\quad  p_z  = p_{\perp}\sinh(y). 
\end{split}
\end{equation}
It is convenient to use the light-cone coordinates in which four-vectors are given by components $p^{\mu}=(p^+,p^-,\mathbf{p}_{\perp})$ with
\begin{equation}
\label{eq:LCC}
p^{\pm} = \frac{E\pm p_z}{\sqrt{2}}.
\end{equation}
In the light-cone coordinates the scalar product of four-vectors is given with the convention above by  
\begin{equation}
p\cdot k = p^+k^- + p^-k^+ - \mathbf{p}_{\perp}\cdot \mathbf{k}_{\perp}.
\end{equation}
Using Eqs.\ \eqref{eq:LCC} and \eqref{eq:EandpT}, the particle four-momenta can be expressed in terms of transverse momentum and rapidity,
\begin{equation}
\label{eq:4mom}
p^{\mu} = (\frac{p_{\perp}}{\sqrt{2}}e^{y}, \frac{p_{\perp}}{\sqrt{2}}e^{-y}, \mathbf{p}_{\perp}).
\end{equation}
The $(d-1)$-dimensional Lorentz invariant particle phase space element in terms of the particle momentum and energy is given by 
\begin{equation}
\frac{{\rm d}^{d-1}\mathbf{p}}{E} = \frac{{\rm d}p_z}{E}{\rm d}p_{\perp 1}\dots {\rm d}p_{\perp (d-2)}= \frac{{\rm d}p_z}{E}{\rm d}^{d-2}\mathbf{p}_{\perp}.
\end{equation}
In the coordinate system $(y,p_{\perp},\phi)$,
\begin{equation}
\label{eq:ztoE}
{\rm d}y=\frac{{\rm d}p_z}{p_{\perp}\cosh(y)} \overset{\eqref{eq:EandpT}}{=} \frac{{\rm d}p_z}{E},
\end{equation}
and
\begin{equation}
\label{eq:xytoptrans}
{\rm d}^{d-2}\mathbf{p}_{\perp} = p^{d-3}_{\perp}{\rm d}p_{\perp}{\rm d}^{d-3}\phi,
\end{equation}
where the factor ${\rm d}^{d-3}\phi$, which contains all of the angular parts, takes care of the integration in a $(d-3)$-dimensional sphere. Thus, the invariant particle phase space element becomes 
\begin{equation}
\label{eq:phasespace}
\frac{{\rm d}^{d-1}\mathbf{p}}{E} = {\rm d}y{\rm d}p_{\perp}p^{d-3}_{\perp}{\rm d}^{d-3}\phi.
\end{equation}

\section{The $(2\rightarrow 2)$ partonic cross section}

Let us now consider the $(2\rightarrow 2)$ scattering of partons,
\begin{equation}
\label{eq:2to2basicpro}
a_A(p_A) + a_B(p_B) \rightarrow a_1(p_1) + a_2(p_2),
\end{equation} 
where the two incoming partons, $a_A$ and $a_B$, which originate from hadrons $A$ and $B$, respectively, scatter into two other partons $a_1$ and $a_2$. In the hadronic CMS frame the four-momenta of the incoming partons in the light-cone coordinates can be expressed in terms of the longitudinal momentum fraction variables as
\begin{equation}
\label{eq:incomingp}
p^{\mu}_A = (x_A\sqrt{\frac{s}{2}},0,{\bf 0}),\quad\quad \quad p^{\mu}_B = (0,x_B\sqrt{\frac{s}{2}},{\bf 0}).
\end{equation}
Using Eq.\ \eqref{eq:4mom}, we may write the outgoing parton four-momenta as 
\begin{equation}
\begin{split}
p^{\mu}_1 & = (\frac{p_{\perp,2}}{\sqrt{2}}e^{y_1},\frac{p_{\perp,2}}{\sqrt{2}}e^{-y_1},-\mathbf{p}_{\perp,2}),\\
p^{\mu}_2 & = (\frac{p_{\perp,2}}{\sqrt{2}}e^{y_2},\frac{p_{\perp,2}}{\sqrt{2}}e^{-y_2},\mathbf{p}_{\perp,2}),
\end{split}
\end{equation}
where we have used the fact, which follows from transverse-momentum conservation, that $\mathbf{p}_{\perp,2}=-\mathbf{p}_{\perp,1}$. The light-cone momentum conservation $p_A^+ = p_1^+ + p_2^+$ and $p_B^- = p_1^- + p_2^-$ fixes the momentum fractions of the incoming partons as a function of the final-state parton variables as 
\begin{equation}
\begin{split}
x_A & = \frac{p_{\perp,2}}{\sqrt{s}}\left (e^{y_1} + e^{y_2}\right ),\\
x_B & = \frac{p_{\perp,2}}{\sqrt{s}}\left (e^{-y_1} + e^{-y_2}\right ).\\
\end{split}
\end{equation}
We can now write the $(2\rightarrow 2)$ partonic cross section  in $d$ dimensions as \cite{KUSO}
\begin{equation}
\label{eq:2to2cs}
\begin{split}
{\rm d}\hat{\sigma}(2\rightarrow 2) = \frac{\mu_0^{4-d}}{2\hat s}\frac{{\rm d}^{d-1}{\bf p}_1}{(2\pi)^{d-1}2E_1}&\frac{{\rm d}^{d-1}{\bf p}_2}{(2\pi)^{d-1}2E_2} \langle \vert \mathcal{M}(a_Aa_B\rightarrow a_1a_2)\vert^2\rangle\\
&(2\pi)^{d}\delta^{(d)}(p_1^{\mu}+p_2^{\mu}-p_A^{\mu}-p_B^{\mu}),
\end{split}
\end{equation}
where $\hat{s} \equiv (p_A + p_B)^2=x_Ax_Bs$ and the standard parameter $\mu_0$ has a dimension of mass to keep the QCD (bare) coupling $g$ dimensionless.  Using Eq.\ \eqref{eq:phasespace}, we obtain for the outgoing partons $i=1,2$ the Lorentz invariant phase space elements in $d=4-2\epsilon$ dimensions as
\begin{equation}
\frac{{\rm d}^{3-2\epsilon}\mathbf{p}_i}{(2\pi)^{3-2\epsilon}2E_i} = \frac{{\rm d}y_i{\rm d}p_{\perp,i}{\rm d}^{1-2\epsilon}\phi_i}{2(2\pi)^{3-2\epsilon}}p_{\perp,i}^{1-2\epsilon}.
%\overset{\eqref{eq:ztoE}}{=} \frac{{\rm d}y_i{\rm d}^{2-2\epsilon}\mathbf{p}_{\perp,i}}{2(2\pi)^{3-2\epsilon}} \overset{\eqref{eq:xytoptrans}}{=} 
\end{equation}
The invariant matrix elements (scattering amplitudes) squared for the $(2\rightarrow 2)$ partonic processes of Eq.\ \eqref{eq:2to2basicpro}, summed over the final spins and colors and averaged over the initial spins and colors, are given by 
\begin{equation}
\label{eq:2to2melement}
\langle \vert \mathcal{M}(a_Aa_B\rightarrow a_1a_2)\vert^2\rangle =\frac{1}{\omega(a_A)\omega(a_B)}\sum_{\mathclap{\substack{\rm{color}\\ \rm{spin}}}}\vert\mathcal{M}(a_Aa_B\rightarrow a_1a_2)\vert^2,
\end{equation}
%\footnote{Note that when averaging over the gluon polarization states in $d=4-2\epsilon$ dimensions there are $2-2\epsilon$ gluon states instead of usual 2}
where the factors $\omega(a)$ stand for the number of possible spin and color states of a parton of type $a$. In particular, we note that in $d=4-2\epsilon$ dimensions
\begin{equation}
\omega(a) =
\begin{cases}
2(1-\epsilon)V_g \quad \text{for} \quad a=g, \\
2N_c \quad \text{for} \quad a=q,\bar{q},
\end{cases}
\end{equation}
where $N_c=3$ and $V_g=N_c^2-1=8$.

\vspace{0.3cm}

Next, rewriting the delta function in Eq.\ \eqref{eq:2to2cs} as 
\begin{equation}
\begin{split}
\delta^{(4-2\epsilon)}(p_1^{\mu}+p_2^{\mu} -p_A^{\mu}-p_B^{\mu}) = \delta^{(2-2\epsilon)}(\mathbf{p}_{\perp,1}&+\mathbf{p}_{\perp,2})\delta(p_1^{+} + p_2^{+} - p_A^{+})\\
&\times\delta(p_1^{-} + p_2^{-} - p_B^{-}),
\end{split}
\end{equation}
and performing the $\mathbf{p}_{\perp,1}$ integration by using the delta function for the conservation of transverse momentum, we obtain the differential 2-parton production cross section
\begin{equation}
\label{eq:int}
\begin{split}
\frac{{\rm d}\hat{\sigma}(2\rightarrow 2)}{{\rm d[PS]}_2} = &\frac{p_{\perp,2}}{8(2\pi)^2x_Ax_Bs}\left (\frac{p_{\perp,2}}{2\pi\mu_0}\right )^{-2\epsilon}\langle \vert \mathcal{M}(a_Aa_B\rightarrow a_1a_2)\vert^2\rangle \\
&\delta(p_1^{+} + p_2^{+} - p_A^{+}-p_B^{+})\delta(p_1^{-} + p_2^{-} - p_A^{-}-p_B^{-}),
\end{split}
\end{equation}
where we have introduced a compact notation for the $p_{\perp,2}$-reduced 2-parton phase space volume element,
\begin{equation}
{\rm d[PS]}_2 = {\rm d}p_{\perp,2}{\rm d}^{1-2\epsilon}\phi_2{\rm d}y_1{\rm d}y_2.
\end{equation}
Then, at the hadron level the differential 2-parton cross section can be obtained using collinear factorization,
\begin{equation}
\label{eq:hhcstwo}
\begin{split}
\frac{{\rm d}\sigma(2\rightarrow 2)}{{\rm d[PS]}_2}=\sum_{a_A,a_B}\int_{0}^{1}{\rm d}x_A{\rm d}x_B  f_{A,0}(a_A,x_A)f_{B,0}(a_B,x_B)\frac{{\rm d}\hat{\sigma}(2\rightarrow 2)}{{\rm d[PS]}_2},
\end{split}
\end{equation} 
where the functions $f_{A,0}(a_A,x_A)$ and $f_{B,0}(a_B,x_B)$ are the bare ("0") parton distribution functions. Finally, performing the $x_A$ and $x_B$ integrals in Eq.\ \eqref{eq:hhcstwo}, we arrive at our starting formula for inclusive 2-parton production cross section at the hadron level,
\begin{equation}
\label{eq:2to2csfinal}
\begin{split}
\frac{{\rm d}\sigma(2\rightarrow 2)}{{\rm d[PS]}_2}=\sum_{a_A,a_B}&\frac{p_{\perp,2}}{16\pi^2s^2}\left (\frac{p_{\perp,2}}{2\pi\mu_0}\right )^{-2\epsilon}\frac{f_{A,0}(a_A,x_A)}{x_A}\\
&\frac{f_{B,0}(a_B,x_B)}{x_B}\langle \vert \mathcal{M}(a_Aa_B\rightarrow a_1a_2)\vert^2\rangle.
\end{split}
\end{equation}

\section{The $(2\rightarrow 3)$ partonic cross section}

Let us next consider the $(2\rightarrow 3)$ scattering of partons,
\begin{equation}
a_A(p_A) + a_B(p_B) \rightarrow a_1(p_1) + a_2(p_2) + a_3(p_3),
\end{equation} 
where the two incoming partons, $a_A$ and $a_B$, from hadrons $A$ and $B$, respectively, scatter to three other partons $a_1, a_2$ and $a_3$. The four-momenta of the incoming partons are given by Eq.\ \eqref{eq:incomingp} and for the three outgoing partons we can write in the light-cone coordinates
\begin{equation}
p_k^{\mu} = (\frac{p_{\perp,k}}{\sqrt{2}}e^{y_k}, \frac{p_{\perp,k}}{\sqrt{2}}e^{-y_k}, \mathbf{p}_{\perp,k}),
\end{equation}
where $k=1,2,3$. The light-cone momentum conservation, $p_A^+ = p_1^+ +p_2^+ + p_3^+$ and $p_B^- = p_1^- + p_2^- + p_3^-$, again relates the momentum fractions $x_A$ and $x_B$ to the final-state parton rapidities and transverse momenta as
\begin{equation}
\begin{split}
x_A &= \frac{1}{\sqrt{s}}\sum_{k=1}^{3}p_{\perp,k}e^{y_k},\\
x_B &= \frac{1}{\sqrt{s}}\sum_{k=1}^{3}p_{\perp,k}e^{-y_k}.
\end{split}
\end{equation}
Using the steps described in the previous section we can write the $(2\rightarrow 3)$ partonic cross section in $d=4-2\epsilon$ dimensions as
\begin{equation}
\label{eq:pplevel2to3f}
\begin{split}
{\rm d}\hat{\sigma}(2\rightarrow 3)=&\frac{\mu^{4\epsilon}_0}{2\hat{s}}\prod_{i=1}^{3}
\frac{{\rm d}y_i{\rm d}^{2-2\epsilon}\mathbf{p}_{\perp,i}}{8(2\pi)^{5-4\epsilon}}
\langle \vert \mathcal{M}(a_Aa_B\rightarrow a_1a_2a_3)\vert^2\rangle\\
&\times \delta^{(2-2\epsilon)}(\mathbf{p}_{\perp,1}+ \mathbf{p}_{\perp,2}+ \mathbf{p}_{\perp,3})\delta(p_1^{+} + p_2^{+} + p_3^{+}- p_A^{+})\\
&\times \delta(p_1^{-} + p_2^{-} + p_3^{-}- p_B^{-}),
\end{split}
\end{equation}
where the matrix element squared, summed over the final spins and colors and averaged over the initial spins and colors, has the same form as in Eq.\ \eqref{eq:2to2melement},
\begin{equation}
\langle \vert \mathcal{M}(a_Aa_B\rightarrow a_1a_2a_3)\vert^2\rangle =\frac{1}{\omega(a_A)\omega(a_B)}\sum_{\substack{\rm{color}\\ \rm{spin}}} \vert\mathcal{M}(a_Aa_B\rightarrow a_1a_2a_3)\vert^2.
\end{equation}
Performing the $\mathbf{p}_{\perp,1}$ integration in Eq.\ \eqref{eq:pplevel2to3f} by using the delta function for the transverse momentum conservation, we obtain 
\begin{equation}
\begin{split}
\frac{{\rm d}\hat{\sigma}(2\rightarrow 3)}{{\rm d[PS]}_3} = &\frac{p_{\perp,2}p_{\perp,3}}{16(2\pi)^5\hat{s}}\left (\frac{p_{\perp,2}p_{\perp,3}}{(2\pi)^2\mu^2_0}\right )^{-2\epsilon}\langle \vert \mathcal{M}(a_Aa_B\rightarrow a_1a_2a_3)\vert^2\rangle\\
&\delta(p_1^+ + p_2^+ + p_3^+ -p_A^+)\delta(p_1^- + p_2^- + p_3^- - p_B^-),
\end{split}
\end{equation}
where the $p_{\perp,2}p_{\perp,3}$-reduced 3-particle phase space element is
\begin{equation}
{\rm d[PS]}_3 = {\rm d}p_{\perp,2}{\rm d}p_{\perp,3}{\rm d}^{1-2\epsilon}\phi_2{\rm d}^{1-2\epsilon}\phi_3{\rm d}y_3{\rm d}y_2{\rm d}y_1.
\end{equation}
At the hadron level, the invariant $(2\rightarrow 3)$ differential 3-parton production cross section can be written, using again collinear factorization, as 
\begin{equation}
\label{eq:hhcs}
\begin{split}
\frac{{\rm d}\sigma(2\rightarrow 3)}{{\rm d[PS]}_3}=\sum_{a_A,a_B}\int_{0}^{1}{\rm d}x_A{\rm d}x_B  f_{A,0}(a_A,x_A)f_{B,0}(a_B,x_B)
\frac{{\rm d}\hat{\sigma}(2\rightarrow 3)}{{\rm d[PS]}_3}.
\end{split}
\end{equation}
Performing the $x_A$ and $x_B$ integrals in Eq.\ \eqref{eq:hhcs} as before, we obtain 
\begin{equation}
\label{eq:2to3csfinal}
\begin{split}
\frac{{\rm d}\sigma(2\rightarrow 3)}{{\rm d[PS]}_3}=\sum_{a_A,a_B}&\frac{p_{\perp,2}p_{\perp,3}}{8(2\pi)^5s^2}\left (\frac{p_{\perp,2}p_{\perp,3}}{(2\pi)^2\mu^2_0}\right )^{-2\epsilon}\frac{f_{A,0}(a_A,x_A)}{x_A}\\
&\frac{f_{B,0}(a_B,x_B)}{x_B}\langle \vert \mathcal{M}(a_Aa_B\rightarrow a_1a_2a_3)\vert^2\rangle.
\end{split}
\end{equation}
This is our starting form for the inclusive differential 3-parton production cross section at the hadron level.

\section{Scattering processes for the partonic cross sections at NLO}
\label{sec25}

The partonic $(2\rightarrow 2)$ and $(2\rightarrow 3)$ matrix elements squared up to NLO, \textrm{i.e.} $\mathcal{O}(\alpha_s^3)$, include several pieces: The Born-level (LO) squared matrix elements for the $(2\rightarrow 2)$ gluon, quark and anti-quark scattering processes can be obtained by crossing from the four basic quark and gluon scatterings,
\begin{equation}
\begin{split}
qq'\rightarrow qq', \quad qq\rightarrow qq , \quad qg\rightarrow qg, \quad gg\rightarrow gg.   
\end{split}
\end{equation}
In order to get the NLO corrections for these processes we should also consider the virtual contributions, which are described by the additional internal exchange of particles. Thus, for the $(2\rightarrow 2)$ processes in NLO there is the same number of incoming and outgoing partons as in the Born level. In practice, for the $(2\rightarrow 2)$ parton processes considered here the $\mathcal{O}(\alpha_s^3)$ virtual contributions arise from the interference of the one-loop corrected $(2\rightarrow 2)$ matrix element with the Born level matrix element. These interference contributions may present collinear\footnote{Note that we only consider massless partons here.} (CL), soft (infrared, IR) and ultraviolet (UV) divergences. The most sophisticated gauge-invariant way to regulate these singularities is to use dimensional regularization \cite{DIMREG}. In this approach the divergences are dealt with going into $d=4-2\epsilon$ dimensions, where the singularities appear as single pole $1/\epsilon$ or double pole 
$1/\epsilon^2$ forms. However, after all the ultraviolet divergences are removed by the renormalization procedure, typically performed in the $\overline{\textrm{MS}}$ scheme, only the IR and CL sigularities are left. To understand this in detail is the main goal of the theory part of this thesis.

\vspace{0.3cm}

In addition, a full $\mathcal{O}(\alpha_s^3)$ calculation includes UV finite contributions from the $(2\rightarrow 3)$ processes, where an extra real gluon is emitted. Like in the $(2\rightarrow 2)$ case, all $(2\rightarrow 3)$ processes can be again derived from the four basic scatterings:
\begin{equation}
\begin{split}
qq'\rightarrow qq'g, \quad qq\rightarrow qqg, \quad qg\rightarrow qgg, \quad gg\rightarrow ggg,  
\end{split}
\end{equation}
after a proper crossing procedure. Also these processes present IR and CL singularities after their squared matrix elements are integrated over the 3-parton phase space. At the squared amplitude level, the computation of these $(2\rightarrow 3)$ processes is, however, quite straightforward since there are no loops in the Feynman diagrams (\textrm{i.e.} no UV, IR and CL poles originating from the loop-momentum integrations). Thus, we are not discussing the computation of these processes further in this thesis.

\vspace{0.3cm}

The ultraviolet renormalized squared matrix elements for these different pieces were computed in $d=4-2\epsilon$ dimensions first by R.~K.~Ellis and Sexton \cite{ELSE}. In chapter 3, I will demostrate in detail how to compute the $\overline{\textrm{MS}}$ renormalized virtual corrections to the $qq'\rightarrow qq'$ scattering process. All the other $(2\rightarrow 2)$ virtual corrections can be calculated similarly.

\vspace{0.3cm}

Finally, after the UV renormalization the remaining IR and CL singularities should be cancelled between the $(2\rightarrow 2)$ and $(2\rightarrow 3)$ parts. After this, one can calculate the physical and finite $\mathcal{O}(\alpha_s^3)$ corrections for example to jet production \cite{KUSO} cross sections or, as discussed in this thesis, to minijet transverse energy production cross sections \cite{NLOET1}. How to cancel these singularities and how to compute the physical NLO cross sections is discussed in more detail in chapter \ref{EKS}.

\chapter{One-loop virtual corrections to $qq'\rightarrow qq'$ scattering process}
\label{VIRCOR}

In this chapter I present the full calculation of dimensionally regularized and ultraviolet renormalized virtual corrections for the $qq'\rightarrow qq'$ parton scattering process at NLO. The rather complicated final answer is given in the original article \cite{ELSE}. However, as this article does not present the intermediate steps and to the best of my knowledge they are not presented in the literature, I believe it will make justice to present them here. To understand such a calculation in all details, and to learn the techniques involved, was also a big part of my PhD thesis work. Some of the techniques that I use throughout this chapter, mainly how to compute the QCD scalar and tensor integrals in a very efficient way, are shown in Appendix C. Thus, also this Appedix is a very important part of the calculation shown here.

\vspace{0.3cm}

The algebraic complications due to the traces of $\gamma$ matrices, substitutions of Mandelstam variables, and
reduction of loop-integrals to form factors were treated with the help of the Mathematica package
FeynCalc \cite{FEYNCALC}. The results for all relevant Feynman rules of QCD propagators, vertices and QCD color algebra are collected in Appendix A. The relevant rules for the $d$-dimensional spinor algebra are collected in Appendix B. Finally, the computations of all tensoral 2-, 3- and 4-point one-loop integrals are carried out by using the Passarino-Veltman reduction to scalar integrals. This procedure is described in detail in Appendix C.

\section{Definitions and notations}

The process under investigation is the following,
\begin{equation}
\label{eq:theprocess}
q(k_2) + q'(k_1) \rightarrow q(k_3) + q'(k_4),
\end{equation}
where the momentum assignments for the quarks are given in the brackets. All quarks are assumed to be massless, $k_i^2=0$. It is convenient to express all the scalar products of the momenta in terms of the (Lorentz invariant) Mandelstam variables\footnote{Note that we drop the hats in the Mandelstam variables. Note also the order of $k_1$ and $k_2$ here.} defined by
\begin{equation}
\begin{split}
s & = (k_1+k_2)^2 = (k_3+k_4)^2 > 0 ,\\
t & = (k_2-k_3)^2 = (k_1-k_4)^2 \leq 0,\\
u & = (k_1-k_3)^2 = (k_2-k_4)^2 \leq 0.\\
\end{split}
\end{equation} 
Due to the momentum conservation these variables are not independent but fulfill the identity
\begin{equation}
s+t+u = 0.
\end{equation}
The invariant matrix element squared for the process of Eq.\ \eqref{eq:theprocess}, summed and averaged over colors and spins, defines a function $\mathcal{A}(s,t,u)$ as follows,
\begin{equation}
\label{eq:def1}
\langle \vert \mathcal{M}(qq'\rightarrow qq')\vert^2\rangle = \frac{1}{\omega(q)\omega(q')}\mathcal{A}(s,t,u),
\end{equation}
where $\omega(q)=\omega(q')=2N_c$. In terms of the bare (0) dimensional coupling $g$ the function $\mathcal{A}$ has an unrenormalized perturbative expansion which we write as
\begin{equation}
\label{eq:amplitude}
\mathcal{A}_0 = \mathcal{A}^{(\textrm{LO})}_{\textrm{born}}(s,t,u) + \mathcal{A}^{(\textrm{NLO})}_{\textrm{virtual},0}(s,t,u) + \mathcal{O}(g^8),
\end{equation}
where we define 
\begin{equation}
\label{eq:NLOFULLVIR}
\mathcal{A}^{(\textrm{LO})}_{\textrm{born}} = g^4\mathcal{A}^{(\textrm{LO})} \quad \text{and} \quad  \mathcal{A}^{(\textrm{NLO})}_{\textrm{virtual},0} = g^6\mathcal{A}^{(\textrm{NLO})}.
\end{equation}
Applying dimensional regularization the bare coupling $g$ can be replaced by a dimensionless one, $g_0$, by writing $g = g_0\mu_0^{\epsilon}$ (see p.\ 10).

\section{Born level squared amplitude $\mathcal{A}^{(\rm LO)}_{\textrm{born}}$}

\begin{figure}[h!]
\center
\includegraphics[scale=.45]{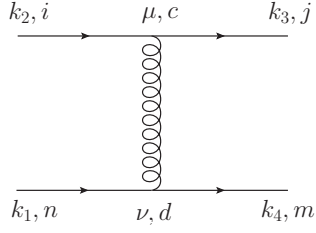}
\caption{Born-level Feynman diagram for $qq'\rightarrow qq'$.}
\label{fig:LOt}
\end{figure}

At the Born level we need to consider only one Feynman diagram in the calculation of $\mathcal{A}^{(\rm LO)}_{\textrm{born}}$. This is depicted in Fig.\ \ref{fig:LOt}. 
By means of the QCD Feynman rules in Appendix \ref{QCDFR}, and using the Feynman gauge, $\xi=1$, the invariant matrix element can be written as
\begin{equation}
i\mathcal{M}_{\rm LO} = \frac{ig_0^2\mu_{0}^{2\epsilon}}{t}(T^c)_{ji}(T^c)_{mn}\biggl [\bar{u}(k_3)\gamma^{\mu}u(k_2) \biggr ]\biggl [\bar{u}(k_4)\gamma_{\mu}u(k_1) \biggr ],
\end{equation}
and the corresponding complex-conjugated matrix element
\begin{equation}
\label{eq:borncopcon}
-i\mathcal{M}_{\rm LO}^{\dagger} = \frac{-ig_0^2\mu_{0}^{2\epsilon}}{t}(T^a)_{ij}(T^a)_{nm}\biggl [\bar{u}(k_2)\gamma^{\nu}u(k_3) \biggr ]\biggl [\bar{u}(k_1)\gamma_{\nu}u(k_4) \biggr ].
\end{equation}
Summing over spins and colors, the squared matrix element becomes
\begin{equation}
\label{eq:LOintres}
\begin{split}
\sum_{\substack{\textrm{spin}\\\textrm{color}}}\vert\mathcal{M}_{\rm LO}\vert^2 = \frac{g_0^4\mu_{0}^{4\epsilon}}{t^2}&\underbrace{\sum_{\textrm{color}}(T^a)_{ij}(T^c)_{ji}(T^a)_{nm}(T^c)_{mn}}_{=~ \mathcal{C}_{\textrm{LO}}}\\
&\times \mathcal{L}^{\mu\nu}_{(q)}(k_3,k_2)\mathcal{L}_{\mu\nu}^{(q)}(k_4,k_1),
\end{split}
\end{equation}
where the quark tensors are defined as
\begin{equation}
\label{eq:qtensor1}
\mathcal{L}^{\mu\nu}_{(q)}(k_3,k_2) = \sum_{\textrm{spin}}\biggl [\bar{u}(k_3)\gamma^{\mu}u(k_2) \biggr ]\biggl [\bar{u}(k_2)\gamma^{\nu}u(k_3) \biggr ],
\end{equation}
and
\begin{equation}
\label{eq:qtensor2}
\mathcal{L}_{\mu\nu}^{(q)}(k_4,k_1) = \sum_{\textrm{spin}}\biggl [\bar{u}(k_4)\gamma_{\mu}u(k_1) \biggr ]\biggl [\bar{u}(k_1)\gamma_{\nu}u(k_4) \biggr ].
\end{equation}
Doing the spin sums using the standard projection operators and solving the quark tensors using Eq.\ \eqref{eq:gammatra}, we find
\begin{equation}
\begin{split}
\mathcal{L}^{\mu\nu}_{(q)}(k_3,k_2)\mathcal{L}_{\mu\nu}^{(q)}(k_4,k_1) &= \textrm{Tr}[\cancel k_3\gamma^{\mu}\cancel k_2\gamma^{\nu}]\textrm{Tr}[\cancel k_4\gamma_{\mu}\cancel k_1\gamma_{\nu}]\\
&= 8t^2\left (\frac{s^2+u^2}{t^2}-\epsilon\right ).
\end{split}
\end{equation}
Furthermore, the sum over the color group generators in Eq.\ \eqref{eq:LOintres} is performed using Eq.\ \eqref{eq:COLORR1}, 
\begin{equation}
\begin{split}
\mathcal{C_{\textrm{LO}}} & = \sum_{\textrm{color}}(T^a)_{ij}(T^c)_{ji}(T^a)_{nm}(T^c)_{mn}\\
& = \sum_{\textrm{color}}\mathrm{Tr}(T^aT^c)\underbrace{\mathrm{Tr}(T^aT^c)}_{=~\frac{1}{2}\delta^{ac}} = \sum_{\textrm{color}} \frac{\delta^{aa}}{4} = 2.
\end{split}
\end{equation}
Thus, we obtain for the Born level squared amplitude 
\begin{equation}
\mathcal{A}^{(\rm LO)}_{\textrm{born}} = \sum_{\substack{\textrm{spin}\\\textrm{color}}}\vert\mathcal{M}_{\rm LO}\vert^2  = g_0^4\mu_{0}^{4\epsilon}\mathcal{A}^{(\rm LO)}(s,t,u), 
\end{equation}
where
\begin{equation}
\label{eq:LOcont}
\mathcal{A}^{(\rm LO)}(s,t,u) = 16\left (\frac{s^2+u^2}{t^2}-\epsilon \right ).
\end{equation}

\section{NLO virtual corrections}

\begin{figure}[h!]
\center
\includegraphics[scale=.70]{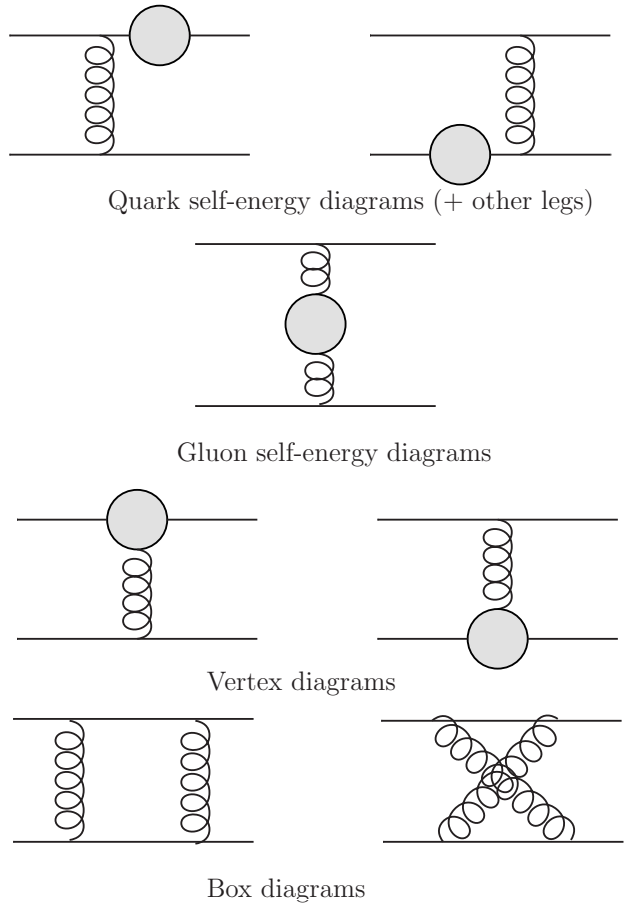}
\caption{Different diagram topologies that arise from the $\mathcal{O}(g^4)$ one-loop QCD virtual corrections.}
\label{fig:virtkuva}
\end{figure}

The QCD virtual correction to the unrenormalized NLO squared amplitude, $\mathcal{A}^{(\textrm{NLO})}_{\textrm{virtual,0}}$, takes into account the interference of the NLO $\mathcal{O}(g^4)$ Feynman diagrams in Fig. \ref{fig:virtkuva} with the Born-level $\mathcal{O}(g^2)$ diagram in Fig.\ \ref{fig:LOt}. 

In practice, we have to calculate the following $\mathcal{O}(g^6)$ terms
\begin{equation}
\begin{split}
 \vert \mathcal{M}\vert^2 & = \vert \mathcal{M}_{\rm LO} + \mathcal{M}_{\rm qSE} + \mathcal{M}_{\rm gSE} + 2\mathcal{M}_{\rm V} + \mathcal{M}_{\rm BOX} \vert^2 + \mathcal{O}(g^8)\\
& = \vert \mathcal{M}_{\rm LO}\vert^2 + \underbrace{2\mathbb{R}e\biggl [\left (\mathcal{M}_{\rm qSE} + \mathcal{M}_{\rm gSE} + 2\mathcal{M}_{\rm V} + \mathcal{M}_{\rm BOX} \right )\mathcal{M}_{\rm LO}^{\dagger}\biggr ]}_{=~ \mathcal{O}(g^6)} + \mathcal{O}(g^8) ,
\end{split}
\end{equation}
where $\mathcal{M}_{\rm qSE}, \mathcal{M}_{\rm gSE}, \mathcal{M}_{\rm V}$ and $\mathcal{M}_{\rm BOX}$ are the $\mathcal{O}(g^4)$ one-loop corrected quark self-energy (qSE), gluon self-energy (gSE), gluon vertex (V), and gluon box (BOX) matrix amplitudes, respectively. Note that we include the factor 2 for the $\mathcal{M}_{\rm V}$, since the two different vertex diagrams shown in Fig.\ \ref{fig:virtkuva} give an identical final answer. Following Eq.\ \eqref{eq:def1}, we define the individual unrenormalized NLO squared amplitude contributions as 
\begin{equation}
\label{eq:SAMPLITUDE}
\mathcal{A}^{(\rm NLO)}_X = 2(1+\delta_{XV})\sum_{\substack{\textrm{spin}\\\textrm{color}}} \mathbb{R}e\left (\mathcal{M}_X\mathcal{M}_{\rm LO}^{\dagger}\right ), ~X=\{\textrm{qSE,gSE,V,BOX}\},
\end{equation}
where $\delta_{XV}=1$ for $X=V$ and otherwise zero. 

\vspace{0.3cm}

Furthermore, since quarks are massless and in the dimensional regularization approach we take the same $\epsilon$ to regularize the UV and IR/CL singularies, all the one-loop diagrams which correct the incoming or outgoing  quark leg (see Fig.\ \ref{fig:virtkuva}) can be set to zero \cite{MUTA} (see also the discussion in Appedix \ref{eq:A0int}). Therefore, the NLO corrections which arise from the one-loop quark self-energy diagrams are directly zero, and we can set $\mathcal{A}^{(\rm NLO)}_{\rm qSE} = 0$.

\vspace{0.3cm}

Thus, the one-loop virtual-corrected and unrenormalized squared NLO amplitude is given by
\begin{equation}
\mathcal{A}^{(\textrm{NLO})}_{\textrm{virtual},0} = \mathcal{A}^{(\textrm{NLO})}_{\textrm{gSE}} + \mathcal{A}^{(\textrm{NLO})}_{\textrm{V}} + \mathcal{A}^{(\textrm{NLO})}_{\textrm{BOX}} = g^6\mathcal{A}^{(\textrm{NLO})}.
\end{equation}

\section{Gluon self-energy at one-loop order}
\label{GSEsection}

\begin{figure}[h!]
\center
\includegraphics[scale=.42]{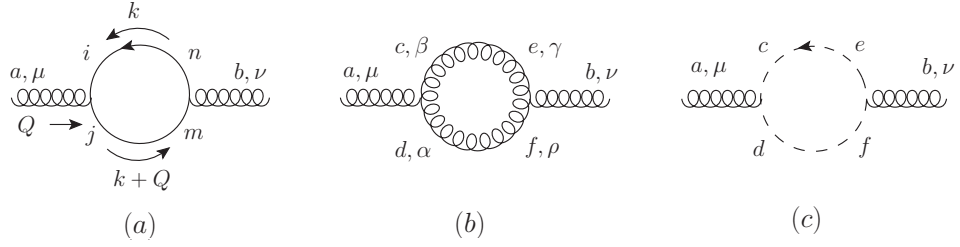}
\caption{The one-loop gluon self-energy diagrams.}
\label{fig:GSE}
\end{figure}

At one-loop order, the gluon self-energy $(i\Pi^{ab}_{\mu\nu})$ correction to the quark-quark scattering is given by the three diagrams shown in Fig. \ref{fig:GSE},
where the diagram (a) is the quark loop contribution $i\Pi^{ab}_{\mu\nu}\large\vert_{(\textrm{q})}$, the diagram (b) the gluon loop contribution $i\Pi^{ab}_{\mu\nu}\large\vert_{(\textrm{gl})}$ and the diagram (c) the ghost loop contribution $i\Pi^{ab}_{\mu\nu}\large\vert_{(\textrm{gh})}$. Because of the gauge invariance in QCD \cite{MUTA}, the gluon self-energy is transverse,
\begin{equation}
\label{eq:ginva}
i\Pi^{ab}_{\mu\nu}(Q^2) = i\delta^{ab}\left (Q^2g_{\mu\nu}-Q_{\mu}Q_{\nu} \right )\Pi(Q^2), \quad i\Pi^{ab}_{\mu\nu}Q^{\mu} = 0.
\end{equation}

\subsection{Quark loop contribution}

Applying the Feynman rules to the vertices and propagators (see Appendix \ref{QCDFR}), we obtain for the quark loop contribution in Fig.\ \ref{fig:GSE} (a) an initial expression to be integrated over the internal momentum $k$:
\begin{equation}
\begin{split}
i\Pi^{ab}_{\mu\nu}\large\vert_{(\textrm{q})}(Q^2) & = (-1)n_qT^a_{ij}T^{b}_{nm}\int_k \mathrm{Tr}\biggl [(ig\gamma_{\mu})\frac{i\delta_{in}\cancel k}{d_0}(ig\gamma_{\nu})\frac{i\delta_{jm}(\cancel k + \cancel Q)}{d_1} \biggr ]\\
& = -g_0^2\mu_{0}^{2\epsilon}n_q\mathrm{Tr}(T^aT^b)\int_{k}\frac{\mathcal{N}^{(\textrm{q})}_{\mu\nu}(k,Q)}{d_0d_1},
\end{split}
\end{equation}
where we use the short-hand notation \eqref{eq:shothand} for the loop-momentum integral, $n_q$ is the number of quark flavours, the factor $(-1)$ reflects the presence of a quark (fermion) loop,
\begin{equation}
d_0 = k^2 + i\delta, \quad\quad d_1 = (k+Q)^2 + i\delta,
\end{equation}
with $\delta \rightarrow 0_+$, and the numerator structure is
\begin{equation}
\label{eq:gsenum}
\begin{split}
\mathcal{N}^{(\textrm{q})}_{\mu\nu}(k,Q) & = \mathrm{Tr}\biggl [\gamma_{\mu}\cancel k\gamma_{\nu}(\cancel k + \cancel Q) \biggr ]\\
& = 4\biggl \{-g_{\mu\nu}\left (k^2 + k_{\alpha}Q^{\alpha} \right ) + 2k_{\mu}k_{\nu} + k_{\mu}Q_{\nu} + k_{\nu}Q_{\mu} \biggr \}.
\end{split}
\end{equation}
The color trace is given by Eq.\ \eqref{eq:COLORR1},
\begin{equation}
\mathrm{Tr}(T^aT^b) = \frac{1}{2}\delta^{ab}.
\end{equation}
Next, using Eq.\ \eqref{eq:gsenum} we identify the rank one and rank two terms $B_{\mu}(Q^2)$ and $B_{\mu\nu}(Q^2)$ as defined in \eqref{eq:tensintgen}. Thus, we immediately get the answer
\begin{equation}
\begin{split}
i\Pi^{ab}_{\mu\nu}\large\vert_{(\textrm{q})}(Q^2)  = -2n_qg_0^2\mu_{0}^{2\epsilon}\delta^{ab}\biggl \{-g_{\mu\nu}Q^{\alpha}B_{\alpha}(Q^2)  + & 2B_{\mu\nu}(Q^2)\\
+ Q_{\nu}B_{\mu}(Q^2) & + Q_{\mu}B_{\nu}(Q^2)\biggr \},
\end{split}
\end{equation}
where the term proportional to $-g_{\mu\nu}k^2$ gives zero, since
\begin{equation}
\int_k\frac{k^2}{d_0d_1} = \int_{\ell}\frac{1}{\ell^2} \overset{\eqref{eq:A0int}}{=} A_0 = 0.
\end{equation}
Furthermore, using the Passarino-Veltman tensor reduction by applying Eqs. \eqref{eq:twoPV}, \eqref{eq:BRANK11} and \eqref{eq:BRANK2}, we obtain
\begin{equation}
\label{eq:qloop}
\begin{split}
i\Pi^{ab}_{\mu\nu}\large\vert_{(\textrm{q})}(Q^2)  =  -2n_qg_0^2\mu_{0}^{2\epsilon}\delta^{ab}\biggl \{Q^2& g_{\mu\nu}\biggl [\frac{1}{2} + \frac{1}{2(1-d)} \biggr ]\\
& -Q_{\mu}Q_{\nu}\biggl [1 + \frac{d}{2(1-d)} \biggr ] \biggr \}B_{0}(Q^2),
\end{split}
\end{equation}
where $B_0(Q^2)$ is given by Eq.\ \eqref{eq:B0final}. Finally, substituting $d=4-2\epsilon$ and expanding the square brackets of Eq. \eqref{eq:qloop} in $\epsilon$ and applying Eq.\ \eqref{eq:B0final}, the final result for the quark loop contribution becomes
\begin{equation}
\label{eq:quarkloopfin}
i\Pi^{ab}_{\mu\nu}\large\vert_{(\textrm{q})}(Q^2)  = i\delta^{ab}\left (Q^2g_{\mu\nu}-Q_{\mu}Q_{\nu} \right )\Pi\large\vert_{(\textrm{q})}(Q^2), 
\end{equation}
where
\begin{equation}
\Pi\large\vert_{(\textrm{q})}(Q^2) = -\frac{2n_qg_0^2}{(4\pi)^2}\left ( \frac{4\pi\mu_{0}^2}{-Q^2-i\delta}\right )^{\epsilon}R_{\Gamma}\biggl \{\frac{1}{3\epsilon} + \frac{5}{9}\biggr \} + \mathcal{O}(\epsilon),
\end{equation}
and $R_{\Gamma}$ is given by Eq.\ \eqref{eq:RGAMMA}.

\subsection{Gluon and ghost loop contributions}

Next, from the Feynman diagram shown in Fig.\ \ref{fig:GSE}~(b), we obtain the gluon loop contribution to the gluon self-energy as
\begin{equation}
\label{eq:gluoncont}
\begin{split}
i\Pi^{ab}_{\mu\nu}\large\vert_{(\textrm{gl})}(Q^2)  &=\frac{g_0^2\mu_{0}^{2\epsilon}}{2}f^{acd}f^{bef}\int_{k}\left (\frac{(-i)g^{\beta\gamma}\delta^{ce}}{d_0}\right )\left (\frac{(-i)g^{\alpha\rho}\delta^{df}}{d_1}\right )\\
& \mathcal{C}_{\mu\beta\alpha}(Q,k,-k-Q)\mathcal{C}_{\nu\gamma\rho}(-Q,-k,k+Q)\\
& = -\frac{g_0^2\mu_{0}^{2\epsilon}}{2}f^{acd}f^{bcd}\int_{k}\frac{\mathcal{N}^{(\textrm{gl})}_{\mu\nu}(k,Q)}{d_0d_1},
\end{split}
\end{equation}
where $1/2$ is the symmetry factor due to the two similar gluon propagators and the $\mathcal{C}$'s are the Lorentz-index and momentum dependent parts of the 3-gluon vertices (see Appendix section \ref{vertices}), 
\begin{equation}
\label{eq:contraction}
\mathcal{N}^{(\textrm{gl})}_{\mu\nu}(k,Q) =  \mathcal{C}_{\mu\beta\alpha}(Q,k,-k-Q){\mathcal{C}_{\nu}}^{\beta\alpha}(-Q,-k,k+Q),
\end{equation}
with
\begin{equation}
\mathcal{C}_{\mu\beta\alpha}(Q,k,-k-Q) = \biggl \{g_{\mu\beta}(Q-k)_{\alpha} + g_{\beta\alpha}(2k+Q)_{\mu} - g_{\mu\alpha}(k+2Q)_{\beta}\biggr \},
\end{equation}
and
\begin{equation}
{\mathcal{C}_{\nu}}^{\beta\alpha}(-Q,-k,k+Q) = \biggl \{{g_{\nu}}^{\beta}(k-Q)^{\alpha} - g^{\beta\alpha}(2k+Q)_{\nu} + {g_{\nu}}^{\alpha}(k+2Q)^{\beta}\biggr \}. 
\end{equation}
The contraction of structure constants in Eq.\ \eqref{eq:gluoncont} can be evaluated using Eq. \eqref{eq:COLORR2},
\begin{equation}
f^{acd}f^{bcd} = N_c\delta^{ab},
\end{equation}
and the numerator structure in Eq.\ \eqref{eq:contraction} can be simplified to
\begin{equation}
\begin{split}
\mathcal{N}^{(\textrm{gl})}_{\mu\nu}(k,Q) = -g_{\mu\nu}\biggl \{5Q^2 & + 2k^2 + 2k_{\alpha}Q^{\alpha} \biggr \} + Q_{\mu}Q_{\nu}(6-d)\\
& + (3-2d)\biggl \{2k_{\mu}k_{\nu} + Q_{\mu}k_{\nu} + Q_{\nu}k_{\mu} \biggr \}.
\end{split}
\end{equation}
Taking the same steps in using the Passarino-Veltman reduction, and performing a few simple rearrangements, we obtain
\begin{equation}
\label{eq:glloop}
\begin{split}
i\Pi^{ab}_{\mu\nu}\large\vert_{(\textrm{gl})}(Q^2) & = -\frac{g_0^2\mu_{0}^{2\epsilon}N_c}{2}\delta^{ab}\biggl \{Q^2g_{\mu\nu}\biggl [\frac{(3-2d)}{2(1-d)}-4 \biggr ]\\
& + Q_{\mu}Q_{\nu}\biggl [(6-d)+\frac{d(3-2d)}{2(d-1)}-(3-2d) \biggr ] \biggr \}B_{0}(Q^2),
\end{split}
\end{equation}
where, as before, $B_0(Q^2)$ is given by Eq.\ \eqref{eq:B0final}. Finally, again expanding the square brackets in Eq.\ \eqref{eq:glloop} in $\epsilon$ and applying Eq.\ \eqref{eq:B0final}, the final result for the gluon loop contribution becomes
\begin{equation}
\label{eq:glloopfin}
\begin{split}
i\Pi^{ab}_{\mu\nu}\large\vert_{(\textrm{gl})}(Q^2)  = i\delta^{ab}\frac{g_0^2N_c}{(4\pi)^2}&\left ( \frac{4\pi\mu_{0}^2}{-Q^2-i\delta}\right )^{\epsilon}R_{\Gamma} \biggl \{ Q^2g_{\mu\nu}\biggl [\frac{19}{12\epsilon}+\frac{58}{18} \biggr ]\\
&-Q_{\mu}Q_{\nu}\biggl [\frac{11}{6\epsilon} + \frac{67}{18} \biggr ]\biggl \} + \mathcal{O}(\epsilon).
\end{split}
\end{equation}
From Eq.\ \eqref{eq:glloopfin}, we can clearly see that, unlike the quark loop, the gluon loop contribution alone does not satisfy the requirement of gauge invariance in Eq.\ \eqref{eq:ginva}. To cure this we need the additional contribution from the ghost loop shown in Fig.\ \ref{fig:GSE}~(c). This is given by
\begin{equation}
\begin{split}
i\Pi^{ab}_{\mu\nu}\large\vert_{(\textrm{gh})}(Q^2) & = (-1)g_0^2\mu_{0}^{2\epsilon}\int_{k} f^{adc}(k+Q)_{\mu}\left (\frac{i\delta^{ce}}{d_0}\right )f^{bef}k_{\nu}\left (\frac{i\delta^{df}}{d_1}\right )\\
& = g_0^2\mu_{0}^{2\epsilon}f^{adc}f^{bcd}\int_{k}\frac{\mathcal{N}^{(\textrm{gh})}_{\mu\nu}(k,Q)}{d_0d_1},
\end{split}
\end{equation}
where the factor of $(-1)$ reflects the presence of a fermion-loop and the numerator structure is
\begin{equation}
\mathcal{N}^{(\textrm{gh})}_{\mu\nu}(k,Q) = k_{\mu}k_{\nu} + k_{\nu}Q_{\mu}.
\end{equation}
The contraction of structure constants takes now the form
\begin{equation}
f^{adc}f^{bcd} = -f^{acd}f^{bcd} = -N_c\delta^{ab}.
\end{equation}
Furthermore, making a few simple rearrangements in the numerator, we obtain
\begin{equation}
\label{eq:ghostloop}
\begin{split}
i\Pi^{ab}_{\mu\nu}\large\vert_{(\textrm{gh})}(Q^2) & = -\delta^{ab}g_0^2\mu_{0}^{2\epsilon}N_c\biggl \{B_{\mu\nu}+Q_{\mu}Q_{\nu}B_1 \biggr \}\\
& = -\delta^{ab}g_0^2\mu_{0}^{2\epsilon}N_c\biggl \{\frac{Q^2g_{\mu\nu}}{4(1-d)} + Q_{\mu}Q_{\nu}\biggl [\frac{d}{4(d-1)}-\frac{1}{2}\biggr ]\biggr \}B_0(Q^2).
\end{split}
\end{equation}
Using the same procedure as before, the final result for the ghost loop contribution becomes
\begin{equation}
\label{eq:ghostloopfin}
\begin{split}
i\Pi^{ab}_{\mu\nu}\large\vert_{(\textrm{gh})}(Q^2)  = i\delta^{ab}\frac{g_0^2N_c}{(4\pi)^2} & \left ( \frac{4\pi\mu_{0}^2}{-Q^2-i\delta}\right )^{\epsilon}R_{\Gamma} \biggl \{ Q^2g_{\mu\nu}\biggl [\frac{1}{12\epsilon}+\frac{4}{18} \biggr ]\\
&-Q_{\mu}Q_{\nu}\biggl [-\frac{1}{6\epsilon} - \frac{5}{18} \biggr ]\biggl \} + \mathcal{O}(\epsilon).
\end{split}
\end{equation}

\vspace{0.3cm}

Summing the gluon and ghost loop contributions in Eqs. \eqref{eq:glloopfin} and \eqref{eq:ghostloopfin} together, we obtain the desired gauge invariant expression
\begin{equation}
\label{eq:glghfinal}
i\Pi^{ab}_{\mu\nu}\large\vert_{(\textrm{gl + gh})}(Q^2) = i\delta^{ab}\left ( Q^2g_{\mu\nu}-Q_{\mu}Q_{\nu}\right )\Pi\large\vert_{(\textrm{gl + gh})}(Q^2),
\end{equation}
where
\begin{equation}
\Pi\large\vert_{(\textrm{gl + gh})}(Q^2) = \frac{g_0^2N_c}{(4\pi)^2}\left ( \frac{4\pi\mu_{0}^2}{-Q^2-i\delta}\right )^{\epsilon}R_{\Gamma} \biggl \{\frac{5}{3\epsilon} + \frac{31}{9}\biggr \} + \mathcal{O}(\epsilon).
\end{equation}

\vspace{0.3cm}

Finally, summing the quark and (gluon + ghost) loop contributions in Eqs. \eqref{eq:quarkloopfin} and \eqref{eq:glghfinal} together, we obtain the full unrenormalized gauge-invariant one-loop gluon self-energy correction:
\begin{equation}
\label{eq:GSEfullp1}
i\Pi^{ab}_{\mu\nu}\large\vert_{(\textrm{1-loop})}(Q^2) = i\delta^{ab}\left ( Q^2g_{\mu\nu}-Q_{\mu}Q_{\nu}\right )\Pi\large\vert_{(\textrm{1-loop})}(Q^2),
\end{equation}
where
\begin{equation}
\label{eq:GSEfullp2}
\begin{split}
\Pi\large\vert_{(\textrm{1-loop})}(Q^2) = \frac{g_0^2}{(4\pi)^2}\left ( \frac{4\pi\mu_{0}^2}{-Q^2-i\delta}\right )^{\epsilon}&R_{\Gamma} \biggl \{\frac{1}{3\epsilon}\biggl [5N_c - 2n_q\biggr ] \\
& - \frac{10n_q}{9} + \frac{31N_c}{9}\biggr \} + \mathcal{O}(\epsilon).
\end{split}
\end{equation}

\subsection{Interference with Born-level}

\begin{figure}[h!]
\center
\includegraphics[scale=.55]{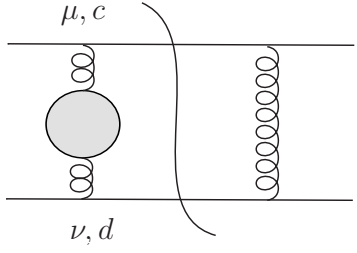}
\caption{The interference of the one-loop gluon self-energy-corrected diagram with the Born one contributing to the quark-quark scattering at NLO.}
\label{fig:GSEINBORN}
\end{figure}

The NLO contribution from the interference of the unrenormalized one-loop gluon self-energy-corrected diagrams, $\mathcal{M}_{\rm gSE}$, with the Born one, $\mathcal{M}_{\rm LO}$, is shown in Fig. \ref{fig:GSEINBORN}.
Using Eq. \eqref{eq:SAMPLITUDE} we get the NLO squared amplitude
\begin{equation}
\label{eq:virtualGSEamp}
\mathcal{A}^{(\textrm{NLO})}_{(\rm gSE)} = \sum_{\substack{\textrm{spin}\\\textrm{color}}}2\mathbb{R}e\left (\mathcal{M}_{\textrm{gSE}}\mathcal{M}_{\textrm{LO}}^{\dagger}\right ), 
\end{equation}
where the matrix element $i\mathcal{M}_{\textrm{gSE}}$ including the full one-loop gluon self-energy can be written as 
\begin{equation}
\begin{split}
i\mathcal{M}_{\textrm{gSE}} = \frac{g^2}{t^2}\underbrace{(T^c)_{ji}(T^d)_{mn}}_{= \mathcal{C}_{(\textrm{gSE})}}\biggl [\bar u(k_3)\gamma^{\mu}u(k_2)\biggl ]i&\Pi^{cd}_{\mu\nu}\large\vert_{(\textrm{1-loop})}(t)\\
&\biggl [\bar u(k_4)\gamma^{\nu}u(k_1)\biggr ].
\end{split}
\end{equation}
Here, $i\Pi^{cd}_{\mu\nu}\large\vert_{(\textrm{1-loop})}(t)$ is again given by Eq.\ \eqref{eq:GSEfullp1} with $Q^2=t$ and $Q_{\mu}=(k_2-k_3)_{\mu}$. The complex-conjugated Born-level matrix element $-i\mathcal{M}_{\textrm{LO}}^{\dagger}$ in Eq.\ \eqref{eq:virtualGSEamp} is given by Eq.\ \eqref{eq:borncopcon}. After some algebra we obtain the interference term
\begin{equation}
\label{eq:intGSEpart}
\begin{split}
\sum_{\substack{\textrm{spin}\\\textrm{color}}}\mathcal{M}_{\textrm{gSE}}\mathcal{M}_{\textrm{LO}}^{\dagger} = \frac{g^4}{t^3}&\mathcal{C}_{(\textrm{gSE+LO})}\mathcal{L}^{\mu\alpha}_{(q)}(k_2,k_3){\mathcal{L}^{\nu}}_{\alpha,(q)}(k_1,k_4)\\
&\biggl \{tg_{\mu\nu}-(k_2-k_3)_{\mu}(k_2-k_3)_{\nu}\biggr \}\Pi\large\vert_{(\textrm{1-loop})}(t),
\end{split}
\end{equation}
where the quark tensors are given by Eqs. \eqref{eq:qtensor1} and \eqref{eq:qtensor2}. The color product, $\mathcal{C}_{(\textrm{gSE+LO})} =\mathcal{C}_{(\textrm{gSE})}\mathcal{C}_{(\textrm{LO})}$, between the gluon self-energy diagram and Born level diagram reads
\begin{equation}
\begin{split}
\mathcal{C}_{(\textrm{gSE+LO})} & = \sum_{\textrm{color}} (T^c)_{ji}(T^c)_{mn}(T^d)_{ij}(T^d)_{nm}\\
& = \sum_{\textrm{color}} \textrm{Tr}(T^cT^d)\textrm{Tr}(T^cT^d) = 2.
\end{split}
\end{equation} 
Furthermore, applying the Dirac equation in Eq.\ \eqref{eq:diraceq}, we find that
\begin{equation}
\mathcal{L}^{\mu\alpha}_{(q)}(k_2,k_3){\mathcal{L}^{\nu}}_{\alpha,(q)}(k_1,k_4)\biggl \{(k_2-k_3)_{\mu}\underbrace{(k_2-k_3)_{\nu}}_{= (k_4-k_1)_{\nu}}\biggr \} =0.
\end{equation}
Finally, substituting the corresponding expression for the $\Pi\large\vert_{(\textrm{1-loop})}(t)$-function from Eq.\ \eqref{eq:GSEfullp2} into Eq.\ \eqref{eq:intGSEpart}, writing 
\begin{equation}
\left (\frac{4\pi\mu_0^2}{-Q^2-i\delta}\right )^{\epsilon} = \left (\frac{4\pi\mu_0^2/Q_s^2}{(-Q^2-i\delta)/Q_s^2}\right )^{\epsilon}
\end{equation}
and expanding $\left (\frac{-Q^2-i\delta}{Q_s^2}\right )^{-\epsilon}$ in $\epsilon$, we obtain
\begin{equation}
\label{eq:finalGSEBORNvir}
\begin{split}
\mathcal{A}^{(\textrm{NLO})}_{(\rm gSE)} = &\frac{g_{0}^6\mu_{0}^{4\epsilon}}{(4\pi)^2} \left (\frac{4\pi\mu_{0}^2}{Q^2_s}\right )^{\epsilon}R_{\Gamma}\mathcal{A}^{(\rm LO)}(s,t,u)\\
&\biggl \{\biggl [\frac{1}{\epsilon}-\ln\left (\bigg\vert \frac{t}{Q^2_s}\bigg\vert\right )\biggr ]\left (\frac{10N_c}{3}-\frac{4n_q}{3}\right ) - \frac{20n_q}{9} + \frac{62N_c}{9}\biggr \}.
\end{split}
\end{equation} 
Here, $\mathcal{A}^{(\rm LO)}(s,t,u)$ is the Born level expression introduced in Eq.\ \eqref{eq:LOcont} and $Q^2_s > 0$ is an arbitrary momentum scale as introduced in \cite{ELSE}. In Eq.\ \eqref{eq:finalGSEBORNvir} it is understood that as only the real part of Eq.\ \eqref{eq:finalGSEBORNvir} is kept, we can substitute
\begin{equation}
\ln\left (\frac{-t-i\delta}{Q^2_s}\right ) \overset{\delta \rightarrow 0_+}{\longrightarrow} \ln\left (\bigg\vert \frac{t}{Q^2_s}\bigg\vert\right ), \quad \text{for} \quad t \leqslant 0.
\end{equation}

\section{One-loop vertex corrections}

\begin{figure}[h!]
\center
\includegraphics[scale=.40]{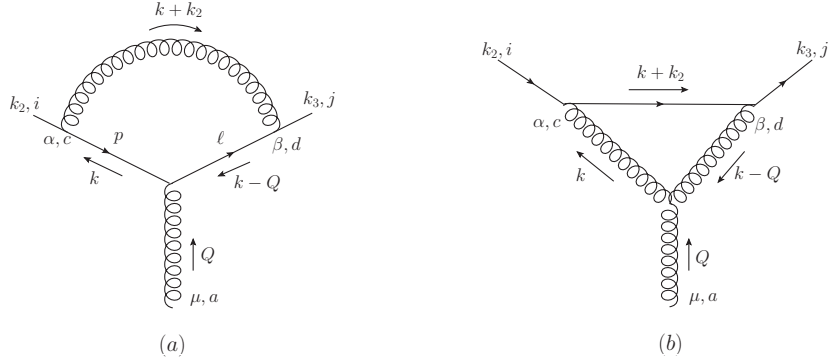}
\caption{One-loop quark-gluon vertex diagrams (a) and (b) contributing to quark-quark scattering.}
\label{fig:verteksit2}
\end{figure}

At one-loop order, the $qqg$-vertex correction $i\Gamma_{\mu}\large\vert_{(\textrm{1-loop})}$ for the quark-quark scattering is given by the two diagrams shown in Fig.\ \ref{fig:verteksit2}.
Applying the Feynman rules for the vertices and propagators of Appendix \ref{QCDFR}, we obtain an initial expression for the vertex correction shown in Fig.\ \ref{fig:verteksit2}~(a) 
\begin{equation}
\label{eq:vertexa}
\begin{split}
i\Gamma_{\mu}^{a}\large\vert_{(a)}(Q^2) & = \int_{k}\bar u(k_3)\biggl [ig(T^d)_{jp}\gamma_{\beta} \biggr ]\frac{i(\cancel k-\cancel Q)}{d_2}\biggl [ig(T^a)_{p\ell}\gamma_{\mu} \biggr ]\frac{i\cancel k}{d_0}\\
&\quad\quad\quad\quad\quad\quad \biggl [ig(T^c)_{\ell i}\gamma_{\alpha} \biggr ]\frac{(-i)g^{\alpha\beta}\delta^{cd}}{d_1}u(k_2)\\
& =  g^3(T^c)_{jp}(T^a)_{p\ell}(T^c)_{\ell i} \int_k \frac{\mathcal{N}_{\mu}^{(a)}(k,Q)}{d_0d_1d_2},
\end{split}
\end{equation}
where the propagator denominators are
\begin{equation}
d_0 = k^2 + i\delta, \quad d_1 = (k+k_2)^2 + i\delta, \quad d_2 = (k-Q)^2 + i\delta
\end{equation}
with $Q = k_3-k_2 = -(k_2 + k_3')$. Here, in order to make direct use of the tensor decomposition of the 3-point $C$ function as discussed in Appendix \ref{Cfunctions}, we have introduced the notation $k_3'=-k_3$. With $k_3'$, all momenta in Fig.\ \ref{fig:verteksit2} are then incoming, as required by the general formulae in Appedinx \ref{SINT}.

\vspace{0.3cm}

Using the Fierz identity for $\textrm{SU}(N_c)$ in Eq.\ \eqref{eq:FIDE}, the contraction of the color matrices in Eq.\ \eqref{eq:vertexa} simplifies to
\begin{equation}
(T^c)_{jp}(T^a)_{p\ell}(T^c)_{\ell i} = \biggr [\frac{1}{2}\delta_{p\ell}\delta_{ji} - \frac{1}{2N_c}\delta_{jp}\delta_{\ell i}\biggl ](T^a)_{p\ell} = -\frac{1}{2N_c}(T^a)_{ji}.
\end{equation}
Furthermore, the numerator structure in Eq.\ \eqref{eq:vertexa} is given by 
\begin{equation}
\label{eq:Vnumerator}
\begin{split}
\mathcal{N}_{\mu}^{(a)}(k,Q) & = \bar u(k_3)\biggl \{\gamma^{\alpha}(\cancel k - \cancel Q)\gamma_{\mu}\cancel k\gamma_{\alpha}\biggr \}u(k_2)\\
& = \bar u(k_3)\left (k^{\beta}k^{\rho} - Q^{\beta}k^{\rho}\right )\biggl \{\gamma^{\alpha}\gamma_{\beta}\gamma_{\mu}\gamma_{\rho}\gamma_{\alpha}\biggr \}u(k_2).
\end{split}
\end{equation}
Following the same procedure as before, we use Eq.\ \eqref{eq:Vnumerator} and identifying the rank one and rank two terms $C_{\mu}(k_2,k_3';Q^2)$ and $C_{\mu\nu}(k_2,k_3';Q^2)$ as defined in Eq.\ \eqref{eq:tensintgen}. Hence, we obtain the answer
\begin{equation}
\label{eq:vertexaim1}
\begin{split}
i\Gamma_{\mu}^{(a)}(Q^2) &= -\frac{g^3}{2N_c}(T^a)_{ji} \bar u(k_3)\biggl [C^{\beta\rho}(k_2,k_3';Q^2)\\
&-Q^{\beta}C^{\rho}(k_2,k_3';Q^2)\biggr ]\biggl \{(4-d)\gamma_{\beta}\gamma_{\mu}\gamma_{\rho} -2\gamma_{\rho}\gamma_{\mu}\gamma_{\beta} \biggr \}u(k_2),
\end{split}
\end{equation}
where the product of $\gamma$ matrices is simplified in $d$ dimensions applying Eq.\ \eqref{eq:gammarel1}. Using Eqs.\ \eqref{eq:threePV} and \eqref{eq:threeOV2} together with the Dirac equation for massless quarks, the contractions of $C^{\beta\rho}, C^{\rho}$ with the $\gamma$ matrices in Eq.\ \eqref{eq:vertexaim1} take a simple form
\begin{equation}
\begin{split}
\bar u(k_3)& C^{\beta\rho}\biggl \{(4-d)\gamma_{\beta}\gamma_{\mu}\gamma_{\rho} -2\gamma_{\rho}\gamma_{\mu}\gamma_{\beta} \biggr \}u(k_2)\\
&  = \biggl [\bar u(k_3)\gamma_{\mu}u(k_2)\biggr ]\underbrace{\biggl \{(2-d)^2C_{00}(Q^2) - (2-d)Q^2C_{12}(Q^2)\biggr \}}_{= -B_0(Q^2)}
\end{split}
\end{equation}
and
\begin{equation}
\begin{split}
\bar u(k_3)&Q^{\beta} C^{\rho}\biggl \{(4-d)\gamma_{\beta}\gamma_{\mu}\gamma_{\rho} -2\gamma_{\rho}\gamma_{\mu}\gamma_{\beta} \biggr \}u(k_2)\\
&  = \biggl [\bar u(k_3)\gamma_{\mu}u(k_2)\biggr ]\underbrace{\biggl \{(4-d)Q^2C_2(Q^2) - 2Q^2C_1(Q^2)\biggr \}}_{=(6-d)B_0(Q^2) + 2Q^2C_0(Q^2)}.
\end{split}
\end{equation}
where we have exploited Eq.\ \eqref{eq:Cranktwoform} for the form factors $C_{00}$ and $C_{12}$, and Eq.\ \eqref{eq:Crankoneform} for $C_1$ and $C_2$. Thus, we obtain 
\begin{equation}
\begin{split}
i\Gamma_{\mu}^{a}\large\vert_{(a)}(Q^2) = \frac{g^3}{2N_c}(T^a)_{ji}\biggl [\bar u(k_3)&\gamma_{\mu}u(k_2)\biggr ]\biggl \{(7-d)B_0(Q^2)\\
& + 2Q^2C_0(Q^2)\biggr \}.
\end{split}
\end{equation}
Finally, expanding in $\epsilon$ and applying Eqs.\ \eqref{eq:B0final} and \eqref{eq:C0final}, the final result for the one-loop vertex correction (a) becomes
\begin{equation}
\label{eq:verter1full}
i\Gamma_{\mu}^{a}\large\vert_{(a)}(Q^2) = i(T^a)_{ji}\biggl [\bar u(k_3)\gamma_{\mu}u(k_2)\biggr ]\Gamma\large\vert_{(a)}(Q^2),
\end{equation}
where
\begin{equation}
\Gamma\large\vert_{(a)}(Q^2) = \frac{g_0^3\mu_0^{\epsilon}}{(4\pi)^2}\left ( \frac{4\pi\mu_{0}^2}{-Q^2-i\delta}\right )^{\epsilon}R_{\Gamma}\frac{1}{2N_c}\biggl \{\frac{2}{\epsilon^2} + \frac{3}{\epsilon} + 8 \biggr \} + \mathcal{O}(\epsilon).
\end{equation}

\vspace{0.3cm}

The vertex diagram shown in Fig.\ \ref{fig:verteksit2}~(b) can be analyzed in the same way. Applying the Feynman rules for the vertices and propagators we obtain:
\begin{equation}
\label{eq:vertexb}
\begin{split}
i\Gamma_{\mu}^{a}\large\vert_{(b)}(Q^2) & = \int_k \bar u(k_3)\biggl [ig(T^f)_{j\ell}\gamma_{\rho} \biggr ]\frac{i(\cancel k + \cancel k_2)}{d_2}\biggl [ig(T^e)_{\ell i}\gamma_{\sigma} \biggr ]\frac{(-i)g^{\rho\beta}\delta^{fd}}{d_3}\\
&gf^{acd}\mathcal{C}_{\mu\alpha\beta}(Q,-k,k-Q)\frac{(-i)g^{\alpha\sigma}\delta^{ce}}{d_0}u(k_2)\\
& = ig^3(T^dT^c)_{ji}f^{acd}\int_{k}\frac{\mathcal{N}_{\mu}^{(b)}(k,Q)}{d_0d_1d_2},
\end{split}
\end{equation}
where the numerator structure reads
\begin{equation}
\label{eq:vertexbnum}
\mathcal{N}_{\mu}^{(b)}(k,Q) = \bar u(k_3)\biggl [\gamma^{\beta}(\cancel k + \cancel k_2)\gamma^{\alpha}\mathcal{C}_{\mu\alpha\beta}(Q,-k,k-Q) \biggr ]u(k_2)
\end{equation}
with
\begin{equation}
\mathcal{C}_{\mu\alpha\beta}(Q,-k,k-Q) = \biggl \{g_{\mu\alpha}(k+Q)_{\beta} - g_{\alpha\beta}(2k-Q)_{\mu} + g_{\mu\beta}(k-2Q)_{\alpha}\biggr \}. 
\end{equation}
The color matrix product can be reduced as follows:
\begin{equation}
\begin{split}
(T^dT^c)_{ji}f^{acd} &= \frac{1}{2}[T^d,T^c]_{ji}f^{acd} + \frac{1}{2}\underbrace{\{T^d,T^c\}_{ji}f^{acd}}_{=0}\\
& = \frac{i}{2}f^{edc}f^{acd} = -\frac{iN_c}{2}(T^a)_{ji},
\end{split}
\end{equation}
where on the first line we have used Eqs.\ \eqref{eq:anticolor} and \eqref{eq:COLORR2}, and on the second line Eq.\ \eqref{eq:COLORR2}. Simplifying the four-vector contractions and applying the Dirac equation in Eq. \eqref{eq:vertexbnum}, we obtain
\begin{equation}
\begin{split}
i\Gamma_{\mu}^{a}\large\vert_{(b)}(Q^2) &= \frac{g^3N_c}{2}(T^a)_{ji}\biggl [\bar u(k_3)\gamma_{\mu}u(k_2) \biggr ]\biggl \{\underbrace{2(d-2)C_{00}(Q^2)}_{=B_{0}(Q^2)}\\
&\quad\quad\quad\quad\quad + \underbrace{2Q^2\left (C_1(Q^2)+C_0(Q^2)\right )}_{=-2B_{0}(Q^2)} \biggr \}\\
& = -\frac{g^3N_c}{2}(T^a)_{ji}\biggl [\bar u(k_3)\gamma_{\mu}u(k_2) \biggr ]B_{0}(Q^2),
\end{split}
\end{equation}
where we have again used Eqs.\ \eqref{eq:Cranktwoform} and \eqref{eq:Crankoneform}. Furthermore, applying Eq.\ \eqref{eq:B0final} our final result for the one-loop vertex correction (b) becomes
\begin{equation}
\label{eq:vertex2full}
i\Gamma_{\mu}^{a}\large\vert_{(b)}(Q^2) = i(T^a)_{ji}\biggl [\bar u(k_3)\gamma_{\mu}u(k_2)\biggr ]\Gamma\large\vert_{(b)}(Q^2),
\end{equation}
where
\begin{equation}
\label{eq:fullverloop}
\Gamma\large\vert_{(b)}(Q^2) = -\frac{g_0^3\mu_0^{\epsilon}}{(4\pi)^2}\left ( \frac{4\pi\mu_{0}^2}{-Q^2-i\delta}\right )^{\epsilon}R_{\Gamma}\frac{N_c}{2}\biggl \{\frac{1}{\epsilon} + 2 \biggr \} + \mathcal{O}(\epsilon).
\end{equation}

\vspace{0.2cm}

Finally, summing the one-loop vertex corrections in Eqs.\ \eqref{eq:verter1full} and \eqref{eq:vertex2full} together, we obtain the full unrenormalized one-loop $qqg$-vertex correction:
\begin{equation}
\label{eq:vertexfullcor}
i\Gamma_{\mu}^{a}\large\vert_{(\textrm{1-loop})}(Q^2) = i(T^a)_{ji}\biggl [\bar u(k_3)\gamma_{\mu}u(k_2)\biggr ]\Gamma\large\vert_{(\textrm{1-loop})}(Q^2),
\end{equation}
where
\begin{equation}
\begin{split}
\Gamma\large\vert_{(\textrm{1-loop})}(Q^2) = \frac{g_0^3\mu_0^{\epsilon}}{(4\pi)^2}\left ( \frac{4\pi\mu_{0}^2}{-Q^2-i\delta}\right )^{\epsilon}R_{\Gamma}&\biggl \{\frac{1}{2N_c}\left (\frac{2}{\epsilon^2}+\frac{3}{\epsilon} + 8\right )\\
- &\frac{N_c}{2}\left (\frac{1}{\epsilon} + 2\right ) \biggr \} + \mathcal{O}(\epsilon).
\end{split}
\end{equation}

\subsection{Interference with Born-level}

\begin{figure}[h!]
\center
\includegraphics[scale=.45]{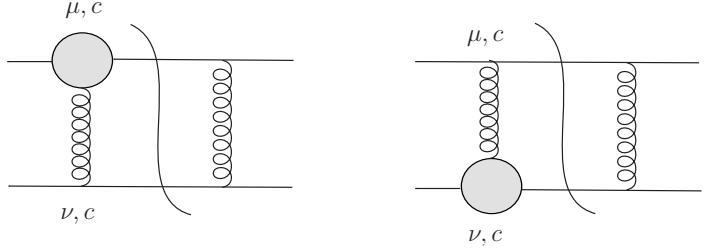}
\caption{The interference of the full unrenormalized one-loop $qqg$-vertex corrected diagrams with the Born one contributing to quark-quark scattering.}
\label{fig:VERTEXINBORN}
\end{figure}

The contributions from the interference of the full unrenormalized one-loop $qqg$-vertex corrected diagrams of Fig.\ \ref{fig:verteksit2} (a) and (b) with the Born one are shown in Fig.\ \ref{fig:VERTEXINBORN}.
The corresponding contribution to the squared NLO amplitude is given by
\begin{equation}
\label{eq:virtualGSE}
\mathcal{A}^{(\textrm{NLO})}_{(\rm V)} = 2\sum_{\substack{\textrm{spin}\\\textrm{color}}}2\mathbb{R}e\left (\mathcal{M}_{\textrm{V}}\mathcal{M}_{\textrm{LO}}^{\dagger}\right ), 
\end{equation}
where the amplitude $i\mathcal{M}_{\textrm{V}}$ for the full one-loop $qqg$-vertex-corrected diagram can be written as 
\begin{equation}
\begin{split}
i\mathcal{M}_{\textrm{V}} = i\frac{g}{t}\underbrace{(T^c)_{ij}(T^c)_{mn}}_{=\mathcal{C}_{(\textrm{V})}}\biggl [\bar u(k_3)\gamma_{\mu}u(k_2)\biggr ]\biggl [\bar u(k_4)\gamma^{\mu}u(k_1)\biggr ]i\Gamma\large\vert_{(\textrm{1-loop})}(t),
\end{split}
\end{equation}
where $i\Gamma^{c}_{\mu}\large\vert_{(\textrm{1-loop})}(t)$ is determined by Eq.\ \eqref{eq:vertexfullcor} with  $Q^2=t$. After some algebra we obtain 
\begin{equation}
\label{eq:vertint}
\begin{split}
\sum_{\substack{\textrm{spin}\\\textrm{color}}}\mathcal{M}_{\textrm{V}}\mathcal{M}_{\textrm{LO}}^{\dagger} = g^3\mathcal{C}_{(\textrm{V+LO})}\mathcal{L}_{\mu\alpha,(q)}(k_3,k_2)&\mathcal{L}^{\mu\alpha}_{(q)}(k_4,k_1)\Gamma\large\vert_{(\textrm{1-loop})}(t),
\end{split}
\end{equation}
where the quark tensors are given by Eqs. \eqref{eq:qtensor1} and \eqref{eq:qtensor2}, and the color matrix product, $\mathcal{C}_{(\textrm{V+LO})}=\mathcal{C}_{(\textrm{V})}\mathcal{C}_{(\textrm{LO})}$, between the vertex diagram and Born-level diagram reads
\begin{equation}
\begin{split}
\mathcal{C}_{(\textrm{V+LO})} & = \sum_{\textrm{color}} (T^c)_{ji}(T^c)_{mn}(T^d)_{ij}(T^d)_{nm}\\
& = \sum_{\textrm{color}} \textrm{Tr}(T^cT^d)\textrm{Tr}(T^cT^d) = 2.
\end{split}
\end{equation} 
Finally, substituting the corresponding expression for the $\Gamma\large\vert_{(\textrm{1-loop})}(t)$-function from Eq. \eqref{eq:fullverloop}, into Eq.\ \eqref{eq:vertint}, we arrive at
\begin{equation}
\label{eq:finalvertexBORNvir}
\begin{split}
\mathcal{A}^{(6)}_{(\rm V)} = \frac{g_{0}^6\mu_{0}^{4\epsilon}}{(4\pi)^2} \left (\frac{4\pi\mu_{0}^2}{Q^2_s} \right )^{\epsilon}&R_{\Gamma}\mathcal{A}^{(\rm LO)}(s,t,u)\frac{2}{N_c}\biggl \{\frac{2}{\epsilon^2}+ \left (3-N_c^2\right )\frac{1}{\epsilon} \\
& - \frac{2}{\epsilon}\ln\left (\bigg\vert\frac{t}{Q^2_s}\bigg\vert\right ) + (N_c^2-3)\ln\left (\bigg\vert\frac{t}{Q^2_s}\bigg\vert\right )\\
& + \ln^2\left(\bigg\vert\frac{t}{Q^2_s} \bigg\vert\right ) + 8 -2N_c^2 \biggr \},\\
\end{split}
\end{equation}
where again $Q_s$ is an arbitrary mass scale and in Eq.\ \eqref{eq:finalvertexBORNvir} it is understood that only the real part is kept, so that we can again replace 
\begin{equation}
\ln\left (\frac{-t-i\delta}{Q^2_s}\right ) \overset{\delta\rightarrow 0_+}{\longrightarrow} \ln\left (\bigg\vert \frac{t}{Q^2_s}\bigg\vert\right ),
\end{equation}
and 
\begin{equation}
\ln^2\left (\frac{-t-i\delta}{Q^2_s}\right )  \overset{\delta\rightarrow 0_+}{\longrightarrow} \ln^2\left (\bigg\vert\frac{t}{Q^2_s}\bigg\vert\right ), 
\end{equation}
for $t \leqslant 0$ and $Q^2_s>0$.

\section{One-loop box corrections}

\begin{figure}[h!]
\center
\includegraphics[scale=.45]{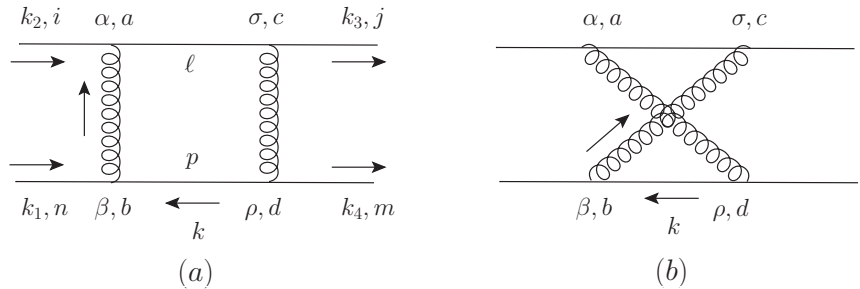}
\caption{The one-loop QCD box diagrams contributing to quark-quark scattering.}
\label{fig:boxes}
\end{figure}

Finally, the last and most complicated one-loop QCD corrections for the $qq'$-scattering are the two UV-finite QCD box diagrams: the direct (dr) box (a) and the crossed (cr) box (b) shown in Fig.\ \ref{fig:boxes}. 
Applying the Feynman rules for the vertices and propagators of Appendix \ref{QCDFR}, for the direct box we obtain
\begin{equation}
\label{eq:drbox}
\begin{split}
i\mathcal{M}^{(\textrm{dr})}_{\textrm{BOX}} &  = \int_k \bar u(k_3)\biggl [ig(T^c)_{j\ell}\gamma^{\sigma} \biggr ]\frac{i(\cancel k + \cancel k_1 + \cancel k_2)}{d_2}\biggl [ig(T^a)_{\ell i}\gamma^{\alpha}\biggr ]u(k_2)\\
&\quad\quad\quad\bar u(k_4)\biggl [ig(T^d)_{mp}\gamma^{\rho}\biggr ]\frac{(-i)\cancel k}{d_0}\biggl [ig(T^b)_{pn}\gamma^{\beta}\biggr ]u(k_1)\\
&\quad\quad\quad\quad\quad\quad\frac{(-i)g_{\alpha\beta}\delta^{ab}}{d_1}\frac{(-i)g_{\sigma\rho}\delta^{cd}}{d_3}\\
& = -g^4\underbrace{(T^cT^a)_{ji}(T^cT^a)_{mn}}_{=\mathcal{C}^{(\textrm{dr})}_{\text{BOX}}}\int_k\frac{\mathcal{N}_{\textrm{BOX}}^{\textrm{dr}}(k,k_i)}{d_0d_1d_2d_3},
\end{split}
\end{equation}
where the denominator factors are
\begin{equation}
\begin{split}
d_0 & = k^2 + i\delta,\\
d_1 & = (k+k_1)^2 + i\delta,\\
d_2 &= (k+k_1+k_2)^2 + i\delta,\\
d_3 & = (k+ k_1+k_2+k_3')^2+i\delta,
\end{split}
\end{equation}
and again $k_3' = -k_3$. The numerator structure in Eq.\ \eqref{eq:drbox} is given by
\begin{equation}
\label{eq:boxnum}
\begin{split}
\mathcal{N}_{\textrm{BOX}}^{\textrm{dr}}(k,k_i) = \biggl [\bar u(k_3)&\gamma^{\sigma}\gamma^{\mu}\gamma^{\alpha}u(k_2)\biggr ]\biggl [\bar u(k_4)\gamma_{\sigma}\gamma^{\nu}\gamma_{\alpha}u(k_1)\biggr ]\\
&\biggl \{k_{\mu}k_{\nu} + k_{\nu}(k_1+k_2)_{\mu} \biggr \}.
\end{split}
\end{equation}
Again, using Eq.\ \eqref{eq:boxnum} we identify the rank one $D_{\mu}$ and rank two $D_{\mu\nu}$ terms as defined in Eq.\ \eqref{eq:tensintgen}, and thus we find quite compact expression for the direct box matrix amplitude
\begin{equation}
\label{eq:drboxmel}
\begin{split}
i\mathcal{M}^{(\textrm{dr})}_{\textrm{BOX}} = -g^4\mathcal{C}^{(\textrm{dr})}_{\text{BOX}}& \biggl [\bar u(k_3)\gamma^{\sigma}\gamma^{\mu}\gamma^{\alpha}u(k_2)\biggr ]\biggl [\bar u(k_4)\gamma_{\sigma}\gamma^{\nu}\gamma_{\alpha}u(k_1)\biggr ]\\
&\biggl \{D_{\mu\nu}(k_1,k_2,k_3';) +D_{\nu}(k_1,k_2,k_3';)(k_1+k_2)_{\mu}\biggr \}.
\end{split}
\end{equation}
%where
%\begin{equation}
%\mathcal{C}^{(\textrm{dr})}_{\text{BOX}} = (T^cT^a)_{ji}(T^cT^a)_{mn}.
%\end{equation}

\subsection{Interference with Born level}

\begin{figure}[h!]
\center
\includegraphics[scale=.45]{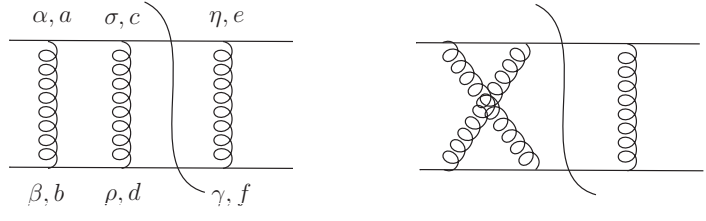}
\caption{The interference of the direct and crossed box diagrams with the Born diagram.}
\label{fig:boxesint}
\end{figure}

The contributions from the interference of the NLO direct and crossed box diagrams (a) and (b) with the Born amplitude are shown in Fig.\ \ref{fig:boxesint}. 
For the direct box diagram, the squared NLO amplitude is given by
\begin{equation}
\mathcal{A}^{(\textrm{NLO})}_{(\textrm{BOX, dr})} = \sum_{\substack{\textrm{spin}\\\textrm{color}}}2\mathbb{R}e\left (\mathcal{M}^{(\textrm{dr})}_{\textrm{box}}\mathcal{M}_{\textrm{LO}}^{\dagger} \right ).
\end{equation}
After some algebra, the product of the direct box and Born-level matrix elements, summed over spin and color can be cast in the form
\begin{equation}
\begin{split}
 &\sum_{\substack{\textrm{spin}\\\textrm{color}}}\mathcal{M}^{(\textrm{dr})}_{\textrm{box}}\mathcal{M}_{\textrm{LO}}^{\dagger} =\frac{ig^6}{t}\mathcal{C}^{(\textrm{dr})}_{(\textrm{BOX+LO})}\mathcal{L}^{\sigma\mu\alpha\eta}(k_3,k_2)\mathcal{L}_{\sigma\alpha\eta}^{\nu}(k_4,k_1)\\
&\biggl \{D_{\mu\nu}(k_1,k_2,k_3';s,t) +D_{\nu}(k_1,k_2,k_3';s,t)(k_1+k_2)_{\mu}\biggr \},
\end{split}
\end{equation}
where
\begin{equation}
\begin{split}
\mathcal{L}^{\sigma\mu\alpha\eta}(k_3,k_2)&=\sum_{\textrm{spin}}\biggl [\bar u(k_3)\gamma^{\sigma}\gamma^{\mu}\gamma^{\alpha}u(k_2)\biggr ]\biggl [\bar u(k_2)\gamma^{\eta}u(k_3)\biggr ]\\
& =\textrm{Tr}\biggl [\cancel k_3\gamma^{\sigma}\gamma^{\mu}\gamma^{\alpha}\cancel k_2\gamma^{\eta} \biggr ],
\end{split}
\end{equation}
\begin{equation}
\begin{split}
\mathcal{L}_{\sigma\alpha\eta}^{\nu}(k_4,k_1)&=\sum_{\textrm{spin}}\biggl [\bar u(k_4)\gamma_{\sigma}\gamma^{\mu}\gamma_{\alpha}u(k_1)\biggr ]\biggl [\bar u(k_1)\gamma_{\eta}u(k_4)\biggr ]\\
& =\textrm{Tr}\biggl [\cancel k_4\gamma_{\sigma}\gamma^{\nu}\gamma_{\alpha}\cancel k_1\gamma_{\eta} \biggr ].
\end{split}
\end{equation}
Using Eqs.\ \eqref{eq:COLORR1} and \eqref{eq:COLORR2} the color matrix product simplifies to
\begin{equation}
\begin{split}
\mathcal{C}^{(\textrm{dr})}_{(\textrm{BOX+LO})} & = \sum_{\textrm{color}}(T^cT^a)_{ji}(T^e)_{ij}(T^cT^a)_{mn}(T^e)_{nm}\\
& = \sum_{\textrm{color}}\textrm{Tr}(T^cT^aT^e)\textrm{Tr}(T^cT^aT^e)\\
& = \frac{1}{16}\sum_{\textrm{color}}\left (d^{cae}d^{cae} + 2id^{cae}f^{cae} -f^{cae}f^{cae}\right )\\
& = \frac{1}{16}\sum_{\textrm{color}} \left (\frac{5}{3}-3\right )\delta^{cc} = -\frac{2}{3}.
\end{split}
\end{equation}

As the next step, we define two functions $\mathcal{I}^{(1)}$ and $\mathcal{I}^{(2)}$ as 	
\begin{equation}
\begin{split}
\mathcal{I}^{(1)}(s,t,u) & = \frac{1}{t}\textrm{Tr}\biggl [\cancel k_3\gamma^{\sigma}\gamma^{\mu}\gamma^{\alpha}\cancel k_2\gamma^{\eta} \biggr ]\textrm{Tr}\biggl [\cancel k_4\gamma_{\sigma}\gamma^{\nu}\gamma_{\alpha}\cancel k_1\gamma_{\eta} \biggr ]\\
& \quad\quad\quad\quad\quad\quad\quad\times \biggl \{(k_1+k_2)_{\mu}D_{\nu}(k_1,k_2,k_3';s,t)\biggr \},\\
\mathcal{I}^{(2)}(s,t,u) & = \frac{1}{t}\textrm{Tr}\biggl [\cancel k_3\gamma^{\sigma}\gamma^{\mu}\gamma^{\alpha}\cancel k_2\gamma^{\eta} \biggr ]\textrm{Tr}\biggl [\cancel k_4\gamma_{\sigma}\gamma^{\nu}\gamma_{\alpha}\cancel k_1\gamma_{\eta} \biggr ]\\
&\quad\quad\quad\quad\quad\quad\quad\times \biggl \{D_{\mu\nu}(k_1,k_2,k_3';s,t)\biggr \}.
\end{split}
\end{equation}
The calculation of these two $\mathcal{I}$ functions, which include rank one and rank two 4-point tensor integrals and two traces of six $\gamma$ matrices, is very complicated. Both of these terms contain a tedious amount of Dirac $\gamma$ algebra together with form factor products in $d$ dimensions, and finally the expressions should be expanded in $\epsilon$ and be simplified in terms of $s,t,u$ and single and double logarithms. However, thanks to the detailed Appedix \ref{Dfunctions}, where I explicitly derived the analytical expressions for the $D_{\mu}$ and $D_{\mu\nu}$ form factors, the whole computation procedure above can be performed quite easily with FeynCalc. Thus, also a very important part of this calculation here, was the Mathematica code that I wrote from scratch. This code calculates the $\mathcal{I}_{(1/2)}$ functions analytically and simplifies very efficiently the final answers in terms of $s,t,u$ and $\epsilon$.

\vspace{0.3cm}

Thus, introducing the sum 
\begin{equation}
\mathcal{I}^{(\rm dr)}(s,t,u) = \mathcal{I}^{(1)}(s,t,u)  + \mathcal{I}^{(2)}(s,t,u),
\end{equation}
we obtain the result, in $d=4-2\epsilon$ dimensions as,
\begin{equation}
\begin{split}
\mathcal{I}^{(\rm dr)}(s,t,u) = \frac{i}{(4\pi)^2}\left (\frac{4\pi}{Q^2_s}\right )^{\epsilon}R_{\Gamma}\mathcal{A}^{(\rm LO)}(s,t,u)\mathcal{B}^{(\rm dr)}(s,t,u),
\end{split}
\end{equation}
where
\begin{equation}
\label{eq:diroxcoef}
\begin{split}
&\mathcal{B}^{\rm dr}(s,t,u) = -\frac{2}{\epsilon^2} +\frac{2}{\epsilon}\ln\left (\frac{-s-i\delta}{Q^2_s}\right ) -\ln^2\left (\frac{-s-i\delta}{Q^2_s}\right )\\
& + \left (1+\frac{1}{2}\frac{s^2-u^2}{s^2+u^2}\right )\biggl \{\pi^2 + \ln^2\left (\frac{-t-i\delta}{Q^2_s}\right ) + \ln^2\left (\frac{-s-i\delta}{Q^2_s}\right )\\
& -2\ln\left (\frac{-t-i\delta}{Q^2_s}\right )\ln\left (\frac{-s-i\delta}{Q^2_s}\right ) \biggr \}\\
& + \left (\frac{ut}{s^2+u^2}\right ) \biggl \{\ln\left (\frac{-t-i\delta}{Q^2_s}\right )-\ln\left (\frac{-s-i\delta}{Q^2_s}\right )\biggr \}.
\end{split}
\end{equation}
Since $s > 0$, $t \leq 0$ and $Q^2_s>0$, the real part of $\mathcal{B}^{\rm dr}(s,t,u)$ is 
\begin{equation}
\label{eq:diroxcoefreal}
\begin{split}
&\mathbb{R}e\biggl [\mathcal{B}^{\rm dr}(s,t,u)\biggr ] = -\frac{2}{\epsilon^2} +\frac{2}{\epsilon}\ln\left (\bigg\vert \frac{s}{Q^2_s}\bigg\vert\right ) + \ln^2\left (\bigg\vert\frac{t}{Q^2_s}\bigg\vert\right ) + \pi^2\\
& + \frac{1}{2}\left (\frac{s^2-u^2}{s^2+u^2}\right )\biggl \{\ln^2\left (\bigg\vert\frac{t}{Q^2_s}\bigg\vert\right ) + \ln^2\left (\bigg\vert\frac{s}{Q^2_s}\bigg\vert\right )\\
&-2\ln\left (\bigg\vert\frac{t}{Q^2_s}\bigg\vert\right )\ln\left (\bigg\vert\frac{s}{Q^2_s}\bigg\vert\right )\biggr \}\\
& + \left (\frac{ut}{s^2+u^2}\right )\biggl \{\ln\left (\bigg\vert\frac{t}{Q^2_s}\bigg\vert\right ) - \ln\left (\bigg\vert\frac{s}{Q^2_s}\bigg\vert\right )\biggr \}\\
& -2\ln\left (\bigg\vert\frac{t}{Q^2_s}\bigg\vert\right )\ln\left (\bigg\vert\frac{s}{Q^2_s}\bigg\vert\right ).
\end{split}
\end{equation}
In Eq.\ \eqref{eq:diroxcoefreal} it is understood that since only the real part is kept
\begin{equation}
\ln\left (\frac{-x-i\delta}{Q^2_s}\right ) \overset{\delta\rightarrow 0_+}{\longrightarrow} \ln\left (\bigg\vert \frac{x}{Q^2_s}\bigg\vert\right ) \quad \text{for} \quad x=s~\text{or}~t,
\end{equation} 
and
\begin{equation}
\begin{split}
\ln^2\left (\frac{-t-i\delta}{Q^2_s}\right )  &\overset{\delta\rightarrow 0_+}{\longrightarrow} \ln^2\left (\bigg\vert\frac{t}{Q^2_s}\bigg\vert\right ),\\ 
\ln^2\left (\frac{-s-i\delta}{Q^2_s}\right )  &\overset{\delta\rightarrow 0_+}{\longrightarrow} \ln^2\left (\bigg\vert\frac{s}{Q^2_s}\bigg\vert\right )-\pi^2. 
\end{split}
\end{equation}
Thus, we obtain
\begin{equation}
\label{eq:drboxfinal}
\begin{split}
\mathcal{A}^{(\textrm{NLO})}_{(\rm BOX, dr)} = \frac{-2g_0^6\mu_0^{4\epsilon}}{(4\pi)^2}\left (\frac{4\pi\mu_0^2}{Q^2_s}\right )^{\epsilon}&R_{\Gamma}\mathcal{A}^{(\rm LO)}(s,t,u)\\
&\times \mathcal{C}^{(\textrm{dr})}_{(\text{BOX+LO})}\mathbb{R}e\biggl [\mathcal{B}^{\rm dr}(s,t,u)\biggr ].
\end{split}
\end{equation}
Similarly, the contribution from the crossed box interference term is obtained from Eq.\ \eqref{eq:drboxfinal} by making the following substitutions:
\begin{enumerate}
\item Replace $k_2\rightarrow k_3, k_3 \rightarrow -k_2$, {\rm i.e.} $(s,u)\rightarrow (-u,s)$,
\item Multiply by an overall minus sign, which follows from the fact that a quark has been switched between the incoming and outgoing states, 
\item Change the color factor as $\mathcal{C}^{(\textrm{dr})}_{(\text{BOX+LO})}\rightarrow \mathcal{C}^{(\textrm{cr})}_{(\text{BOX+LO})} = 7/3$.
\end{enumerate}
Thus, recalling that $\mathcal{A}^{(\rm LO)}(-u,t,s) = \mathcal{A}^{(\rm LO)}(s,t,u)$, we find 
\begin{equation}
\label{eq:crboxfinal}
\begin{split}
\mathcal{A}^{(\textrm{NLO})}_{(\rm BOX, cr)} = \frac{2g_0^6\mu_0^{4\epsilon}}{(4\pi)^2}\left (\frac{4\pi\mu_0^2}{Q^2_s}\right )^{\epsilon}&R_{\Gamma}\mathcal{A}^{(\rm LO)}(s,t,u)\\
& \times \mathcal{C}^{(\textrm{cr})}_{(\text{BOX+LO})}\mathbb{R}e\biggl [\mathcal{B}^{\rm dr}(-u,t,s)\biggr ],
\end{split}
\end{equation}
where, again for $u \leqslant 0$, $t \leqslant 0$ and $s > 0$,
\begin{equation}
\label{eq:crosxcoefreal}
\begin{split}
&\mathbb{R}e\biggl [\mathcal{B}^{\rm dr}(-u,t,s)\biggr ] = -\frac{2}{\epsilon^2} +\frac{2}{\epsilon}\ln\left (\bigg\vert \frac{u}{Q^2_s}\bigg\vert\right ) + \ln^2\left (\bigg\vert \frac{t}{Q^2_s}\bigg\vert \right ) + \pi^2\\
& - \frac{1}{2}\left (\frac{s^2-u^2}{s^2+u^2}\right )\biggl \{\ln^2\left (\bigg\vert \frac{t}{Q^2_s}\bigg\vert \right ) + \ln^2\left (\bigg\vert \frac{u}{Q^2_s}\bigg\vert \right )\\
&-2\ln\left (\bigg\vert\frac{t}{Q^2_s}\bigg\vert\right )\ln\left (\bigg\vert\frac{u}{Q^2_s}\bigg\vert\right )\biggr \}\\
& + \left (\frac{st}{s^2+u^2}\right )\biggl \{\ln\left (\bigg\vert\frac{t}{Q^2_s}\bigg\vert\right ) - \ln\left (\bigg\vert\frac{u}{Q^2_s}\bigg\vert\right )\biggr \}\\
& -2\ln\left (\bigg\vert\frac{t}{Q^2_s}\bigg\vert\right )\ln\left (\bigg\vert\frac{u}{Q^2_s}\bigg\vert\right ).
\end{split}
\end{equation}
Finally, summing the direct and crossed box contributions in Eqs. \eqref{eq:drboxfinal} and \eqref{eq:crboxfinal} together, the full one-loop virtual box correction to the squared NLO amplitude can be cast in the following form: 
\begin{equation}
\label{eq:boxfinal}
\begin{split}
\mathcal{A}^{(\textrm{NLO})}_{\rm BOX} = 2\frac{g_0^6\mu_0^{4\epsilon}}{(4\pi)^2}&\left (\frac{4\pi\mu_0^2}{Q^2_s}\right )^{\epsilon}R_{\Gamma}\mathcal{A}^{(\rm LO)}(s,t,u)\biggl \{-\frac{6}{\epsilon^2}\\
& + \frac{1}{\epsilon}\biggl [\frac{4}{3}\ln\left (\bigg\vert \frac{s}{Q^2_s}\bigg\vert\right ) + \frac{14}{3}\ln\left (\bigg\vert \frac{u}{Q^2_s}\bigg\vert\right ) \biggr ]\\
&\quad\quad\quad + 3\ln^2\left (\bigg\vert\frac{t}{Q^2_s}\bigg\vert\right ) + 3\pi^2 + \mathcal{B}(s,t,u)\biggr \},
\end{split}
\end{equation}
where the function $\mathcal{B}$ is defined by
\begin{equation}
\label{eq:Bfunction}
\begin{split}
\mathcal{B}(s,t,u) = &\frac{1}{2}\left (\frac{s^2-u^2}{s^2+u^2}\right )\biggl \{-\frac{5}{3}\ln^2\left (\bigg\vert\frac{t}{Q^2_s}\bigg\vert\right )\\
& + \frac{2}{3}\ln^2\left (\bigg\vert\frac{s}{Q^2_s}\bigg\vert\right )- \frac{7}{3}\ln^2\left (\bigg\vert\frac{u}{Q^2_s}\bigg\vert\right )\\
& -\ln\left (\bigg\vert\frac{t}{Q^2_s}\bigg\vert\right )\biggl [\frac{4}{3}\ln\left (\bigg\vert\frac{s}{Q^2_s}\bigg\vert\right ) - \frac{14}{3}\ln\left (\bigg\vert\frac{u}{Q^2_s}\bigg\vert\right )\biggr ]\biggr \}\\
& +\frac{2}{3}\left (\frac{ut}{s^2+u^2}\right )\biggl \{\ln\left (\bigg\vert\frac{t}{Q^2_s}\bigg\vert\right ) - \ln\left (\bigg\vert\frac{s}{Q^2_s}\bigg\vert\right )\biggr \}\\
& +\frac{7}{3}\left (\frac{st}{s^2+u^2}\right )\biggl \{\ln\left (\bigg\vert\frac{t}{Q^2_s}\bigg\vert\right ) - \ln\left (\bigg\vert\frac{u}{Q^2_s}\bigg\vert\right )\biggr \}\\
& -\ln\left (\bigg\vert\frac{t}{Q^2_s}\bigg\vert\right )\biggl \{\frac{4}{3}\ln\left (\bigg\vert\frac{s}{Q^2_s}\bigg\vert\right ) + \frac{14}{3}\ln\left (\bigg\vert\frac{u}{Q^2_s}\bigg\vert\right )\biggr \}.
\end{split}
\end{equation}

\subsection{Full result for the unrenormalized $qq'\rightarrow qq'$  virtual correction}

At this point, we are ready to write the full unrenormalized result for the $qq'\rightarrow qq'$ virtual NLO correction defined in Eq.\ \eqref{eq:NLOFULLVIR}. Summing the self-energy, vertex and box contributions from Eqs.\ \eqref{eq:finalGSEBORNvir}, \eqref{eq:finalvertexBORNvir} and \eqref{eq:boxfinal} together, we obtain
\begin{equation}
\begin{split}
\mathcal{A}^{(\textrm{NLO})}_{\rm virtual,0} & = \mathcal{A}^{(\textrm{NLO})}_{\rm gSE} + \mathcal{A}^{(\textrm{NLO})}_{\rm V} + \mathcal{A}^{(\textrm{NLO})}_{\rm BOX} = g_0^6\mu_0^{6\epsilon}\mathcal{A}^{(\textrm{NLO})},
\end{split}
\end{equation}
where 
\begin{equation}
\label{eq:FINALUNR}
\mathcal{A}^{(\textrm{NLO})} = \frac{1}{8\pi^2}\left (\frac{4\pi}{Q^2_s}\right )^{\epsilon}R_{\Gamma}\mathcal{A}^{\rm LO}(s,t,u)\mathcal{H}^{(\textrm{NLO})}_{\textrm{virtual,0}}(s,t,u).
\end{equation}
Here the unrenormalized NLO kernel $\mathcal{H}^{(\textrm{NLO})}_{\textrm{virtual,0}}$ for the $qq'\rightarrow qq'$ scattering process is given by 
\begin{equation}
\begin{split}
\mathcal{H}^{(\textrm{NLO})}_{\textrm{virtual,0}} = & -\frac{1}{\epsilon^2}\left (6-\frac{2}{N_c}\right ) + \frac{1}{\epsilon}\biggl \{\frac{4}{3}\ln\left (\bigg\vert \frac{s}{Q^2_s}\bigg\vert \right ) + \frac{14}{3}\ln\left (\bigg\vert \frac{u}{Q^2_s}\bigg\vert  \right )\\
& - \frac{2}{N_c}\ln\left (\bigg\vert \frac{t}{Q^2_s}\bigg\vert  \right )\biggr \} + \frac{1}{\epsilon}\biggl \{\frac{2N_c^2+9}{3N_c}-\frac{4T_R}{3}\biggr \}\\
& - \ln\left (\bigg\vert \frac{t}{Q^2_s}\bigg\vert \right )\left (\frac{2N_c^2+9}{3N_c}\right ) + \ln^2\left (\bigg\vert\frac{t}{Q^2_s}\bigg\vert\right )\left (\frac{3N_c+1}{N_c}\right ) \\
& + \mathcal{B}(s,t,u) + 3\pi^2 -\frac{20T_R}{9} +  \frac{13N_c^3+72}{9N_c},
\end{split}
\end{equation}
where $T_R = n_q/2$ and the function $\mathcal{B}(s,t,u)$ is given by Eq.\ \eqref{eq:Bfunction}.

\section{Ultraviolet renormalization}

The one-loop virtual corrections we have computed\footnote{Recall that the box corrections are UV-finite.} are ultraviolet divergent and the result in Eq.\ \eqref{eq:FINALUNR} includes $1/\epsilon$ poles which must be renormalized away. In practice, renormalization is performed as follows\footnote{See also \cite{BOJAK}.}: First, in terms of the bare and dimensionless coupling $g_0$, the unrenormalized squared amplitude $\mathcal{A}$ in Eq.\ \eqref{eq:amplitude} has the perturbative expansion
\begin{equation}
\mathcal{A}_0 = g_0^4\mu_0^{4\epsilon}\mathcal{A}^{(\textrm{LO})} + g_0^6\mu_0^{6\epsilon}\mathcal{A}^{(\textrm{NLO})} + \mathcal{O}(g_0^8).
\end{equation}
Next, replacing the bare couping $g = g_0\mu_0^{\epsilon}$ with the renormalized running coupling $\tilde{g}_R = g_R\mu_R^{\epsilon}$ ($g_R$ is dimensionless and tilde for the dimensionful renormalized coupling at $d=4-2\epsilon$) evaluated at the renormalization scale $\mu_R$,
\begin{equation}
g = Z_g(\mu_R)\tilde{g}_R,
\end{equation}
we  obtain the renormalized perturbative expansion:
\begin{equation}
\label{eq:renorma}
\mathcal{A}_R = g_R^4\mu_R^{4\epsilon}\mathcal{A}^{\textrm{LO}}\biggl [Z_g^4 + g_R^2\mu_R^{2\epsilon}\left (\frac{\mathcal{A}^{\textrm{NLO}}}{\mathcal{A}^{\textrm{LO}}}\right )Z_g^2\biggr ] + \mathcal{O}(g_R^8),
\end{equation}
where $\mathcal{A}_R$ is the renormalized squared amplitude.  The value of the renormalization constant $Z_g$ can be calculated perturbatively. In the $\overline{\textrm{MS}}$ scheme \cite{BFUNC,MUTA},
\begin{equation}
Z_g(\mu_R)^2 = 1 - \left (\frac{\alpha_s}{2\pi}\right )\frac{\beta_0}{\hat{\epsilon}} + \mathcal{O}(\alpha_s^2),
\end{equation}
where 
\begin{equation}
\frac{1}{\hat{\epsilon}} = (4\pi)^{\epsilon}R_{\Gamma}\frac{1}{\epsilon},
\end{equation}
the renormalized strong coupling $\alpha_s = g_R^2/(4\pi)$ and
\begin{equation}
\beta_0 = \frac{11N_c-4T_R}{6}.
\end{equation}
The parameter $\beta_0$ is the first coefficient in the perturbative expansion of the QCD $\beta$-function which provides the $\mu_R$ dependence of the renormalized strong running coupling $\alpha_s$ as
\begin{equation}
\beta = \mu_R^2\frac{\partial \alpha_s}{\partial \mu_R^2} = -\alpha_s^2\left (\frac{\beta_0}{2\pi}\right ) + \cdots,
\end{equation}
where $\alpha_s = \alpha_s(\mu_R^2)$. Hence, the asymptotic behaviour at one-loop level can be written as
\begin{equation}
\alpha_s(\mu_R^2) = \frac{2\pi}{\beta_0\ln\left (\frac{\mu_R^2}{\Lambda_{\textrm{QCD}}^2}\right )},
\end{equation}
where $\Lambda_{\textrm{QCD}}$ marks the scale where perturbation theory definitely breaks down since there is an unphysical pole in the perturbative expansion at $\mu_R \rightarrow \Lambda_{\rm QCD}$.

\vspace{0.3cm}

Next, using the fixed ($k=1,2,\dots$) order series expansion of $Z_g$, 
\begin{equation}
Z_g(\mu_R)^{2k} = 1 - k\left (\frac{\alpha_s}{2\pi}\right )\frac{\beta_0}{\hat{\epsilon}} + \mathcal{O}(\alpha_s^2),
\end{equation}
and substituting this expression into Eq.\ \eqref{eq:renorma}, we find
\begin{equation}
\begin{split}
\mathcal{A}_R(s,t,u) = g_R^4&\mu_R^{4\epsilon}\mathcal{A}^{(\textrm{LO})}(s,t,u)\biggl [1 - 2\left (\frac{\alpha_s}{2\pi}\right )\frac{\beta_0}{\hat{\epsilon}}\\
& + \left (\frac{\alpha_s}{2\pi}\right )\left (\frac{4\pi\mu_R^2}{Q^2_s}\right )^{\epsilon}R_{\Gamma}\mathcal{H}^{(\textrm{NLO})}_{\textrm{virtual,0}}(s,t,u) \biggr ] + \mathcal{O}(g_R^8).
\end{split}
\end{equation}
Finally, rewriting 
\begin{equation}
\frac{\beta_0}{\hat{\epsilon}} = \left (\frac{4\pi\mu_R^2}{Q^2_s}\right )^{\epsilon}R_{\Gamma}\biggl [\frac{\beta_0}{\epsilon} + \beta_0\ln\left (\frac{Q^2_s}{\mu_R^2}\right )\biggr ],
\end{equation}   
we get the fully UV renormalized amplitude 
\begin{equation}
\label{eq:FINALANS}
\begin{split}
\mathcal{A}_R(s,t,u) = g_R^4\mu_R^{4\epsilon}&\mathcal{A}^{(\textrm{LO})}(s,t,u)\biggl [1 
 + \left (\frac{\alpha_s}{2\pi}\right )\left (\frac{4\pi\mu_R^2}{Q^2_s}\right )^{\epsilon}R_{\Gamma}\\
 & \times \mathcal{H}^{(\textrm{NLO})}_{\textrm{virtual,R}}(s,t,u) \biggr ] + \mathcal{O}(g_R^8),
\end{split}
\end{equation}
where the UV renormalized NLO kernel $\mathcal{H}^{(\textrm{NLO})}_{\textrm{virtual,R}}$ for the $qq'\rightarrow qq'$ scattering process is given by
\begin{equation}
\begin{split}
\mathcal{H}^{(\textrm{NLO})}_{\textrm{virtual,R}} & = -\frac{2\beta_0}{\epsilon} + 2\beta_0\ln\left (\frac{\mu_R^2}{Q^2_s}\right ) + \mathcal{H}^{(\textrm{NLO})}_{\textrm{virtual,0}}\\
 &= -\frac{1}{\epsilon^2}\left (6-\frac{2}{N_c}\right ) +  \frac{1}{\epsilon}\biggl \{\frac{2N_c^2+9-33N_c}{3N_c}\biggr \} \\
& + \frac{1}{\epsilon}\biggl \{\frac{4}{3}\ln\left (\bigg\vert \frac{s}{Q^2_s}\bigg\vert \right ) + \frac{14}{3}\ln\left (\bigg\vert \frac{u}{Q^2_s}\bigg\vert  \right )- \frac{2}{N_c}\ln\left (\bigg\vert \frac{t}{Q^2_s}\bigg\vert  \right )\biggr \} \\
& - \ln\left (\bigg\vert \frac{t}{Q^2_s}\bigg\vert \right )\left (\frac{2N_c^2+9}{3N_c}\right ) + \ln^2\left (\bigg\vert\frac{t}{Q^2_s}\bigg\vert\right )\left (\frac{3N_c+1}{N_c}\right ) \\
& + T_R\biggl \{\frac{4}{3}\biggl [\ln\left (\bigg\vert \frac{t}{Q^2_s}\bigg\vert \right )-\ln\left (\frac{\mu_R^2}{Q^2_s}\right )\biggr ] - \frac{20}{9}\biggr \} \\
& + \mathcal{B}(s,t,u) + \frac{11N_c}{3}\ln\left (\frac{\mu_R^2}{Q^2_s}\right ) + 3\pi^2+  \frac{13N_c^3+72}{9N_c}.
\end{split}
\end{equation}
Remarkably, the final result presented above fully agrees with the corresponding result, Eq.\ (2.9), in the original work by Ellis and Sexton  \cite{ELSE}.

\vspace{0.3cm}

In the next chapter, I will show how to compute the UV renormalized and IR and CL safe physical cross sections, based on the discussion here and in chapter 2.

\chapter{Computation of physical cross sections at NLO}
\label{EKS}

In this chapter, I will describe how to calculate physical cross sections for partonic scattering processes at NLO in pQCD. In practice, I will illustrate how to combine the UV renormalized $(2\rightarrow 2)$ virtual and $(2\rightarrow 3)$ real $\mathcal{O}(\alpha_s^3)$ IR and CL singular squared amplitudes, which were computed in $d=4-2\epsilon$ in \cite{ELSE}, into physical cross sections. The whole procedure is based on the subtraction method whose detailed documentation can be found in \cite{KUSO}. Again aiming at a self-contained study, in this chapter I briefly review the main features of this method in order to explain the cancellation of the singularities in sufficient detail. For clarity, I mostly keep the notation of \cite{KUSO}, and also otherwise closely follow the discussion in \cite{KUSO}.

\vspace{0.3cm}

I should also mention that in the algorithms used to define an observable high-$p_T$ jet, the typical way to arrive at an IR and CL safe NLO jet cross section is to introduce a finite size for the jet cone, see e.g. the discussion in \cite{JETCONE}. However, in this thesis we consider the production of minijets \cite{NLOET1,NLOET2} (whose $p_T\sim 1-3~{\rm GeV}$) which are not directly observable as final-state jets. Therefore, in our case, jet algorithms are not directly applicable but we need a more general definition for the physical (IR/CL safe) quantity we want to compute. The subtraction method discussed in \cite{KUSO} is ideal for our purposes.

\section{Physical cross section}
\label{PX}

At order $\mathcal{O}(\alpha_s^3)$, the cross section $\mathcal{I}_X$ for a physical quantity $X$ in a hadronic (or nuclear) collision can be written as 
\begin{equation}
\label{eq:jetcross}
\begin{split}
\mathcal{I}_X &= \frac{1}{2!}\int {\rm d[PS]}_2 \frac{{\rm d}\sigma(2\rightarrow 2)}{{\rm d[PS]}_2}\mathcal{S}_{X,2}(p_1^{\mu},p_2^{\mu})\\
& \quad\quad\quad\quad\quad\quad + \frac{1}{3!}\int {\rm d[PS]}_3 \frac{{\rm d}\sigma(2\rightarrow 3)}{{\rm d[PS]}_3}\mathcal{S}_{X,3}(p_1^{\mu},p_2^{\mu},p_3^{\mu}),\\
%& = \mathcal{I}(2\rightarrow 2) + \mathcal{I}(2\rightarrow 3),
\end{split}
\end{equation}
where the differential parton production cross sections ${\rm d}\sigma(2\rightarrow n)/{{\rm d[PS]}_n}$ at hadron level are defined in Eqs.\ \eqref{eq:2to2csfinal} and \eqref{eq:2to3csfinal} and the counting factors $1/n!$ need to be introduced when we treat all of the final state partons as though they were identical.
The measurement functions $\mathcal{S}_{X,2}$ and $\mathcal{S}_{X,3}$ which depend on the four-momenta of the final state partons specify the physical observable which in principle can be measured. An example of such an observable is jet production considered in \cite{KUSO}. If the observable is such that it is not sensitive to long-distance (small momentum exchange) physics, it can be described by pQCD, provided that the measurement functions are safe from the collinear and soft singularities. Hence, the mathematical requirements for $\mathcal{S}_{X,2}$ and $\mathcal{S}_{X,3}$ are that $\mathcal{S}_{X,3}$ should reduce to $\mathcal{S}_{X,2}$ when two of the outgoing partons become collinear or one of the outgoing partons becomes soft. For instance if $p_3^{\mu} \uparrow\uparrow p_2^{\mu}$
\begin{equation}
\label{eq:cri1}
\mathcal{S}_{X,3}(p_1^{\mu},(1-\lambda)p_2^{\mu},\lambda p_2^{\mu}) = \mathcal{S}_{X,2}(p_1^{\mu},p_2^{\mu}),
\end{equation}
where $0 \leq \lambda \leq 1$ and where $\lambda \rightarrow 0$ shows the soft limit $p_3^{\mu}=0$. The same should happen for example if $p_3^{\mu}\uparrow\uparrow p_{A}^{\mu}(p_B^{\mu})$
\begin{equation}
\label{eq:cri2}
\begin{split}
\mathcal{S}_{X,3}(p_1^{\mu},p_2^{\mu},\lambda p_{A,B}^{\mu})  = \mathcal{S}_{X,2}(p_1^{\mu},p_2^{\mu}).
\end{split}
\end{equation}
Here, we also assume that the above equations for $\mathcal{S}_{X,3}$ and $\mathcal{S}_{X,2}$ are satisfied in the different permutations of the four-momenta $p_i^{\mu} \uparrow\uparrow p_j^{\mu}$ and $p_i^{\mu} \uparrow\uparrow p_A^{\mu}(p_B^{\mu})$ for $i,j=1,2,3$.

\section{Singularities of the $(2\rightarrow 2)$ cross section}

In this section we discuss how Kunszt and Soper in \cite{KUSO} wrote the $(2\rightarrow 2)$ cross section into terms containing IR and CL  $1/\epsilon, 1/\epsilon^2$ poles and terms that are finite as $\epsilon \rightarrow 0$. The leading idea here is that these singular terms are cancelled against similar terms in the $(2\rightarrow 3)$ part of the cross section.

\vspace{0.3cm}

First, by substituting the expression for the $(2\rightarrow 2)$ differential 2-parton production cross section of Eq.\ \eqref{eq:2to2csfinal} into the $(2\rightarrow 2)$ part of the cross section in Eq.\ \eqref{eq:jetcross}, we find
\begin{equation}
\label{eq:permatel}
\begin{split}
\mathcal{I}(2\rightarrow 2)_X = \frac{1}{2!}&\int  {\rm d[PS]}_2 \sum_{\{a\}_2}\frac{p_{\perp,2}}{16\pi^2s^2}\left (\frac{p_{\perp,2}}{2\pi\mu}\right )^{-2\epsilon}\frac{f_{A,0}(a_A,x_A)}{x_A}\\
&\frac{f_{B,0}(a_B,x_B)}{x_B}\langle \vert \mathcal{M}(a_Aa_B\rightarrow a_1a_2)\vert^2\rangle\mathcal{S}_{X,2}(p_1^{\mu},p_2^{\mu}),
\end{split}
\end{equation}
where we denote $\{a\}_2  \equiv \{a_A,a_B,a_1,a_2\}$ and the $\mu$ corresponds now to the renormalized coupling $g_R$. Next, the UV renormalized squared amplitudes (of which the $\mathcal{A}_{\rm R}(s,t,u)$ in Eq.\ \eqref{eq:FINALANS} is an example) can be conveniently written in terms of functions named $\psi^{(4)}$ and $\psi^{(6)}$:
\begin{equation}
\label{eq:matper}
\begin{split}
\langle \vert \mathcal{M}(a_Aa_B\rightarrow a_1a_2)&\vert^2\rangle = \frac{(4\pi\alpha_s^2)}{\omega(a_A)\omega(a_B)}\biggl \{\psi^{(4)}(\{a\}_2,\{p^{\mu}\}_2)\\
& + \frac{\alpha_s}{2\pi}\left (\frac{4\pi\mu^2}{Q^2_s}\right )R_{\Gamma}\psi^{(6)}(\{a\}_2,\{p^{\mu}\}_2) + \mathcal{O}(\alpha_s^2) \biggr \},
\end{split}
\end{equation}
where we denote $\{p^{\mu}\}_2 \equiv \{p_A^{\mu},p_B^{\mu},p_1^{\mu},p_2^{\mu}\}$ and the quantities $R_{\Gamma}$ and $Q^2_s$ are defined in Eqs.\ \eqref{eq:RGAMMA} and \eqref{eq:finalGSEBORNvir}. The functions $\psi^{(4)}$ and $\psi^{(6)}$ contain the $(2\rightarrow 2)$ Born-level $\mathcal{O}(\alpha_s^2)$ matrix elements squared and their one-loop $\mathcal{O}(\alpha_s^3)$ virtual corrections, respectively, summed over the initial and final spin and color. The analytical expressions for these functions are nicely presented in \cite{KUSO}. 

\vspace{0.3cm}

In order to match the parton distribution functions to the squared NLO amplitudes, one needs to introduce also the renormalized parton distributions $f_i(a_A,x_A)$ defined in the $\overline{\rm MS}$ scheme \cite{KUSO}
\begin{equation}
\label{eq:modparton}
\begin{split}
f_{i,0}(a_i,x_i) = \sum_{a'_i}\int_{x_i}^{1}\frac{{\rm d}z}{z} &f_i(a'_i,\frac{x_i}{z})\biggl \{\delta_{a'_ia_i}\delta(1-z)\\
& + \frac{(4\pi)^{\epsilon}}{\epsilon\Gamma(1-\epsilon)}\left (\frac{\alpha_s}{2\pi}\right )\mathcal{P}_{a_i/a'_i}(z) + \mathcal{O}(\alpha_s^2)\biggr \},
\end{split}
\end{equation}
where the $\mathcal{P}_{a_i/a'_i}(z)$ are the standard $\epsilon$-independent LO Altarelli-Parisi kernels given in \cite{ALTPAR} and the index $i$ denotes the type of the hadron A or B. Note that these "modified" PDFs do not have the scale dependence yet, since we only care about the singular parts. However, in the end, one must, also add the scale dependence to these distributions by including the standard $\propto -\left (\frac{\alpha_s}{2\pi}\right )\mathcal{P}_{a_i/a'_i}(z)\log(\mu_{\rm F}^2/\mu^2)$ term in Eq.\ \eqref{eq:modparton}, where the factorization scale $\mu_{\rm F}^2$ is of the order of the physical scale in the process \cite{AEM}.

\vspace{0.3cm}

Substituting the expressions in Eqs.\ \eqref{eq:matper} and \eqref{eq:modparton} into the $\mathcal{I}(2\rightarrow 2)_X$  cross section formula of Eq.\ \eqref{eq:permatel}, we obtain a sum of three terms:
\begin{equation}
\mathcal{I}(2\rightarrow 2)_X = \int{\rm d[PS]}_2\biggl \{\mathcal{I}_{\textrm{BORN}}(2\rightarrow 2) + \mathcal{I}_{\textrm{CT}}(2\rightarrow 2) + \mathcal{I}_{\textrm{HO}}(2\rightarrow 2) \biggr \},
\end{equation}
where the term 
\begin{equation}
\begin{split}
\mathcal{I}_{\textrm{BORN}}(2\rightarrow 2) = \alpha_s^2 \frac{p_{\perp,2}}{2s^2}\left (\frac{p_{\perp,2}}{2\pi\mu}\right )^{-2\epsilon}&\sum_{\{a\}_2}\mathcal{L}(a_A,a_B,x_A,x_B)\\
&\psi^{(4)}(\{a\}_2,\{p^{\mu}\}_2)\mathcal{S}_{X,2}(p_1^{\mu},p_2^{\mu}),
\end{split}
\end{equation}
is the Born-level (BORN) integrand \cite{KUSO} and the information of the parton luminosity and the averaging factors $\omega(a_{A,B})$ are contained in the function $\mathcal{L}$ defined by
\begin{equation}
\mathcal{L}(a_A,a_B,x_A,x_B) = \frac{f_A(a_A,x_A)}{\omega(a_A)x_A}\frac{f_{B}(a_B,x_B)}{\omega(a_B)x_B}.
\end{equation}
The term denoted $\mathcal{I}_{\rm CT}$ now contains the contribution which originates from the $\overline{\rm MS}$ definition of the PDFs in Eq.\ \eqref{eq:modparton}:
\begin{equation}
\begin{split}
\mathcal{I}_{\textrm{CT}}&(2\rightarrow 2) = \alpha_s^2 \frac{p_{\perp,2}}{2s^2}\left ( \frac{p_{\perp,2}}{2\pi\mu}\right )^{-2\epsilon}\sum_{\{a\}_2}\mathcal{F}(\{a,a',x\}_{A,B})\\
&\frac{(4\pi)^{\epsilon}}{\epsilon\Gamma(1-\epsilon)}\left (\frac{\alpha_s}{2\pi}\right )\psi^{(4)}(\{a\}_2,\{p^{\mu}\}_2)\mathcal{S}_{X,2}(p_1^{\mu},p_2^{\mu}),
\end{split}
\end{equation}
where the function $\mathcal{F}$ is given by
\begin{equation}
\begin{split}
\mathcal{F}(\{a,a',x\}_{A,B})& = \sum_{a'_A}\int_{x_A}^{1}\frac{{\rm d}z}{z} \frac{\omega(a'_A)}{z\omega(a_A)}\mathcal{L}(a'_A,a_B,\frac{x_A}{z},x_B)\mathcal{P}_{a_A/a'_A}(z)\\
& + \sum_{a'_B}\int_{x_B}^{1}\frac{{\rm d}z}{z} \frac{\omega(a'_B)}{z\omega(a_B)}\mathcal{L}(a_A,a'_B,x_A,\frac{x_B}{z})\mathcal{P}_{a_B/a'_B}(z),
\end{split}
\end{equation}
and we denote
\begin{equation}
\{a,a',x\}_{A,B} = \{a_A,a_B,a'_A,a'_B,x_A,x_B\}.
\end{equation}
The last term, $\mathcal{I}_{\rm HO}$, then contains the $\mathcal{O}(\alpha_s^3)$ terms of Eq.\ \eqref{eq:matper}, so that to this order we can simply substitute $f_{i,0} \rightarrow f_i$, arriving at
\begin{equation}
\begin{split}
\mathcal{I}_{\textrm{HO}}(2\rightarrow 2) = \alpha_s^3\frac{p_{\perp,2}}{4\pi s^2}\left (\frac{16\pi^3\mu^4}{p_{\perp,2}^2Q^2_s}\right )^{\epsilon}&R_{\Gamma}\sum_{\{a\}_2}\mathcal{L}(a_A,a_B,x_A,x_B)\\
\psi^{(6)}(\{a\}_2,\{p^{\mu}\}_2)\mathcal{S}_{X,2}(p_1^{\mu},p_2^{\mu}).
\end{split}
\end{equation}
The function $\psi^{(4)}$ is finite and the singular structure of $\psi^{(6)}$ has the simple general form \cite{KUSO}:
\begin{equation}
\label{eq:singbeh}
\begin{split}
\psi^{(6)} = \psi^{(4)}\biggl \{&-\frac{1}{\epsilon^2}\sum_{n}\mathcal{C}(a_n) - \frac{1}{\epsilon}\sum_{n}\gamma(a_n)\biggr \}\\
& + \frac{1}{2\epsilon}\sum_{\substack{m,n\\ n\neq m}}\ln\left (\frac{2p_n\cdot p_m}{Q^2}\right )\psi_{mn}^{(4,c)}(\{a\}_2,\{p^{\mu}\}_2)\\
& + \psi_{\textrm{NS}}^{(6)}(\{a\}_2,\{p^{\mu}\}_2),
\end{split}
\end{equation}
where $n,m=A,B,1,2$, and the $p_{n(m)}$ denote the external parton momenta. The $\epsilon$-dependent functions $\psi^{(4)}_{mn}$ are related to $\psi^{(4)}$ as discussed in \cite{KUSO}.

\section{Singularities of the $2\rightarrow 3$ cross section}

In this section, we inspect how one can cleverly decompose the $\mathcal{I}(2\rightarrow 3)_X$ part of the cross section into terms that are finite as $\epsilon \rightarrow 0$, and terms that become infinite as $\epsilon \rightarrow 0$ \cite{KUSO}, and sketch how the singularities of the $(2\rightarrow 2)$ and $(2\rightarrow 3)$ parts cancel in the end.

\vspace{0.3cm}

First, by substituting the expression for the $(2\rightarrow 3)$ differential 3-parton production cross section of Eq.\ \eqref{eq:2to3csfinal} into the $\mathcal{I}(2\rightarrow 3)_X$ part of the cross section in Eq.\ \eqref{eq:jetcross}, we obtain
\begin{equation}
\label{eq:2to3partv1}
\begin{split}
\mathcal{I}(2\rightarrow 3)_X = \frac{1}{2!}&\int  {\rm d[PS]}_3  \sum_{\{a\}_3}\frac{p_{\perp,2}p_{\perp,3}}{8(2\pi)^5s^2}\left (\frac{p_{\perp,2}p_{\perp,3}}{(2\pi)^2\mu^2}\right )^{-2\epsilon}\\
&\frac{f_{A,0}(a_A,x_A)}{x_A}\frac{f_{B,0}(a_B,x_B)}{x_B}\langle \vert \mathcal{M}(a_Aa_B\rightarrow a_1a_2a_3)\vert^2\rangle\\
&\Theta(p_{\perp,3} < p_{\perp,1})\Theta(p_{\perp,3} < p_{\perp,2})\mathcal{S}_3(p_1^{\mu},p_2^{\mu},p_3^{\mu}),
\end{split}
\end{equation}
where we denote $\{a\}_3 = \{a_A,a_B,a_1,a_2,a_2\}$. In this expression we have introduced two theta functions, $\Theta(p_{\perp,3} < p_{\perp,1})$ and $\Theta(p_{\perp,3} < p_{\perp,2})$, which appear since the parton labelled as 3 can be taken to be the one having smallest transverse momentum. This also cancels a factor 3 in the original prefactor 1/3!.

\vspace{0.3cm}

The squared $(2\rightarrow 3)$ amplitude in Eq.\ \eqref{eq:2to3partv1} is singular whenever two partons are collinear $(p_i^{\mu}\uparrow\uparrow p_j^{\mu})$ or one parton is soft $(p_i^{\mu}=0)$. Since now, with the above choice,
\begin{equation}
\label{eq:integreg}
p_{\perp,3} < p_{\perp,1}, \quad p_{\perp,3} < p_{\perp,2},
\end{equation}
and also due to the transverse momentum conservation, the only singularities that occur in the integration region of Eq.\ \eqref{eq:integreg} are the cases where the parton 3 is soft or when the parton 3 is collinear with any of the partons $A,B,1,2$. Thus, we can write the $(2\rightarrow 3)$ squared amplitude as
\begin{equation}
\label{eq:decomppsi}
\langle \vert \mathcal{M}(a_Aa_B\rightarrow a_1a_2a_3)\vert^2\rangle = \frac{(4\pi\alpha_s^3)}{\omega(a_A)\omega(a_B)}\Psi(\{a\}_3,\{p^{\mu}\}_3),
\end{equation}
where we denote $\{p^{\mu}\}_3 = \{p_A^{\mu},p_B^{\mu},p_1^{\mu},p_2^{\mu},p_3^{\mu}\}$, and where the function $\Psi$ can be expressed as,
\begin{equation}
\label{eq:decompsi2}
\Psi(\{a\}_3,\{p^{\mu}\}_3) = \sum_{\substack{m,n}} \frac{p_m\cdot p_n}{(p_m\cdot p_3)(p_n\cdot p_3)}\Psi_{mn}(\{a\}_3,\{p^{\mu}\}_3).
\end{equation}
In the expression above the singular factor $1/(p_m\cdot p_3)(p_n\cdot p_3)$ is extracted from the function $\Psi$ as explained in \cite{KUSO}, and the functions $\psi_{mn}$, which are symmetric in $n$ and $m$, are the coefficients of the singular terms in the squared $(2\rightarrow 3)$ scattering amplitude. As discussed in \cite{KUSO}, it is possible to construct the functions $\Psi_{mn}$ using the results of \cite{ELSE}. Next, in order to separate the collinear and soft singularities (and then also match these singularities to the $(2\rightarrow 2)$ part) the singular factor is rewritten in the following form \cite{KUSO},
\begin{equation}
\frac{1}{(p_m\cdot p_3)(p_n\cdot p_3)} = \frac{1}{(p_m+p_n)\cdot p_3(p_m\cdot p_3)} + \frac{1}{(p_m+p_n)\cdot p_3(p_n\cdot p_3)}.
\end{equation}
Using the expression above, the squared amplitude in Eq.\ \eqref{eq:decomppsi} can be decomposed according to Eq.\ \eqref{eq:decompsi2} into four terms
\begin{equation}
\langle \vert \mathcal{M}(a_Aa_B\rightarrow a_1a_2a_3)\vert^2\rangle = \langle \vert \mathcal{M}\vert^2\rangle_{A} + \langle \vert \mathcal{M}\vert^2\rangle_{B} + \langle \vert \mathcal{M}\vert^2\rangle_{1} + \langle \vert \mathcal{M}\vert^2\rangle_{2} , 
\end{equation}
where
\begin{equation}
\begin{split}
\langle \vert \mathcal{M}\vert^2\rangle_{n} = \frac{(4\pi\alpha_s^3)}{\omega(a_A)\omega(a_B)}\sum_{\substack{m=\{n\}_2\\ m\neq n}}S_{mn}\Psi_{mn}(\{a\}_3,\{p^{\mu}\}_3)
\end{split}
\end{equation}
and we denote $\{n\}_2 = \{A,B,1,2\}$. Here, the factor 
\begin{equation}
S_{mn} = \frac{(p_n\cdot p_m)}{p_n\cdot p_3(p_n+p_m)\cdot p_3}
\end{equation}
contains a $1/(p_n\cdot p_3)$ collinear singularity and a soft singularity for parton 3. Now, we can write the $(2\rightarrow 3)$ part of the cross section as
\begin{equation}
\mathcal{I}(2\rightarrow 3)_X = \mathcal{I}(2\rightarrow 3)_A + \mathcal{I}(2\rightarrow 3)_B + \mathcal{I}(2\rightarrow 3)_1 + \mathcal{I}(2\rightarrow 3)_2,
\end{equation}
where, for example, the last term becomes \cite{KUSO}
\begin{equation}
\mathcal{I}(2\rightarrow 3)_2 = \int {\rm d[PS]}_3 \frac{F_2(y_1,p_{\perp,2},\phi_2,p_{\perp,3},y_3,\phi_3)}{p_{\perp,3}\biggl [\cosh(y_2-y_3)-\cos(\phi_2-\phi_3) \biggr ]},
\end{equation}
and the function $F_2$ is given by
\begin{equation}
\label{eq:F2def}
\begin{split}
F_2 &= \Theta(p_{\perp,3} < p_{\perp,2})\Theta(p_{\perp,3} < p_{\perp,1})\frac{\alpha_s^3}{2(2\pi)^2s^2}\left ( \frac{p_{\perp,2}}{(2\pi)^2\mu}\right )^{-2\epsilon}\\
&\sum_{\{a\}_3} \mathcal{L}(a_A,a_B,a_1,a_2)\sum_{m=A,B,1}\frac{p_{\perp,3}(p_2\cdot p_m)\Psi_{mn}}{(p_2+p_m)\cdot p_3}\mathcal{S}_{X,3}(p_1^{\mu},p_2^{\mu},p_3^{\mu}).
\end{split}
\end{equation}
Here, the divergent factor $1/(p_2\cdot p_3)$ in the $(2\rightarrow 3)$ squared amplitude is rewritten in terms of the integration variables $y_2,y_3,\phi_2$ and $\phi_3$ as
\begin{equation}
\begin{split}
p_2\cdot p_3 & = p_2^+p_3^- + p_2^-p_3^+ - \mathbf{p}_{\perp,2}\cdot \mathbf{p}_{\perp,3}\\
& = p_{\perp,2}p_{\perp,3}\biggl [\cosh(y_2-y_3)-\cos(\phi_2-\phi_3) \biggr ].
\end{split}
\end{equation}
The three other terms $\mathcal{I}(2\rightarrow 3)_A$, $\mathcal{I}(2\rightarrow 3)_B$, and $\mathcal{I}(2\rightarrow 3)_1$ can be treated independently and in a similar manner, as explained in \cite{KUSO}. 

\vspace{0.3cm}

Next, each of the terms $\mathcal{I}(2\rightarrow 3)_n$ is decomposed into terms that are divergent and terms that are finite as $\epsilon \rightarrow 0$. In the sigular terms the integration over the phase space of the third parton can be performed analytically.  For example, let us consider the case of $\mathcal{I}(2\rightarrow 3)_2$, where this goal is achieved by inserting zero in Eq.\ \eqref{eq:F2def}, \textrm{i.e.} rewriting the function $F_2$ in the following way (see also \cite{NLOET2}): 
\begin{equation}
F_2(y_1,p_{\perp,2},\phi_2,p_{\perp,3},y_3,\phi_3) =  F_2(\textrm{soft}) + F_2(\textrm{collinear}) + F_2(\textrm{finite}),
\end{equation}
where
\begin{equation}
\begin{split}
F_2(\textrm{soft}) &= F_2(y_1,p_{\perp,2},y_2,\phi_2,0,y_3,\phi_3)\Theta(p_{\perp,3} < \frac{p_{\perp,2}}{2}),\\
F_2(\textrm{collinear}) &= F_2(y_1,p_{\perp,2},y_2,\phi_2,p_{\perp,3},y_2,\phi_2)\\
&-F_2(y_1,p_{\perp,2},y_2,\phi_2,0,y_2,\phi_2)\Theta(p_{\perp,3} < \frac{p_{\perp,2}}{2}),\\
F_2(\textrm{finite}) &= F_2(y_1,p_{\perp,2},y_2,\phi_2,p_{\perp,3},y_3,\phi_3)\\
& - F_2(y_1,p_{\perp,2},y_2,\phi_2,0,y_3,\phi_3)\Theta(p_{\perp,3} < \frac{p_{\perp,2}}{2})\\
& -F_2(y_1,p_{\perp,2},y_2,\phi_2,p_{\perp,3},y_2,\phi_2)\\
& + F_2(y_1,p_{\perp,2},y_2,\phi_2,0,y_2,\phi_2)\Theta(p_{\perp,3} < \frac{p_{\perp,2}}{2}).
\end{split}
\end{equation}
Here, in the "soft" term $p_{\perp,3}= 0$ in $F_2$ and $\Theta(p_{\perp,3} < \frac{p_{\perp,2}}{2})$ has been introduced to set (by hand) an upper bound for the $p_{\perp,3}$ integral. Similarly, in the "collinear" term  $y_3,\phi_3$ are set equal to $y_2,\phi_2$ in $F_2$ and a theta function above is included. The "finite" term is finite as $\epsilon \rightarrow 0$ and the phase space integration can be evaluated numerically in four dimensions, $\epsilon \rightarrow 0$, provided that the measurement function $\mathcal{S}_{X,3}$ is IR and CL safe as discussed in section \ref{PX}. Thus, we obtain the following decomposition
\begin{equation}
\mathcal{I}(2\rightarrow 3)_2 = \mathcal{I}(\textrm{soft})_2 + \mathcal{I}(\textrm{collinear})_2 + \mathcal{I}(\textrm{finite})_2.  
\end{equation}
Since the above procedure can be performed similarly to other three terms $\mathcal{I}(2\rightarrow 3)_A$, $\mathcal{I}(2\rightarrow 3)_B$, and $\mathcal{I}(2\rightarrow 3)_1$, see the details in \cite{KUSO}, we may write the full decomposition as
\begin{equation}
\mathcal{I}(2\rightarrow 3)_X = \sum_{\ell =\{n\}_2} \biggl [ \mathcal{I}(\textrm{soft})_{\ell} + \mathcal{I}(\textrm{collinear})_{\ell} + \mathcal{I}(\textrm{finite})_{\ell}\biggr ].
\end{equation}
Here, the phase space integration over the singular $(2\rightarrow 3)$ "soft" and "collinear" terms can be performed analytically. After the integration we are left with several terms that contain $1/\epsilon$ and $1/\epsilon^2$ poles and have $(2\rightarrow 2)$ kinematics \cite{KUSO}. Some of these terms then cancel each other and the remaining singular terms cancel against identical terms in $\mathcal{I}(2\rightarrow 2)_X$. This "term by term" cancellation procedure is nicely demonstrated in Table 1 of \cite{NLOET2}. 

\vspace{0.3cm}

Thus, the final result of the non-trivial exercise performed in \cite{KUSO} is that the physical cross section can be computed using 
\begin{equation}
\mathcal{I}_X = \mathcal{I}(2\rightarrow 2, \textrm{net})_X + \mathcal{I}(2\rightarrow 3, \textrm{net})_X, 
\end{equation}
where the notation "net" refers to removing the $1/\epsilon$ and $1/\epsilon^2$ terms by using the subtraction method and then setting $\epsilon \rightarrow 0$. The remaining finite terms are evaluated by multidimensional numerical integration.

\vspace{0.5cm}

Next, we turn to the phenomenological part of the thesis, where we apply the techniques learned here -- the UV-renormalized squared matrix amplitudes, the subtraction method with IR/CL-safe measurement functions to calculate minijet transverse energy production in $A+A$ collisions at the LHC and RHIC.

\chapter{The original EKRT model}
\label{oldEKRT}

In this chapter I will briefly review the key features of the original EKRT model \cite{EKRT}, which combines collinearly factorized pQCD minijet production with gluon saturation applying an uncertainty principle related geometrical saturation criterion for the produced minijets.

\section{Parton production and saturation}
\label{saturation}

In the original EKRT model \cite{EKRT}, the initial parton production above a minimum $p_T$ scale $p_0$ in a mid-rapidity unit $\Delta y$ in central $A+A$ collisions is computed by applying collinearly factorized pQCD \cite{EKL}. The leading idea of the model is that the low-$p_T$ parton (minijet, dominantly gluons) production is controlled by saturation, fusions, among the produced gluons. In this approach the saturation of initial parton production is conjectured to take place when the high-$p_T$ partons, which are produced before the low-$p_T$ partons, start to fill up the available average transverse area $\pi R_A^2$ of the 
central collision and eventually at some $p_T$ the whole transverse region is filled with the produced partons.
Further gluon production at lower $p_T$ is then not significant as the produced gluons just fuse with the harder ones. Thus, at saturation the scale $p_0 = p_{\textrm{sat}}$ is a solution of a saturation criterion \cite{EKRT}
\begin{equation}
\label{eq:orginalsat}
N_{AA}(p_0,A,\sqrt{s_{NN}},\Delta y)\frac{\pi}{p_0^2} = \kappa\pi R_A^2,
\end{equation}
where the factor $\pi/p_0^2$ is the effective transverse area occupied by each produced parton and $R_A$ is the nuclear radius which is calculated from the expression (see the Appendix of \cite{EKL})
\begin{equation}
R_A=1.12A^{1/3}-0.86A^{-1/3}. 
\end{equation}
The proportionality constant $\kappa$, which might also contain some power(s) of $\alpha_s$, is set to unity. In the saturation equation above, the key quantity is the number of minijets, $N_{AA}$, produced above a transverse momentum scale $p_{T} = p_0 \gg \Lambda_{\textrm{QCD}}$ into the rapidity acceptance window $\Delta y$. For a given collision energy  $\sqrt{s_{NN}}$ and nuclear mass number $A$ the number of produced minijets is obtained from \cite{EKRT,ETEKRT2}
\begin{equation}
N_{AA}(p_0,A,\sqrt{s_{NN}},\Delta y) = T_{AA}(\mathbf{0})\sigma\langle N\rangle_{p_0,\Delta y},
\end{equation}
where the collision geometry is given by the standard nuclear overlap function 
\begin{equation}
T_{AA}(\mathbf{b}) = \int {\rm d}^2\mathbf{s}T_A(\mathbf{s})T_A(\mathbf{s}-\mathbf{b}),
\end{equation}
with $\mathbf{b}$ denoting the impact parameter and $\mathbf{s}=(x,y)$ the transverse coordinate. The function $T_A(\mathbf{s})$ is the nuclear thickness function, which is an integral over the longitudinal coordinate $z$ of the nuclear density function $n_A(\mathbf{s},z)$,
\begin{equation}
T_A(\mathbf{s}) = \int_{-\infty}^{\infty} {\rm d}z n_A(\mathbf{s},z),
\end{equation}
where the nuclear density is parametrized with the Woods-Saxon profile
\begin{equation}
n_A(\mathbf{s},z) = \frac{n_0}{\exp\left (\frac{\sqrt{\vert \mathbf{s}\vert^2 + z^2} - R_A}{d}\right ) + 1},
\end{equation}
with $n_0=0.17~\rm fm^{-3}$ and $d=0.54~\rm fm$ \cite{EKL}. The quantity $\sigma\langle N\rangle_{p_0,\Delta y}$ is the perturbatively computable minijet cross section with a rapidity acceptance $\Delta y$. According to collinear factorization and leading-order pQCD the minijet cross section can be defined as \cite{ETEKRT2,PHD1}
\begin{equation}
\label{eq:Ndistri}
\begin{split}
\sigma\langle N\rangle_{p_0,\Delta y} = \int_{p_0^2} {\rm d}p_{\textrm{T}}^2\biggl \{\int_{\Delta y}{\rm d}y_1\int{\rm d}y_2 &+ \int{\rm d}y_1\int_{\Delta y}{\rm d}y_2\biggr \}\\
&\times \sum_{\langle k\ell \rangle}\frac{1}{1 + \delta_{k\ell}}\frac{{\rm d}\sigma^{AA \rightarrow k\ell + X}}{{\rm d}p_{\textrm{T}}^2{\rm d}y_1{\rm d}y_2},
\end{split}
\end{equation}
where the inclusive cross section for producing partons of flavours $k$ and $\ell$ is given by
\begin{equation}
\label{eq:Ndist2}
\begin{split}
\frac{{\rm d}\sigma^{AA \rightarrow k\ell + X}}{{\rm d}p_{\textrm{T}}^2{\rm d}y_1{\rm d}y_2} = K\sum_{ij}x_1f_{i/A}(x_1,\mu_{\rm F}^2)x_2f_{j/A}(x_2,\mu_{\rm F}^2)\frac{{\rm d}\hat{\sigma}(ij\rightarrow k\ell)}{{\rm d}\hat{t}}.	
\end{split}
\end{equation}
The DGLAP-evolved \cite{ALTPAR,DGLAP1,DGLAP2,DGLAP3} nuclear parton distribution functions (nPDFs) $f_{i/A}$ and $f_{j/A}$ are defined for each parton flavor $i$ and $j$ in terms of nuclear modifications $R_{i/A}(x,\mu_{\rm F}^2)$ and the corresponding free proton PDFs $f_{i/p}(x,\mu_{\rm F}^2)$ such that
\begin{equation}
f_{i/A}(x,\mu_{\rm F}^2) = R_{i/A}(x,\mu_{\rm F}^2)f_{i/p}(x,\mu_{\rm F}^2).
\end{equation}
The factor $K$ in Eq.\ \eqref{eq:Ndist2} accounts for the NLO corrections. In the original EKRT model this factor was introduced only in an effective sense, since the number of produced partons is well defined only in LO pQCD.

\vspace{0.3cm}

The original EKRT setup \cite{EKRT} exploited the LO GRV94 PDFs \cite{GRV} combined with the nuclear effects from the LO EKS98 parametrization \cite{EKS981,EKS982}. Also a constant $K=2$ was assumed in \cite{EKRT}. In the later EKRT setups \cite{ETEKRT1,ETEKRT2}, a $\sqrt{s_{NN}}$-dependent $K$ was introduced on the basis of the NLO computation of minijet transverse energy $E_T$ production (see section \ref{ETNLOCOMP}) \cite{NLOET1,NLOET2}.

\vspace{0.3cm}

The original EKRT model \cite{EKRT}, when coupled to ideal hydrodynamics, was successful in predicting the scaling laws for particle multiplicities in central $A+A$ collisions, ${\rm d}N/{\rm d}y \sim s_{NN}^{0.19\dots 0.20}A^{0.92}$. The later, more detailed setups were dramatically successful in predicting the LHC charged-particle multiplicity \cite{ETEKRT1} (measured a decate later) as well as describing the identical particle spectra at RHIC \cite{ETEKRT2,ENRR}. In spite of these successes, there are, however, a number of certain shortcomings in the original EKRT model. These are summarized in \cite{PHD2}, and discussed futher in chapter 6.

\section{Minijet transverse energy and initial energy density}
\label{ied}

\begin{figure}[h!]
\center
\includegraphics[scale=.45]{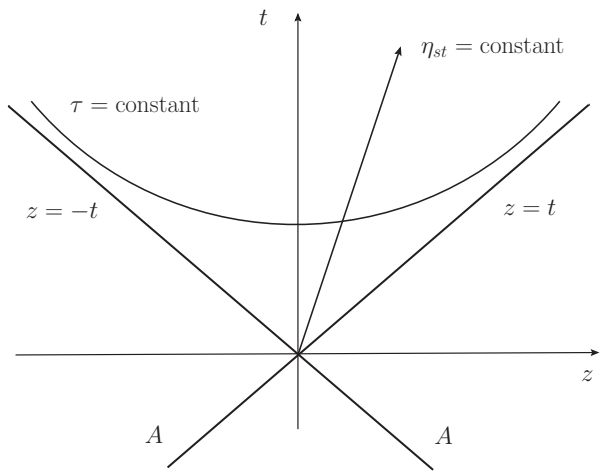}
\caption{$A+A$ collision in the $(z,t)$ plane, where the thick lines are the trajectories of the colliding nuclei, the light cones.}
\label{fig:etatau}
\end{figure}

Once the saturation scale is obtained as the solution $p_0=p_{\textrm{sat}}(\sqrt{s_{NN}},A)$ of Eq.\ \eqref{eq:orginalsat}, the amount of transverse energy $E_T$ carried by the minijets into the rapidity window $\Delta y$ can be computed as \cite{EKRT}
\begin{equation}
E_T(p_0,A,\sqrt{s_{NN}},\Delta y) = T_{AA}(\mathbf{0})\sigma\langle E_T\rangle_{p_0,\Delta y},
\end{equation}
where the first moment of the semi-inclusive $E_T$ distribution $\sigma\langle E_T\rangle_{p_0,\Delta y}$ to the LO accuracy \cite{EKL} is obtained from Eq.\ \eqref{eq:Ndistri} by integrating over the transverse momenta with a one extra weight of $p_{\textrm{T}}$ \cite{PHD1}:
\begin{equation}
\sigma\langle E_T\rangle_{p_0,\Delta y} = \int_{p_0^2}{\rm d}p_{T}^2p_{T}\frac{{\rm d}(\sigma\langle N\rangle_{p_0,\Delta y})}{{\rm d}p_{T}^2}.
\end{equation}
The pQCD calculation of the minijet $E_T$ production is formulated in the momentum space and a connection to spatial initial energy density needs to be established. At ultrarelativistic energies the nuclei are strongly contracted and the produced system forms a state where the forward/backward regions are in a very strong longitudinal expansion. Hence, originally motivated by the expectation of a longitudinal boost-invariance \cite{BOOSTINV}, it is convenient to use the light-cone coordinates $\tau$ and $\eta_{\textrm{st}}$ defined as
\begin{equation}
\tau = \sqrt{t^2-z^2},\quad\quad \eta_{\textrm{st}} = \frac{1}{2}\ln\left (\frac{t+z}{t-z}\right ),
\end{equation}
with $t = \tau\cosh(\eta_{\textrm{st}})$ and $z=\tau\sinh(\eta_{\textrm{st}})$. Here, the coordinate $\tau$ is the longitudinal proper time and $\eta_{\textrm{st}}$ is the space-time rapidity. The lines of constant $\tau$ and constant $\eta_{\textrm{st}}$ are sketched in Fig.\ \ref{fig:etatau}.
Since the typical partons at saturation have a longitudinal spread of $\Delta z \sim 1/p_{\textrm{sat}} \ll 1~{\rm fm}$, one can consider the collision region as a point in the $z$-direction and assume that the minijet rapidity is the same as the space-time rapidity, $y = \eta_{\textrm{st}}$. On the basis of the uncertainty principle the formation time is taken to be the inverse of the saturation scale, $\tau_0 = 1/p_{\textrm{sat}}$. Therefore, the initial energy density $\epsilon(\tau_0)$ and  $\tau_0$, which can be used as initial conditions for hydrodynamics, are obtained from the computed minijet $E_T(p_0=p_{\textrm{sat}})$ by assuming that the system thermalizes basically at formation $\tau_0 = 1/p_{\textrm{sat}}$.

\vspace{0.3cm}

In the original EKRT setup, where hydrodynamics with only 1-dimensional longitudinal Bjorken scaling flow was considered \cite{EKRT}, the averaged initial energy density, $\langle \epsilon \rangle$, of the produced minijet plasma in central $A$+$A$ collisions was obtained through the Bjorken estimate \cite{BOOSTINV}:
\begin{equation}
\langle \epsilon \rangle = \frac{E_{T}(p_0=p_{\textrm{sat}})}{\Delta V(\tau_0)},
\end{equation}
where the minijets in the mid-rapidity unit $\Delta y$ occupy a volume $\Delta V(\tau_0) = \pi R_A^2\tau_0\Delta y$ at the formation time $\tau_0 = 1/p_{\textrm{sat}}$. In the later, more detailed EKRT setups \cite{ETEKRT1,ETEKRT2}, where the pQCD calculation included the $\sqrt{s_{NN}}$-dependent $K$ factor \cite{NLOET1,NLOET2}, the saturated minijet initial conditions served as input for ideal (1+1)-dimensional hydrodynamics once a binary-collision (BC) transverse profile \cite{BCANDWN} for the initial energy density,
\begin{equation}
\label{eq:iedensityfor}
\epsilon(\mathbf{s},\tau_0) = \frac{{\rm d}E_T(p_0 = p_{\textrm{sat}})}{{\rm d}V(\tau_0)} = T_A(\mathbf{s})T_A(\mathbf{s})\frac{\sigma\langle E_T\rangle_{p_{\textrm{sat}},\Delta y}}{\tau_0\Delta y},
\end{equation}
was assumed. In Eq.\ \eqref{eq:iedensityfor} the transverse profile of the initial energy density is extracted by differentiating the overlap function $T_{AA}$ with respect to ${\rm d}^2\mathbf{s}$ in most central $(\mathbf{b} =\mathbf{0})$ $A+A$ collisions, and the differential volume element at $\tau=\tau_0$ and $\eta \approx y \approx 0$ is given by
\begin{equation}
\begin{split}
{\rm d}V(\tau)  = {\rm d}z{\rm d}^2{\mathbf{s}} = \left ({\rm d}\tau\sinh(\eta) + \tau\cosh(\eta){\rm d}\eta\right ){\rm d}^2{\mathbf{s}} \approx \tau_0\Delta y {\rm d}^2{\mathbf{s}}.
\end{split}
\end{equation}

\section{Computation of minijet $E_T$ in NLO}
\label{ETNLOCOMP}

As discussed in \cite{NLOET1,NLOET2}, the $E_T$ production for the pQCD minijet calculation can be extended to NLO in an infrared and collinear safe manner. This calculation is based on collinear factorization and the subtraction method \cite{KUSO} discussed in some detail in chapter \ref{EKS}. In this section I briefly explain the main features of this formulation.

\subsection{Measurement functions for $E_T$}

\begin{figure}[h!]
\center
\includegraphics[scale=.43]{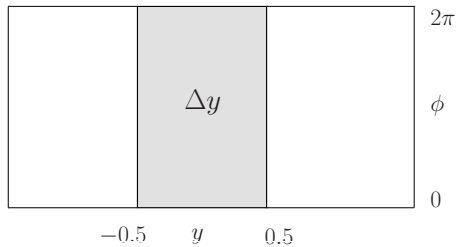}
\caption{The rapidity acceptance region $\Delta y$ in the $(y,\phi)$-plane.}
\label{fig:ycuts}
\end{figure}

In a hard scattering of partons at NLO, we can have one, two, three or zero minijets in a mid-rapidity region $\Delta y$, defined in the $(y,\phi)$-plane as
\begin{equation}
\Delta y: \quad \vert y\vert \leq 0.5, \quad 0 \leq \phi \leq 2\pi,
\end{equation}
where $\phi$ is the azimuthal angle and $y$ the rapidity, see Fig.\ \ref{fig:ycuts}. 
As only massless partons are considered, the transverse energy $E_T$ entering $\Delta y$ can be defined as a sum of the absolute values $p_{T,i}$ of the transverse momenta of those partons whose rapidities fall in $\Delta y$:
\begin{equation}
E_T = \epsilon(y_1)p_{T,1} + \epsilon(y_2)p_{T,2} + \epsilon(y_3)p_{T,3},
\end{equation}
where the step function $\epsilon(y_i)$ is defined as
\begin{equation}
\epsilon(y_i) \equiv \begin{cases}
1 \quad \textrm{if} \quad y_i\in\Delta y\\
0 \quad \textrm{otherwise}.
\end{cases}
\end{equation}
In the LO and NLO $(2\rightarrow 2)$ cases, the transverse momenta are equal in magnitude, $p_{T,1} = p_{T,2} = p_{T}$. Thus the perturbative scatterings can in this case be simply defined to be those with large enough transverse momentum, $p_{T} \geq p_0 \gg \Lambda_{\textrm{QCD}}$, or equivalently,
\begin{equation}
\label{eq:2to2hardness}
p_{T,1} + p_{T,2} \geq 2p_0.
\end{equation}
This can be straightforwardly generalized to the NLO $(2\rightarrow 3)$ case as
\begin{equation}
p_{T,1} + p_{T,2} + p_{T,3} \geq 2p_0.
\end{equation}
By combining the definition of $E_T$ in $\Delta y$ and the definition of a hard perturbative scattering discussed above, the measurement function $\mathcal{S}_2$ for the minijet $E_T$ in $\Delta y$ can now be written down for the $(2\rightarrow 2)$ scatterings as
\begin{equation}
\label{eq:ETS2}
\mathcal{S}_2 = \Theta(p_{T,1} + p_{T,2} \geq 2p_0) \delta(E_T -\biggl \{\epsilon(y_1)p_{T,1} + \epsilon(y_2)p_{T,2} \biggr \} )
\end{equation}
and for the $(2\rightarrow 3)$ scatterings correspondingly
\begin{equation} 
\label{eq:ETS3}
\begin{split}
\mathcal{S}_3 = \Theta(p_{T,1} & + p_{T,2}  + p_{T,3} \geq 2p_0)\\
&\times \delta(E_T -\biggl \{\epsilon(y_1)p_{T,1} + \epsilon(y_2)p_{T,2} + \epsilon(y_3)p_{T,3} \biggr \} ),
\end{split}
\end{equation}
where the functions denoted by $\Theta$ and $\delta$ are the usual step and delta functions, respectively.

\subsection{IR/CL-safe minijet $E_T$}

As discussed in chapter 4, the measurement functions define the physical, IR/CL-safe, quantity to compute. For the measurement functions $\mathcal{S}_2$ and $\mathcal{S}_3$ in Eqs. \eqref{eq:ETS2} and \eqref{eq:ETS3}, which clearly are infrared and collinear safe (\textrm{cf.} Eq. \eqref{eq:cri1}), the physical quantity to compute is the semi-inclusive $E_T$ distribution of minijets in a rapidity interval $\Delta y$ in $N+N$ (or $A+A$) collisions \cite{EKL,NLOET1,NLOET2}
\begin{equation}
\label{eq:sietdi}
\frac{{\rm d}\sigma}{{\rm d}E_T}\bigg\vert_{p_0,\Delta y} = \frac{{\rm d}\sigma}{{\rm d}E_T}\bigg\vert_{p_0,\Delta y}^{(2\rightarrow 2)} +  \frac{{\rm d}\sigma}{{\rm d}E_T}\bigg\vert_{p_0,\Delta y}^{(2\rightarrow 3)},
\end{equation}
where
\begin{equation}
\begin{split}
\frac{{\rm d}\sigma}{{\rm d}E_T}\bigg\vert_{p_0,\Delta y}^{(2\rightarrow 2)} & = \frac{1}{2!}\int{\rm d[PS]}_2\frac{{\rm d}(2\rightarrow 2)}{{\rm d[PS]}_2}\mathcal{S}_2(p_1^{\mu},p_2^{\mu}),\\
\\
\frac{{\rm d}\sigma}{{\rm d}E_T}\bigg\vert_{p_0,\Delta y}^{(2\rightarrow 3)} &= \frac{1}{3!}\int{\rm d[PS]}_3\frac{{\rm d}(2\rightarrow 3)}{{\rm d[PS]}_3}\mathcal{S}_3(p_1^{\mu},p_2^{\mu},p_3^{\mu}).
\end{split}
\end{equation}
Integrating the delta functions away in Eq.\ \eqref{eq:sietdi} we obtain the first moment of the semi-inclusive $E_T$ distribution in NLO (needed for Eq.\ \eqref{eq:iedensityfor}) as
\begin{equation}
\begin{split}
\sigma\langle E_T\rangle_{p_0,\Delta y}  \equiv \int_{0}^{\sqrt{s}}{\rm d}E_T E_T\frac{{\rm d}\sigma}{{\rm d}E_T}\bigg\vert_{p_0,\Delta y} = \sigma\langle E_T\rangle_{p_0,\Delta y}^{(2\rightarrow 2)} + \sigma\langle E_T\rangle_{p_0,\Delta y}^{(2\rightarrow 3)},
\end{split}
\end{equation}
where
\begin{equation}
\begin{split}
\sigma\langle E_T\rangle_{p_0,\Delta y}^{(2\rightarrow 2)} & = \frac{1}{2!}\int{\rm d[PS]}_2\frac{{\rm d}(2\rightarrow 2)}{{\rm d[PS]}_2}\tilde{\mathcal{S}}_2(p_1^{\mu},p_2^{\mu}),\\
\\
\sigma\langle E_T\rangle_{p_0,\Delta y}^{(2\rightarrow 3)} &= \frac{1}{3!}\int{\rm d[PS]}_3\frac{{\rm d}(2\rightarrow 3)}{{\rm d[PS]}_3}\tilde{\mathcal{S}}_3(p_1^{\mu},p_2^{\mu},p_3^{\mu}).
\end{split}
\end{equation}
Above, the measurement functions for the $\sigma\langle E_T\rangle_{p_0,\Delta y}$ have been written as
\begin{equation}
\tilde{\mathcal{S}}_2(p_1^{\mu},p_2^{\mu}) = \biggl \{\epsilon(y_1) + \epsilon(y_2)\biggr \}p_{T,2}\Theta(p_{T,2} \geq p_0)
\end{equation}
and
\begin{equation}
\begin{split}
\tilde{\mathcal{S}}_3(p_1^{\mu},p_2^{\mu},p_3^{\mu}) = \biggl \{\epsilon(y_1)p_{T,1}& + \epsilon(y_2)p_{T,2} + \epsilon(y_3)p_{T,3}\biggr \}\\
& \times \Theta(p_{T,1} + p_{T,2} + p_{T,3} \geq 2p_0).
\end{split}
\end{equation}
Naturally, also these measurement functions fulfill the requirements for infrared and collinear safety. Thus, the quantity $\sigma\langle E_T\rangle_{p_0,\Delta y}$ is a well defined NLO quantity to compute.

\vspace{0.3cm}

Finally, we should also note that the renormalization scale $\mu_R$ in the strong coupling $\alpha_s(\mu_R)$ and the factorization scale $\mu_F$ in the PDFs have to be chosen in such a way that the scales for the $(2\rightarrow 3)$ terms reduce to those for the terms with $(2\rightarrow 2)$ kinematics in the IR and CL limits. This is done by fixing $\mu_F = \mu_R = \mu$  to be equal, where the scale $\mu$ is set to be proportional to the hardness of the collision, \textrm{i.e.} to the total $p_{T}$ produced in the hard process, regardless of the parton being in $\Delta y$ or not:
\begin{equation}
\begin{split}
2\rightarrow 2: &\quad \mu = N_{\mu}\left (p_{T,1} + p_{T,2}\right )/2 = N_{\mu}p_{T},\\
2\rightarrow 3: &\quad \mu = N_{\mu}\left (p_{T,1} + p_{T,2} + p_{T,3}\right )/2,
\end{split}
\end{equation}
where $N_{\mu}$ is a constant of the order of unity.

\vspace{0.3cm}

As explained in section \ref{ied}, the initial energy density which provides the initial conditions for the hydrodynamical evolution, is computed from the IR/CL-safe NLO quantity $\sigma\langle E_T\rangle_{\Delta y,p_0}$ at $p_0=p_{\textrm{sat}}$. However, the saturation scale $p_{\textrm{sat}}$, which gives the formation (and thermalization) time $\tau_0=1/p_{\textrm{sat}}$, is determined on the basis of the minijet number in LO with a rather ad hoc $K$ factor. Thus, it would be clearly more consistent to formulate the saturation criterion in terms of the produced $E_T$ instead of $N_{AA}$. Also, since the NLO minijet $E_T$ computation will affect the hydrodynamical initial energy densities, and especially since the minijet $E_T$ is not a directly measurable observable, we should study whether there is any extra freedom in defining the IR/CL-safe measurement functions $\tilde{\mathcal{S}}_2$ and $\tilde{\mathcal{S}}_3$. These are the main questions I will discuss and answer in the next chapter.

\chapter{The NLO-improved EKRT model}
\label{EKRS}

In this chapter I briefly summarize the main ingredients of the NLO-improved computation of minijet $E_T$ production and the new formulation of the minijet saturation in $E_T$ \cite{PHD2,PHD3}. Also the main numerical results for the computed minijet initial state are discussed.

\section{NLO-improved computation of minijet $E_T$ production}

\subsection{New set of measurement functions}

As already stated in section \ref{ETNLOCOMP}, the formulation of the measurement function $\tilde{\mathcal{S}}_2$ in LO and NLO cases corresponds to $E_T$ of at least equal to $p_0$ to enter into $\Delta y$ from each $(2\rightarrow 2)$ subprocess. In the NLO $(2\rightarrow 3)$ case one may encounter subprocesses which fulfill the requirement of being perturbative $\sum_i p_{T,i} \geq 2p_0$, but contribute less than $p_0$ of $E_T$ into $\Delta y$. Thus, a possible further element in defining the measurement function $\tilde{\mathcal{S}}_3$ is that in the $(2\rightarrow 3)$ case we may still restrict the amount of the minimum $E_T$ at $\Delta y$ in an infrared and collinear safe way.

\vspace{0.3cm}

For example, in the $(2\rightarrow 2)$ case, the non-zero $E_T$ in $\Delta y$ is always larger than $p_0$, since
\begin{equation}
E_T = \epsilon(y_1)p_{T_1} + \epsilon(y_2)p_{T_2} \overset{\eqref{eq:2to2hardness}}{=} \left (\epsilon(y_1) + \epsilon(y_2)\right )p_{T} \geq p_0,
\end{equation}
for $y_1\in\Delta y$ or $y_2\in\Delta y$ (which includes also the case $y_1,y_2\in\Delta y$). Similarly, in the $(2\rightarrow 3)$ case we can have hard processes where two partons, say 1 and 2, with transverse momenta $p_{T,1} = p_{T,2}=p_{T,\textrm{hard}}$, falling outside $\Delta y$ and one soft parton 3 with transverse momentum $p_{T,3} \ll p_{T,\textrm{hard}} $ inside $\Delta y$. Thus the amount of $E_T$ in $\Delta y$ is 
\begin{equation}
E_T = \epsilon(y_1)p_{T,1} + \epsilon(y_2)p_{T,2} + \epsilon(y_3)p_{T,3} = p_{T,3} \ll p_{T,\textrm{hard}}. 
\end{equation}
At the infrared (soft) limit $p_{T,3} = 0$, we obviously have no $E_T$ in $\Delta y$ and the $(2\rightarrow 2)$ limit is correctly recovered. We could also require for the $(2\rightarrow 3)$ case the $E_T$ in $\Delta y$ be at least $p_0$, as always in the $(2\rightarrow 2)$ case. As originally discussed in \cite{NLOET2}, also this latter example is an equally well infra-red and collinear safe case.

\vspace{0.3cm}

The above two examples are only special cases of a possible definition of the $E_T$ in $\Delta y$. In fact any minimum amount of $E_T \in [0,p_0]$ contains an equally good infrared and collinear safe restriction for the $E_T$ in $\Delta y$ which relaxes back to the $(2\rightarrow 2)$ case in the infrared and collinear limits. Thus by combining the possible restrictions of $E_T$ discussed above and the original definitions of hard scattering and $E_T$ in $\Delta y$ discussed in chapter \ref{oldEKRT}, the new set of infrared and collinear safe measurement functions $\tilde{\mathcal{S}}_2$ and $\tilde{\mathcal{S}}_3$ can now be written down as \cite{PHD2}
\begin{equation}
\begin{split}
\tilde{\mathcal{S}}_2(p_1^{\mu},p_2^{\mu}) = \biggl \{\epsilon(y_1) + \epsilon(y_2)\biggr \}p_{T,2}\Theta(p_{T,2} \geq p_0),\\
\end{split}
\end{equation} 
and 
\begin{equation}
\begin{split}
\tilde{\mathcal{S}}_3(p_1^{\mu},p_2^{\mu},p_3^{\mu};\beta) =  &\biggl \{\epsilon(y_1)p_{T,1} + \epsilon(y_2)p_{T,2} + \epsilon(y_3)p_{T,3} \biggr \}\\
& \times \Theta \left (p_{T,1} + p_{T,2} + p_{T,3} \geq 2p_0 \right )\\
& \times \Theta \left (\epsilon(y_1)p_{T,1} + \epsilon(y_2)p_{T,2} + \epsilon(y_3)p_{T,3} \geq \beta p_0 \right ),
\end{split}
\end{equation} 
where the hardness parameter $\beta \in [0,1]$ defines the minimum $E_T$ required in the interval $\Delta y$. Since the new set of measurement functions above are infrared and collinear safe the first moment of the $E_T$ distribution, defined in Eq.\ \eqref{eq:sietdi}, 
\begin{equation}
\begin{split}
\sigma\langle E_T\rangle_{p_0,\Delta y,\beta}  = \frac{1}{2!}&\int{\rm d[PS]}_2\frac{{\rm d}(2\rightarrow 2)}{{\rm d[PS]}_2}\tilde{\mathcal{S}}_2(p_1^{\mu},p_2^{\mu})\\
& + \frac{1}{3!}\int{\rm d[PS]}_3\frac{{\rm d}(2\rightarrow 3)}{{\rm d[PS]}_3}\tilde{\mathcal{S}}_3(p_1^{\mu},p_2^{\mu},p_3^{\mu};\beta)
\end{split}
\end{equation}
is still a well defined NLO quantity to compute. As pointed out in \cite{PHD2}, a priori we do not know what value for $\beta$ we should use; this is a phenomenological parameter which we obtain by comparing to the data.

\subsection{Improved PDFs and their nuclear modifications}

As a straightforward improvement of the original NLO $E_T$ calculation, where only the LO nPDFs were used, we now, in \cite{PHD2}, apply the NLO EPS09 nPDFs \cite{EPS09} and NLO CTEQ6M parton distributions \cite{CTEQ6M}. Using the error sets of the EPS09 analysis we can study also the propagation of the nPDF uncertainties into the computed $E_T$ (see next section). In addition, we can also extend the new NLO $E_T$ calculation consistently to non-central $A+A$ collisions with the new NLO EPS09s impact-parameter dependent nPDFs \cite{EPS09s}, as discussed in \cite{PHD3}.

\subsection{Numerical implementation}

For the numerical studies I have implemented all these NLO improvements into our group's NLO minijet program \cite{KTPHD}, which is partly built on top of the Ellis-Kunszt-Soper NLO jet code \cite{KUSO}. In this code the numerical integrations have been implemented via the Fortran NAG library \cite{NAG} and performed on the parallel JYFL computer cluster. The finite four- and six-dimensional integrals over the two- and three-particle phase space, respectively, introduced in the subtraction procedure (see chapter 4) are performed with the Monte-Carlo integration subroutine of NAG. I also developed our code further so that it now runs significantly faster than before. Thus the pQCD part of the calculation of the minijet $E_T$ production is now performed for the very first time genuinely and consistently to NLO.

\section{Average saturation in minijet $E_T$}

As discussed in section \ref{saturation} the formulation of the original EKRT model with saturation of the number of produced minijets is problematic, since the number of produced minijets cannot be defined in a manner which would be infrared and collinear safe in NLO pQCD. Also it has not been clear whether an explicit $\alpha_s$ should appear in Eq.\ \eqref{eq:orginalsat} if it describes a fusion of the produced minijets (gluons). Furthermore, the saturation criterion in Eq.\ \eqref{eq:orginalsat} is extensive in $\Delta y$ on left-hand side but apparently not, or at least not obviously, on the right-hand side.

\vspace{0.3cm}

To improve the formulation of minijet saturation we take the following new angle in interpreting the minijet saturation. Instead of a saturation of the number of produced final state gluons we formulate the saturation in terms of the transverse energy $E_T$ production which is, as already described, CL and IR safe quantity in NLO pQCD. In this case the $E_T$ production is expected to cease when the $(3\rightarrow 2)$ and higher-order partonic processes start to dominate over the conventional $(2\rightarrow 2)$ processes. We thus require that at saturation\footnote{Saturation is assumed transversally non-local, averaged, here.}, the rapidity densities of the $E_T$ production fulfill the condition
\begin{equation}
\label{eq:newsatcri}
\frac{{\rm d}E_T}{{\rm d}y}(2\rightarrow 2) \sim \frac{{\rm d}E_T}{{\rm d}y}(2\rightarrow 3).
\end{equation}
We can get the needed scaling laws by writing
\begin{equation}
\label{eq:scalelaw}
\begin{split}
\frac{{\rm d}E_T}{{\rm d}y}(n\rightarrow 2) & \sim \int {\rm d}^2\mathbf{s}\otimes (T_Ag)^n\otimes \sigma\langle E_T\rangle(n\rightarrow 2)\\
& \sim \pi R_A^2 \otimes (T_Ag)^n p_0^{4-2n} \otimes \left (\frac{\alpha_s^n}{p_0^2}\right )p_0,
\end{split}
\end{equation}  
where we have assigned a factor  $\pi R_A^2$ for the transverse integration ${\rm d}^2\mathbf{s}$, the  factor $T_Ag$ for each of the incoming gluons\footnote{Here $g$ denotes the gluon PDFs and $T_A \sim A/(\pi R_A^2)$.}, the scale $p_0^{4-2n}$ to compensate for the $\rm fm^{-2}$ dimension of the extra $T_A$ in the $(3\rightarrow 2)$ case, and the factor $(\alpha_s^n/p_0^2)p_0$ for the $E_T$ cross-section $\sigma\langle E_T\rangle(n\rightarrow 2)$. Substituting Eq.\ \eqref{eq:scalelaw} into the saturation condition in Eq.\ \eqref{eq:newsatcri}, we obtain
\begin{equation}
\pi R_A^2(T_Ag)^2\left (\frac{\alpha_s^2}{p_0^2}\right )p_0 \sim \pi R_A^2(T_Ag)^3 \frac{1}{p_0^2}\left (\frac{\alpha_s^3}{p_0^2}\right )p_0.
\end{equation}
This leads to a scaling $T_Ag \sim p_0^2/\alpha_s$ for the gluon density probed at saturation $p_0=p_{\textrm{sat}}$. Feeding this scaling law back to the saturation condition in Eq.\ \eqref{eq:newsatcri} we arrive at the following geometrical-like saturation criterion for the average minijet $E_T$ produced in a central $A+A$ collision:
\begin{equation}
\label{eq:newavesat}
E_T(p_0,A,\sqrt{s_{NN}},\Delta y,\beta) = K_{\textrm{sat}}R_A^2p_0^3\Delta y,
\end{equation}
where 
\begin{equation}
E_T(p_{0},A,\sqrt{s_{NN}},\Delta y,\beta) = T_{AA}(\mathbf{0})\sigma\langle E_T\rangle_{p_{0},\Delta y,\beta},
\end{equation}
and the factor $K_{\textrm{sat}}$ is an $\alpha_s$-independent proportionality constant of the order of one. In Eq.\ \eqref{eq:newavesat} we see see that no explicit $\alpha_s$ appears and that the rapidity interval $\Delta y$ now correctly appears also on the right-hand side. Furthermore, the parameter $K_{\textrm{sat}}$ is a priori not known but needs to be determined from the data.

\subsection{Average initial energy density}

Once the saturation momentum scale $p_0 = p_{\textrm{sat}}(\sqrt{s_{NN}},A;K_{\textrm{sat}},\beta)$ fulfilling the new averaged saturation criterion above is found and the parameters of the NLO minijet $E_T$ calculation, $\beta$ and $K_{\textrm{sat}}$, are fixed, the initial QCD-matter energy density with the BC profile \cite{BCANDWN} is constructed as in Eq.\ \eqref{eq:iedensityfor}, 
\begin{equation}
\begin{split}
\epsilon(\mathbf{s},\tau_0) = \frac{{\rm d}E_T}{{\rm d}^2\mathbf{s}}\frac{1}{\tau_0\Delta y} = T_A(\mathbf{s})T_A(\mathbf{s})\frac{\sigma\langle E_T\rangle_{p_{\textrm{sat}},\Delta y,\beta}}{\tau_0\Delta y}.
\end{split}
\end{equation}
Here, since the transverse profile is not fixed by the pQCD calculation, one could also use the wounded nucleon (WN) transverse profiles from the optical Glauber model \cite{BCANDWN}. In this case, the WN profile for the initial energy density in central $A+B$ collisions is given by
\begin{equation}
\epsilon(\mathbf{s}) = K_{\textrm{WN}}\biggl [T_A(\mathbf{s})\exp\left (-\sigma_{NN}T_B(\mathbf{s})\right ) + T_B(\mathbf{s})\exp\left (-\sigma_{NN}T_A(\mathbf{s})\right )\biggr ],
\end{equation} 
where $\sigma_{NN}$ is the cross section for inelastic $N+N$ collisions and the overall normalization constant $K_{\textrm{WN}}$ is fixed so that we have the same amount of transverse energy, $E_T(p_0 = p_{\textrm{sat}})$,
as with the BC profile. Both the BC and WN profiles were studied in \cite{PHD2}.

\section{Local saturation in minijet $E_T$}

To compute also the transverse profiles for the initial energy densities, we need to introduce a local saturation criterion for the minijet $E_T$ production in $A$+$A$ collisions \cite{PHD3}. In the context of the original EKRT model this was done in \cite{BCANDWN,EKT1}. Here, and in \cite{PHD3}, we follow this localization procedure, only for $E_T$ and not for $N_{AA}$. Generalizing Eq.\ \eqref{eq:newavesat} to non-zero impact parameters $(\mathbf{b}\neq 0)$ and requiring 
\begin{equation}
\label{eq:newsatcriloc}
\frac{{\rm d}E_T}{{\rm d}^2\mathbf{s}{\rm d}y}(2\rightarrow 2) \sim \frac{{\rm d}E_T}{{\rm d}^2\mathbf{s}{\rm d}y}(2\rightarrow 3)
\end{equation}
as discussed in \cite{PHD3}, leads to the following local saturation criterion,
\begin{equation}
\label{eq:locsatnew}
\frac{{\rm d}E_T}{{\rm d}^2\mathbf{s}}(p_0,\sqrt{s_{NN}},A,\mathbf{s},\mathbf{b};\beta) = \frac{K_{\textrm{sat}}}{\pi}p_0^3\Delta y,
\end{equation} 
where the NLO minijet $E_T$ production in $\Delta y$, for arbitrary $\mathbf{s}$ and $\mathbf{b}$, is given by
\begin{equation}
\frac{{\rm d}E_T}{{\rm d}^2\mathbf{s}}(p_0,\dots;\beta) = T_A(\mathbf{s}_1)T_A(\mathbf{s}_2)\sigma\langle E_T\rangle _{p_0,\Delta y,\beta}.
\end{equation}

\begin{figure}[h!]
\center
\includegraphics[scale=1.4]{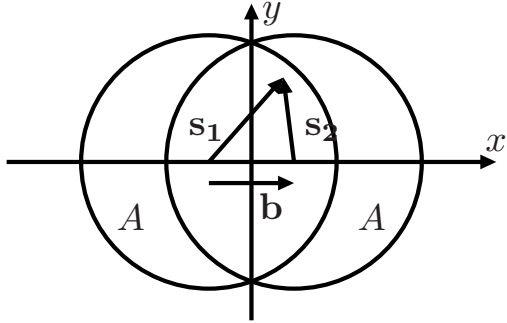}
\vspace{-0.3cm}
\caption{$(A+A)$ collision geometry in the transverse coordinate plane.}
\label{fig:col}
\end{figure}

The collision geometry with $\mathbf{s}_1=\mathbf{s}-\mathbf{b}/2$ and $\mathbf{s}_1=\mathbf{s}+\mathbf{b}/2$  is shown in Fig.\ \ref{fig:col}. 
Here we choose the coordinate system  at $\mathbf{b} \neq 0$ so that $\mathbf{b} = b_x\mathbf{e}_x$ and the centers of the nuclei are at $x=b_x/2$ and $-b_x/2$.
For given $K_{\textrm{sat}}, \beta$ and $\sqrt{s_{NN}}$, one can solve Eq.\ \eqref{eq:locsatnew} for $p_0 = p_{\textrm{sat}}(\sqrt{s_{NN}},A,\mathbf{s},\mathbf{b};K_{\textrm{sat}},\beta)$, and obtain the total minijet ${\rm d}E_T(p_{\textrm{sat}})/{\rm d}^2\mathbf{s}$ produced in the mid-rapidity region $\Delta y$.

\vspace{0.3cm}

For nuclei with realistic transverse profiles, the solution $p_0 = p_{\textrm{sat}}$ of the local saturation criterion above depends explicitly on the transverse coordinates through the $\mathbf{s}$ dependence of the EPS09s nPDFs. It, however, turns out that in practice $p_{\textrm{sat}}$ depends on $\mathbf{s}$ only through the thickness function product $T_AT_A$, because the $\mathbf{s}$ dependence of the nPDFs is quite weak near the centres of the nuclei, which is the most relevant region in this calculation, \textrm{i.e.} the region where $p_{\textrm{sat}}$ is so large that we can still trust pQCD. 
\begin{figure}[h!]
\center
\includegraphics[scale=.35]{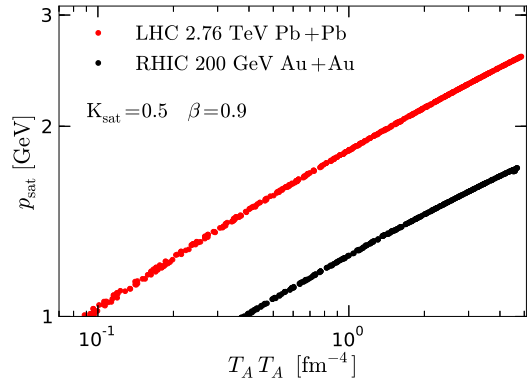}
\vspace{-0.5cm}
\caption{Examples of the saturation momenta
obtained for the LHC and RHIC $A+A$ collisions, as functions of $T_AT_A$. Figure from \cite{PHD3}.}
\label{fig:fitEKRT}
\end{figure}
Figure \ref{fig:fitEKRT} shows an example of $p_{\textrm{sat}}$ as a function of $T_AT_A$ from calculations with fixed $K_{\textrm{sat}}, \beta$ and  $\mathbf{b} = (0, 2.35, 7.87, 10.6)~{\rm fm}$ corresponding to four different centrality classes (obtained from the Optical Glauber model) in $\sqrt{s_{NN}}=2.76~{\rm TeV}$ \textrm{Pb+Pb} collisions at the LHC and $\sqrt{s_{NN}}=200~{\rm GeV}$ \textrm{Au+Au} collisions at RHIC. As we can see, all the different cases collapse onto the same curve, indicating that $p_{\textrm{sat}}$ indeed depends only on the product $T_AT_A$ up to a very good approximation.

\section{Local initial energy density}
\label{locaini}

Once the solution $p_0 = p_{\textrm{sat}}$ is known, the local initial energy density profile is obtained as
\begin{equation}
\epsilon(\mathbf{s},\tau_{\textrm{sat}}) = \frac{{\rm d}E_T}{{\rm d}^2\mathbf{s}}\frac{1}{\tau_{\textrm{sat}}\Delta y} = \frac{K_{\textrm{sat}}}{\pi}p_{\textrm{sat}}^4,
\end{equation}  
where the formation time of the minijet plasma at each point $\mathbf{s}$ is given by $\tau_{\textrm{sat}}=1/p_{\textrm{sat}}(\sqrt{s_{NN}},A,\mathbf{s},\mathbf{b};\beta,K_{\textrm{sat}})$. However, for the hydrodynamical evolution we need the initial stage at a fixed $\tau_0$. For this technical reason we need to first evolve the energy density at all points to the same fixed $\tau_0$. Our strategy is as follows: First, we set a minimum saturation scale $p_{\textrm{sat}}^{\textrm{min}} = 1~{\rm GeV} \gg \Lambda_{\textrm{QCD}}$, and assume that we can still trust the pQCD calculation here. This corresponds to a maximum formation time $\tau_0 = 1/p_{\textrm{sat}}^{\textrm{min}} \sim 0.2~{\rm fm}$. Second, the pre-thermal evolution from $\tau_{\textrm{sat}}$ to $\tau_0$ is obtained (at each point $\mathbf{s}$) using either the Bjorken free streaming (FS)
\begin{equation}
\epsilon(\mathbf{s},\tau_0) = \epsilon(\mathbf{s},\tau_{\textrm{sat}})\left (\frac{\tau_{\textrm{sat}}}{\tau_0}\right ) 
\end{equation}
or the Bjorken hydrodynamic scaling solution (BJ)
\begin{equation}
\epsilon(\mathbf{s},\tau_0) = \epsilon(\mathbf{s},\tau_{\textrm{sat}})\left (\frac{\tau_{\textrm{sat}}}{\tau_0}\right )^{4/3}. 
\end{equation}
We take these two limits to represent the uncertainty in the pre-thermali\-zation evolution: In the FS case the transverse energy is preserved and in the BJ case a maximum amount of the transverse energy is reduced by the longitudinal pressure. Finally, the region below $p_{\textrm{sat}}^{\textrm{min}}=1~{\rm GeV}$ is considered as a boundary. In this region, the energy density is then obtained by using an interpolation, where the FS/BJ-evolved pQCD energy density is smoothly connected to the BC profile at the dilute edge. For more details see \cite{PHD3}.

\vspace{0.3cm}

\begin{figure}[h!]
\center
\includegraphics[scale=.33]{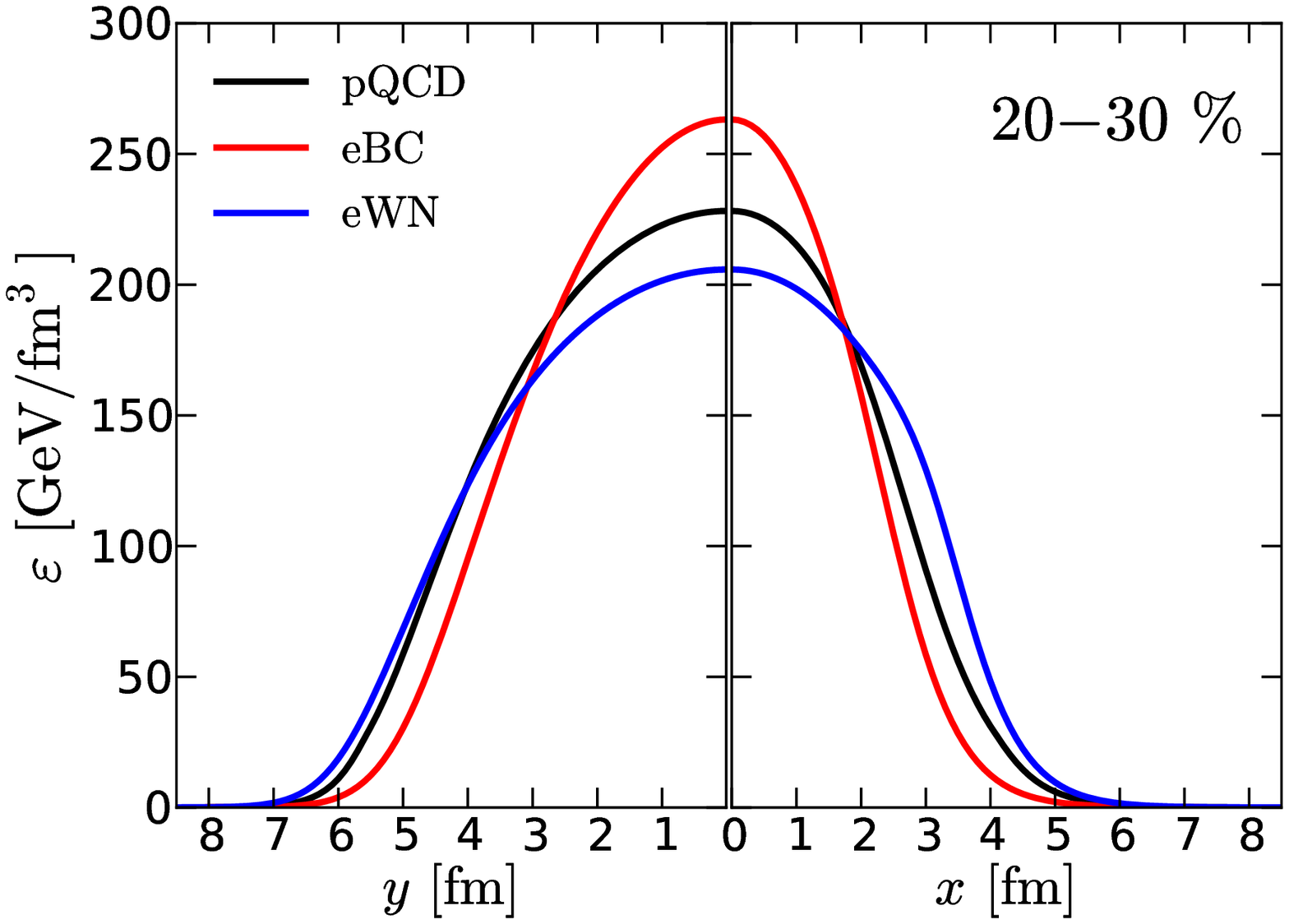}
\includegraphics[scale=.33]{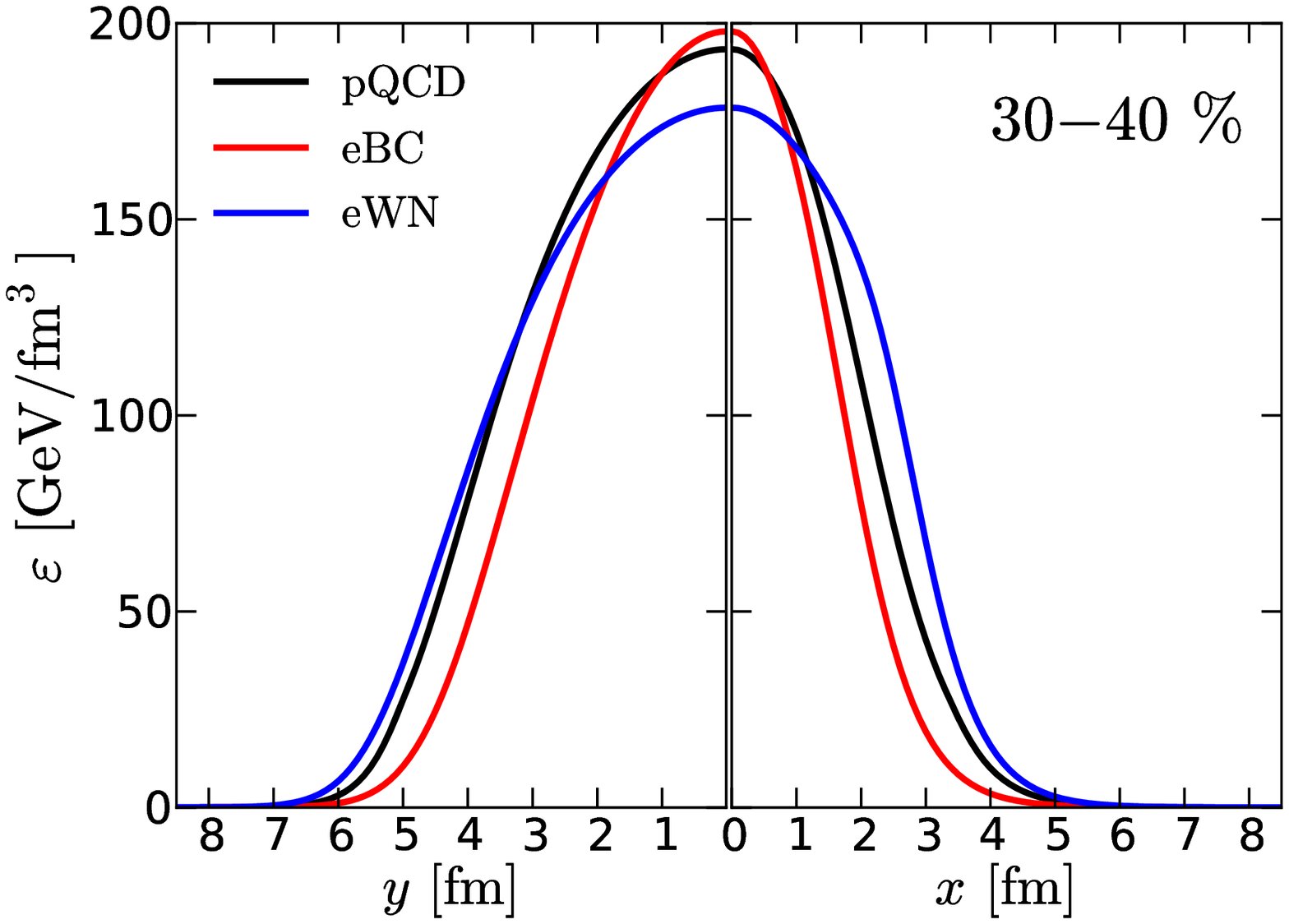}
\vspace{-0.3cm}
\caption{Computed energy density profiles in the $x$-and $y$-directions for selected centralities in Pb+Pb collisions at the LHC. Figures are based on the results in \cite{PHD3}.}
\label{fig:initialstate}
\end{figure}

The resulting initial profiles for the \textrm{Pb+Pb} collisions at the LHC, $\sqrt{s_{NN}}=2.76~{\rm TeV}$, are shown in Figs.\ \ref{fig:initialstate} for selected centralities. These figures show the computed energy density profiles in the $x$- and $y$-directions with $\beta = 0.9$, $K_{\textrm{sat}}=0.69$ and BJ pre-thermal evolution. The comparison between the computed pQCD + saturation (labeled as "pQCD") profile with the eBC and eWN profiles \cite{BCANDWN} (which are normalized to the initial total entropy in central collisions) is also shown. As seen in the figure, the pQCD + saturation based profiles lie between the eBC and eWN in a non-trivial way. 
\begin{figure}[h!]
\center
\includegraphics[scale=.33]{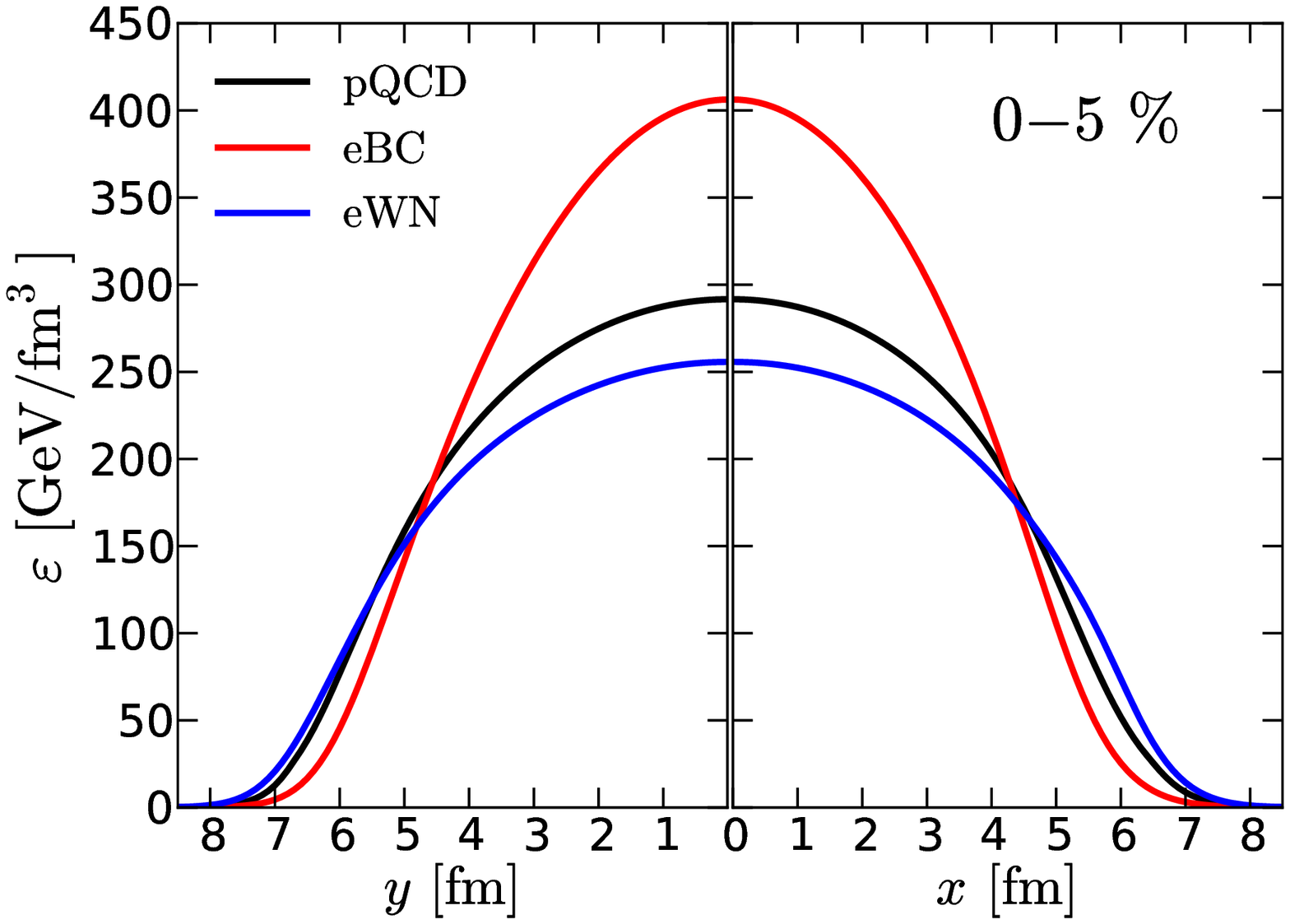}
\includegraphics[scale=.33]{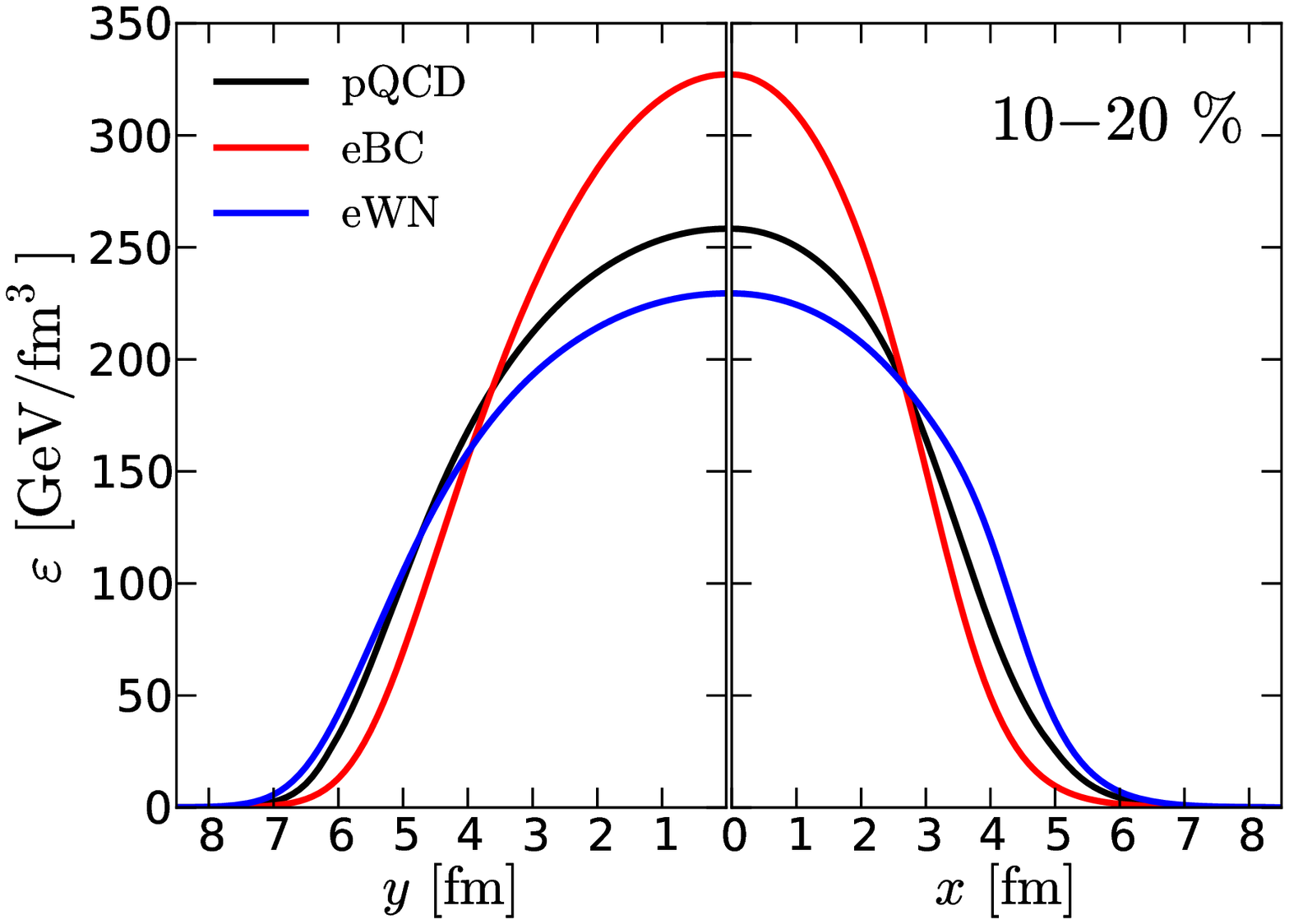}
%\caption{(Color online) }
\label{fig:initial}
\end{figure}

\vspace{0.3cm}

\begin{figure}[h!]
\center
\includegraphics[scale=.38]{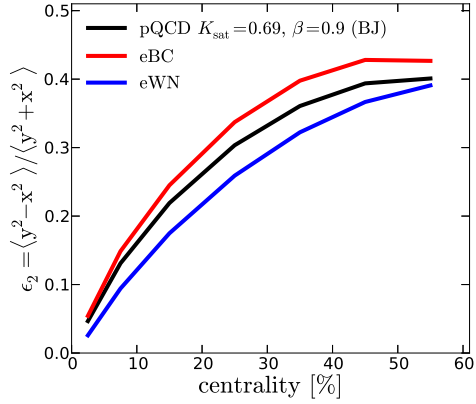}
\vspace{-0.3cm}
\caption{Computed spatial eccentricity as a function of centrality, for \textrm{Pb+Pb} collisions at the LHC.}
\label{fig:ecce}
\end{figure}

In non-central collisions the computed initial state is azimuthally anisotropic in the transverse plane, as can be seen in Figs.\ \ref{fig:initialstate}, where the computed energy density drops faster in the $x$ direction than in the $y$ direction. This anisotropy can be quantified by the spatial eccentricity defined as \cite{ECCE}
\begin{equation}
\epsilon_2 \equiv \frac{\langle y^2-x^2\rangle}{\langle y^2 + x^2\rangle} \equiv \frac{\int {\rm d}^2\mathbf{s}\epsilon(\mathbf{s},\tau_0)(y^2-x^2)}{\int {\rm d}^2\mathbf{s}\epsilon(\mathbf{s},\tau_0)(y^2+x^2)}.
\end{equation}
This spatially anisotropic distribution of the initial QCD-matter leads to anisotropic pressure gradients which will generate an anisotropic transverse flow field in the system (see chapter \ref{HYDRO}). The computed spatial eccentricity for the LHC Pb+Pb case as a function of centrality is shown in Fig.\ \ref{fig:ecce}.  

As seen in the figure, the pQCD eccentricity is between the eBC and eWN eccentricities. These results nicely demonstrate the necessity to perform a \emph{dynamical} calculation, as we have presented here, for the initial conditions.

\chapter{Relativistic hydrodynamics}
\label{HYDRO}

In this chapter I briefly introduce the hydrodynamic modeling of high energy nuclear collisions and present the main ingredients of our ideal and viscous hydrodynamical simulations, which we have used to evolve the QCD-matter initial state given by the NLO-improved EKRT model, and convert it to the final state particles.

\section{Equations of motion}

In general, the basic equations describing the hydrodynamical evolution of a relativistic fluid are the local conservation laws of energy and momentum, which can be expressed by the formula \cite{HYDROY1,DIRK}
\begin{equation}
\label{eq:HYDROY1}
\partial_{\mu}T^{\mu\nu}(x) = 0.
\end{equation}
Here, the quantity $T^{\mu\nu}$ is the energy-momentum tensor, which is a symmetric $(T^{\mu\nu} = T^{\nu\mu})$ $4\times 4$ rank-two tensor, and the four-vector $x=(t,\mathbf{x})$ with $\mathbf{x} \in \mathcal{R}^3$ labels the position of the fluid cell. 

\vspace{0.3cm}

In addition, if the relativistic fluid contains conserved charges, with local charge densities $n_i(x)$, their evolution is expressed by continuity equations
\begin{equation}
\label{eq:HYDROY2}
\partial_{\mu}j_i^{\mu}(x) = 0, \quad (i=1,\dots,n),
\end{equation}
where the local charge currents are expressed by
\begin{equation}
\label{eq:hydrovara}
j_i^{\mu}(x) = n_i(x)u^{\mu}(x) + V_i^{\mu}(x),
\end{equation}
with the local fluid four-velocity 
\begin{equation}
u^{\mu}(x) = \gamma(1,\mathbf{v}) = \gamma(1,v_x,v_y,v_z)
\end{equation}
and Lorentz factor $\gamma = 1/\sqrt{1-\vert \mathbf{v}\vert^2}$. The quantity $V_i^{\mu}(x)$ in Eq.\ \eqref{eq:hydrovara} is the local charge diffusion current, which is given by $V_i^{\mu} = (g^{\mu\nu} - u^{\mu}u^{\nu})j_{i,\nu}$. 

\section{Ideal hydrodynamics}

With an assumption of ideal hydrodynamics where all dissipative effects are neglected and the system is assumed to be in local thermodynamic equilibrium, one can decompose the energy-momentum tensor as \cite{HYDROY1,HUOVRUUS}:
\begin{equation}
\label{eq:HYDROY3}
\begin{split}
T^{\mu\nu} = \left (\epsilon + P\right ) u^{\mu}u^{\nu} - P g^{\mu\nu},
\end{split}
\end{equation}
where $\epsilon=\epsilon(x)$ is the local energy density, $P=P(x)$ is the local pressure and $g^{\mu\nu}$ is the metric tensor. If the only conserved charge density that is taken into account is the local net-baryon number density $n_{\textrm{B}}(x)$, Eq.\ \eqref{eq:HYDROY2} is expressed by
\begin{equation}
\label{eq:HYDROY4}
\partial_{\mu}(n_Bu^{\mu}) = 0.
\end{equation}
In this case, we have three equations \eqref{eq:HYDROY1}, \eqref{eq:HYDROY3} and \eqref{eq:HYDROY4} with 6 independent variables, $\epsilon, P, n_{\textrm{B}}$ and three independent components\footnote{Three unknowns only, since $u_{\mu}u^{\mu}=1$.} of $u^{\mu}$. Since the conservation laws of energy, momentum and the baryon number give only 5 equations, its clear that one still requires an equation of state (EoS) to link the thermodynamic quantities, i.e. $P=P(\epsilon,n_{\textrm{B}})$, in order to close the system of equations. 

\vspace{0.3cm}

Thus, once the initial state $(\epsilon, u^{\mu})$ and EoS are given, the hydrodynamical equations can be solved. However, in the mid-rapidity region of ultrarelativistic heavy-ion collisions at RHIC and LHC, the net-baryon number can be assumed to be very close to zero and we can express the pressure in terms of the energy density alone, $P(\epsilon,n_{\rm B}) = P(\epsilon)$, and neglect Eq.\ \eqref{eq:HYDROY4}.

\section{Longitudinal boost-invariance and initial conditions}

When considering particle production at mid-rapidities in heavy-ion collisions at RHIC and LHC the hydrodynamical equations can be simplified by assuming boost invariance along the beam-direction ($z$-direction), which is maintained by a longitudinal scaling flow velocity $v_z = z/t$ \cite{BOOSTINV}. In practice, this assumption reduces the (3+1)-dimensional hydrodynamics to be numerically (2+1)-dimensional, since now the expansion in the $z$-direction is trivial. In the articles included in this thesis the hydrodynamical equations are solved numerically using the SHArp and Smooth Transport Algorithm (SHASTA)  \cite{SHAHSTA1,SHAHSTA2}.

\vspace{0.3cm}

Due to the boost invariance assumption the energy density and transverse velocity do not depend on the space-time rapidity $\eta_{\textrm{st}}$ but only on the longitudinal proper time $\tau$ and transverse coordinates $x$ and $y$, \textrm{i.e.},
\begin{equation}
\epsilon(\tau,x,y,\eta_{\textrm{st}}) = \epsilon(\tau,x,y)  \quad \text{and} \quad \mathbf{v}_T(\tau,x,y,\eta_{\textrm{st}}) = \mathbf{v}_T(\tau,x,y).
\end{equation}
As initial conditions, the boost-invariant (ideal) hydrodynamic system requires the initial transverse velocity $\mathbf{v}_{T}(x,y,\tau)$ and initial transverse energy density $\epsilon(x,y,\tau)$ at an initial time $\tau_0$. In this thesis the initial transverse velocity is always chosen to be zero.

\section{Viscous hydrodynamics}

In a system\footnote{Here, we again restrict ourselves to a system of no net conserved charges.} which is near local thermodynamic equilibrium but where dissipative effects cannot, however, be neglected, the energy momentum tensor decomposition in Eq.\ \eqref{eq:HYDROY3} leads to additional terms \cite{HYDROY1}
\begin{equation}
T^{\mu\nu}(x) = (\epsilon + P + \Pi)u^{\mu}u^{\nu} - (P+\Pi)g^{\mu\nu} + \pi^{\mu\nu}, 
\end{equation}
where $\Pi$ is the bulk pressure and $\pi^{\mu\nu}$ is the shear stress tensor. 

\vspace{0.3cm}

In the studies \cite{PHD2,PHD3} presented in this thesis we employ the boost invariant viscous hydrodynamical setup \cite{PIEVO1,PIEVO2}, where the bulk pressure $\Pi$ is taken to be zero and the evolution equation of $\pi^{\mu\nu}=T^{\langle \mu\nu\rangle}$ is given by transient relativistic hydrodynamics \cite{PIMUNU1,PIMUNU2,PIMUNU3}
\begin{equation}
\label{eq:pievo}
\tau_{\pi}\dot{\pi}^{\langle \mu\nu\rangle } + \pi^{\mu\nu} = 2\eta\sigma^{\mu\nu} - c_1\pi^{\mu\nu}\theta - \left (c_2{\sigma^{\langle \mu}}_{\lambda} - c_3{\pi^{\langle \mu}}_{\lambda}\right )\pi^{\nu\rangle\lambda},
\end{equation}
where the angular brackets $\langle \rangle$ denote the symmetrized and traceless projection, orthogonal to the fluid four-velocity $u^{\mu}$. The co-moving time derivative $u^{\mu}\partial_{\mu}$ is denoted by the dot, $\eta$ is the shear viscosity coefficient, $\sigma^{\mu\nu} = \partial^{\langle \mu}u^{\nu\rangle}$ is the shear tensor, and $\theta = \partial_{\mu}u^{\mu}$ is the expansion rate. In Eq.\ \eqref{eq:pievo} the $\pi^{\mu\nu}$ is initially set to zero, and the coefficients of the non-linear terms are taken to be \cite{LISA} $c_1 = 4\tau_{\pi}/3$, $c_2 = 10\tau_{\pi}/7$, $c_3 = 9/(70P)$, where $\tau_{\pi} = 5\eta/(\epsilon + P)$. The evolution equations for $\pi^{\mu\nu}$ are solved numerically together with $\partial_{\mu}T^{\mu\nu}=0$. The details of the numerical algorithm can be found in \cite{PIEVO1,PIEVONUM}.

\vspace{0.3cm}

\begin{figure}[h!]
\center
\includegraphics[scale=.45]{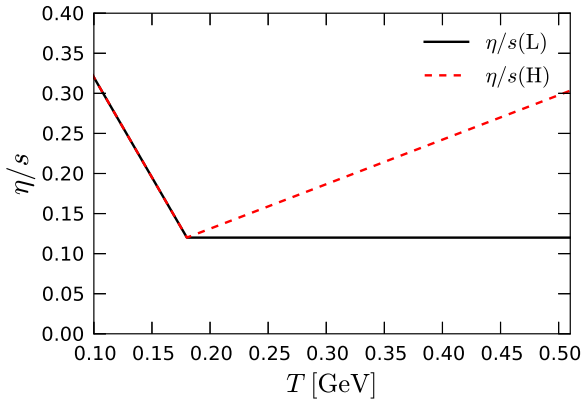}
\caption{Two different parametrizations (H) and (L) of $\eta/s$ as a function of temperature. Figure from \cite{PHD3}.}
\label{fig:etapers}
\end{figure}

In the article \cite{PHD3}, we used a rough but realistic (non-constant) temperature dependent shear viscosity $\eta/s(T)$. In practice, we converged to using the two different parametrizations shown in Fig.\ \ref{fig:etapers}. In both cases we take the minimum of $\eta/s$ to be at $T=180~{\rm MeV}$ and assume the ratio $\eta/s$ to decrease linearly as a function of temperature in the (low-temperature) hadronic phase. For the high-temperature QGP phase we either use linearly increasing (H) or constant (L) function of temperature. 
These parametrizations are shown to reproduce the elliptic flow data (see section \ref{flow}) in \cite{PHD3}.
%%%%%%%%%%%%%%FIGURE%%%%%%%%%%%%%%%%%%%%%%%%%%%%%%%%%%%%
%%%%%%%%%%%%%%%%%%%%%%%%%%%%%%%%%%%%%%%%%%%%%%%%%%%%%%%%%

\section{Equation of state and decoupling}

The equation of state, $P = P(\epsilon)$, where we neglect the baryon number density, is one of the key inputs in hydrodynamic simulations. In this thesis two different EoSs are used:
\begin{itemize}
\item The s95-v1 \cite{EOS1} which combines lattice QCD results at high temperatures and matches to the low-$T$ hadron resonance gas at $T=180~{\rm MeV}$.
\item The s95p-PCE-v1 which is the same as above, but with chemical freeze-out at $T_{\textrm{chem}} =165-175~\rm{MeV}$ \cite{EOS2} .
\end{itemize}

\vspace{0.3cm}

To obtain final hadrons in hydrodynamical simulations we assume that free hadrons are directly emitted from the fluid at a decoupling surface which is a three-dimensional hypersurface usually characterized by a constant kinetic freeze-out temperature $T(x) = T_{\textrm{f}}$. We use the standard Cooper and Frye description \cite{COOPERFRY} to calculate the particle momentum distributions, which is followed by strong and electromagnetic decays of unstable hadrons (for more details, see \cite{RESON} and references therein).

\section{Hadronic observables}
\label{flow}

To analyze the bulk-hadronic observables in terms of the hydrodynamical description, we compute the transverse momentum distributions of identified hadrons $i$ as a function of collision centrality \cite{FOURIER},
\begin{equation}
\label{eq:spectra}
\begin{split}
\frac{{\rm d}N_i(\mathbf{b})}{{\rm d}p_{T}^2{\rm d}y{\rm d}\phi} = \frac{{\rm d}N_i(\mathbf{b})}{{\rm d}p_{T}^2{\rm d}y}\biggl \{1  + 2\sum_{n=1}^{\infty}v_n(y,p_{T};\mathbf{b})\cos(n\phi)\biggr \},
\end{split}
\end{equation}
where $\mathbf{b}$ is the averaged impact parameter corresponding to a given centrality, $y$ is the rapidity and $\phi$ is the azimuthal angle of momentum $\mathbf{p}_T$. The transverse momentum, rapidity and centrality dependent Fourier coefficients $v_n$ quantify the degree of azimuthal anisotropy in the momentum distributions. These coefficients can be calculated for each centrality class as 
\begin{equation}
v_n(y,p_{T};\mathbf{b}) \equiv \left (\frac{{\rm d}N_i(\mathbf{b})}{{\rm d}p_{T}^2{\rm d}y}\right )^{-1}\int_{-\pi}^{\pi}{\rm d}\phi \cos(n\phi)
\frac{{\rm d}N_i(\mathbf{b})}{{\rm d}p_{T}^2{\rm d}y{\rm d}\phi},
\end{equation}
where the first, second and third Fourier harmonics, $v_1, v_2$ and $v_3$, are called direct, elliptic and triangular flow coefficients, respectively. Due to the symmetry of the nuclear overlap region in the average initial state the odd harmonics, $v_1,v_3,\dots$, vanish\footnote{Note that this symmetry is broken if the event-by-event fluctuations are taken into account.}. Also, due to the longitudinal boost symmetry assumed throughout this thesis the remaining coefficients, $v_2,v_4,\dots$, do not depend on the rapidity $y$. Therefore, the transverse momentum distribution in Eq.\ \eqref{eq:spectra} simplifies to
\begin{equation}
\begin{split}
\frac{{\rm d}N_i(\mathbf{b})}{{\rm d}p_{T}^2{\rm d}y{\rm d}\phi} = \frac{{\rm d}N_i(\mathbf{b})}{{\rm d}p_{T}^2{\rm d}y}\biggl \{ 1  + 2v_2&(p_{T};\mathbf{b})\cos(2\phi)\\
& + 2v_4(p_{T};\mathbf{b})\cos(4\phi) + \cdots \biggr \}.
\end{split}
\end{equation}
The most prominent component of the remaining Fourier expansion above is the coefficient $v_2$. The collective transverse flow \cite{OY} leads to anisotropic particle distributions resulting in \textrm{e.g.} a non-zero elliptic flow coefficient. For the non-central collisions, the initial overlap geometry between the colliding nuclei is almond shaped. This causes a larger pressure gradient in the $x$ (impact parameter) direction, than in the $y$ direction. As the transverse flow builds up, this initial spatial anisotropy is then translated into a momentum anisotropy, which can be observed in the measured particle azimuthal distributions.

\vspace{0.3cm}

An important observable in $A+A$ collisions is also the multiplicity, measured in a certain pseudorapidity acceptance. In order to obtain the (flat) rapidity distribution of the total number of particles (multiplicity), ${\rm d}N/{\rm d}y$, from boost-invariant hydrodynamical simulations, the integration in Eq.\ \eqref{eq:spectra} is performed over the $(\phi,p_{T})$ as
\begin{equation}
\label{eq:mult}
\frac{{\rm d}N}{{\rm d}y} = \int {\rm d}p_{T}^2{\rm d}\phi\sum_{i}\frac{{\rm d}N_i}{{\rm d}p_{T}^2{\rm d}y{\rm d}\phi}\bigg\vert_{y=0}.
\end{equation}
To obtain the charged particle multiplicity from Eq.\ \eqref{eq:mult}, one simply excludes all neutral particles from the above sum. The connection between the rapidity $y$ and pseudorapidity $\eta = (1/2)\log\left (\frac{\vert \mathbf{p}\vert + p_z}{\vert \mathbf{p}\vert - p_z}\right )$ is
\begin{equation}
y = \sinh^{-1}\left (\frac{p_{T}}{m_{T,i}}\sinh(\eta)\right ),
\end{equation}
where the transverse mass $m_{T,i}^2 = p_{T,i}^2 + m_{i}^2$. Using this we obtain the pseudorapidity distribution of the total multiplicity as
\begin{equation}
\label{eq:multpseu}
\frac{{\rm d}N}{{\rm d}\eta} = \int{\rm d}p_{T}^2{\rm d}\phi\sum_{i}J_{i}(\eta,p_{T})\frac{{\rm d}N_i}{{\rm d}p_{T}^2{\rm d}y{\rm d}\phi},
\end{equation}
where the quantity $J_i(\eta, p_T) = \partial y/\partial \eta$ is the Jacobian of the coordinate transformation from $y$ to $\eta$. In the boost-invariant approximation, as discussed in \cite{ETEKRT1} we can compute the averaged total multiplicity of particles in a pseudorapidity bin $\Delta \eta$ as
\begin{equation}
\frac{{\rm d}N}{{\rm d}\eta}\bigg\vert_{\Delta \eta} = \frac{2}{\Delta\eta}\int {\rm d}p_{T}^2{\rm d}\phi\sum_{i}\frac{{\rm d}N_i}{{\rm d}p_{T}^2{\rm d}y{\rm d}\phi}\bigg\vert_{y=0}\sinh^{-1}\left (\frac{p_T}{m_{T,i}}\sinh\left (\frac{\Delta \eta}{2}\right )\right ).
\end{equation}

\chapter{Main Results}
\label{results}

In this chapter I discuss the main results of the articles \cite{PHD1,PHD2,PHD3} included in this thesis. The main focus here is on the NLO-improved EKRT framework which is used to compute the initial conditions for hydrodynamical simulations in $\sqrt{s_{NN}} = 200~{\rm GeV}$ \textrm{Au+Au} collisions at RHIC and $\sqrt{s_{NN}} = 2.76~{\rm TeV}$ \textrm{Pb+Pb} collisions at the LHC. I will proceed in chronological order, which also reflects the improvement steps in our modeling.

\section{EKRT framework for computing charged\\ hadron $p_T$ spectra at the LHC}

Let us first discuss the results obtained in \cite{PHD1}, where the updated LO EKRT framework with a parton-medium interaction modeling\footnote{No attempt is made here for a detailed discussion on parton-medium interaction modeling, as the author has not participated in the calculations regarding it.} was used to describe simultaneously the low and high $p_T$-spectra of charged hadrons in most central \textrm{Pb+Pb} collisions at the LHC.

\vspace{0.3cm}

In \cite{PHD1} we first considered the LHC initial state from the viewpoint of the original LO EKRT model. The idea in \cite{PHD1} was to fix the $K$ factor in the LO pQCD minijet $E_T$ calculation in such a way that the measured LHC total charged-hadron multiplicity \cite{Aamodt:2010pb} is reproduced with ideal hydrodynamics. This computation was done by using an updated version of the original EKRT model, where the minijet calculation was performed using the LO CTEQ6L1 PDFs \cite{CTEQ6M} together with LO EPS09 nuclear effects \cite{EPS09}. The results was $K=1.54$, $p_{\textrm{sat}}=1.58~{\rm GeV}$ and $\tau_0=0.12~{\rm fm}$, which was then used to compute $E_T(p_{\textrm{sat}})$ and from this the initial conditions $\tau_0$ and $\epsilon(\mathbf{s}, \tau_0)$ with eBC profile for ideal hydrodynamics. For the ideal hydrodynamic evolution we applied the boost invariant setup with the s95-v1 EoS and setting the kinetic freeze-out at $T_{\textrm{f}}=165~{\rm MeV}$ (see chapter \ref{HYDRO}).

\vspace{0.3cm}

\begin{figure}[h!]
\center
\includegraphics[scale=.55]{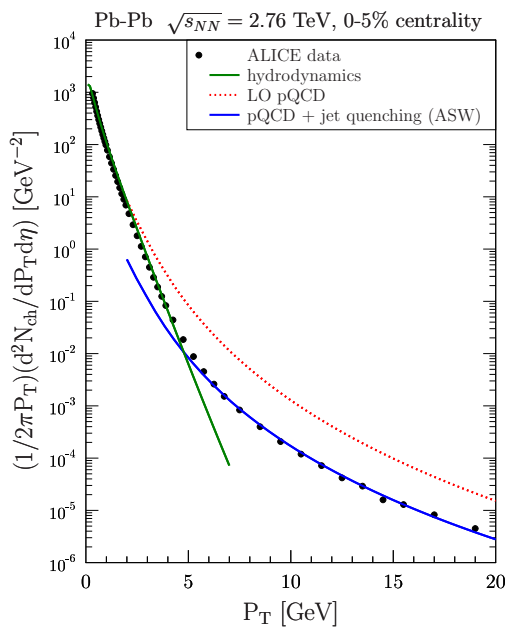}
\caption{Transverse momentum spectrum of charged hadrons in 5\% most central $\sqrt{s_{NN}} = 2.76~{\rm TeV}$ \textrm{Pb+Pb} collisions at the LHC (see text for details) Figure from \cite{PHD1}.}
\label{fig:hadpT}
\end{figure}

Figure\ \ref{fig:hadpT} shows the $p_T$ spectrum of charged hadrons in most central $\sqrt{s_{NN}} = 2.76~{\rm TeV}$ \textrm{Pb+Pb} collisions as measured by the ALICE collaboration \cite{ALICEPT}. Also shown is the comparison with the computed theoretical calculations  using a two component picture: the low $p_T$ region is described by the EKRT model with ideal hydrodynamics whereas in the high $p_T$ region we apply a pQCD + jet quenching framework.
%%%%%%%%%%%%%%FIGURE%%%%%%%%%%%%%%%%%%%%%%%%%%%%%%%%%%%%
%%%%%%%%%%%%%%%%%%%%%%%%%%%%%%%%%%%%%%%%%%%%%%%%%%%%%%%
Figure \ \ref{fig:hadpT} clearly shows that the hydrodynamics with saturated minijet initial conditions (green curve) can describe the measured hadron spectra even up to $4-5~{\rm GeV}$, which is the same applicability region for a hydrodynamical description as predicted in \cite{ETEKRT2}. Also, it is observed that at high $p_T$ the measured hadron spectra are clearly suppressed as compared to the LO pQCD spectrum without jet quenching (red curve) while the results which include the jet quenching (blue curve) nicely describe the region $p_T \gtrsim 5~{\rm GeV}$.
%%%%%%%%%%%%%%FIGURE%%%%%%%%%%%%%%%%%%%%%%%%%%%%%%%%%%%%
\begin{figure}[h!]
\center
\includegraphics[scale=.78]{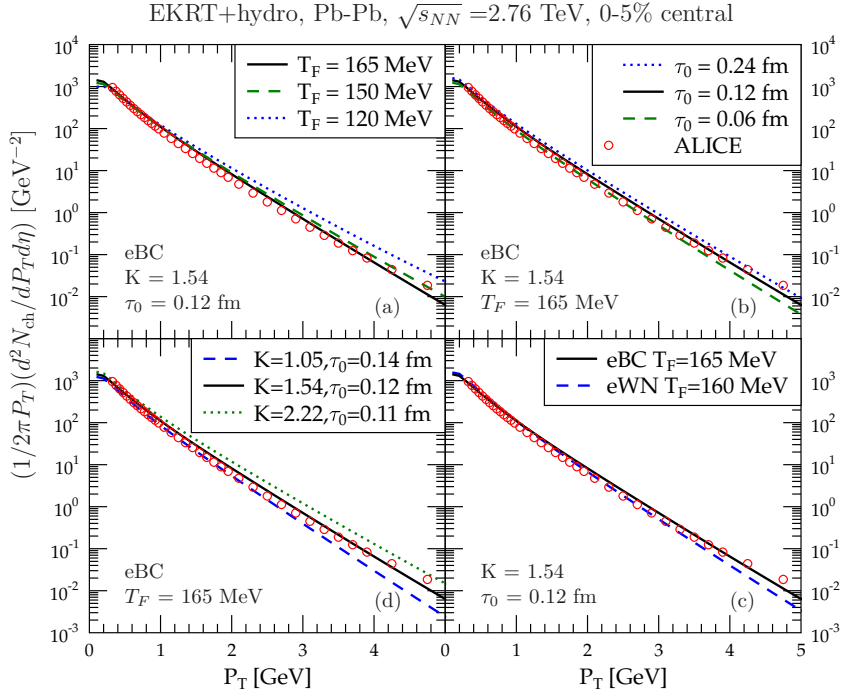}
\caption{Effects of different EKRT model parameters to the charged-hadron $p_T$ spectrum in 5\% most central $\sqrt{s_{NN}} = 2.76~{\rm TeV}$ \textrm{Pb+Pb} collisions at the LHC. Figure from \cite{PHD1}.}
\label{fig:hadpTsys}
\end{figure}
%%%%%%%%%%%%%%%%%%%%%%%%%%%%%%%%%%%%%%%%%%%%%%%%%%%%%%%%%

\vspace{0.3cm}

Next, the uncertainties in the EKRT fit and in the hydrodynamical parameters have been charted in Fig.\ \ref{fig:hadpTsys}. First, in panel (a), the value of $T_f$ is varied between $120-165~{\rm MeV}$, fixing the initial conditions to the default set-up. The results show that the lowest value  $(120~{\rm MeV})$ for the freeze-out temperature is clearly not favored by the data. In panel (b) we study the sensitivity of the results to the different initial times, varying $\tau_0$ by a factor of two. We obtain that a larger $\tau_0$ is slightly disfavored by the data while a smaller $\tau_0$ gives a good agreement with the data. Here we note that, since the computed $E_T$ is kept constant, the multiplicity depends on $\tau_0$ as explained in \cite{PHD1}. The panel (c) shows the sensitivity of the hadronic $p_T$ spectrum to the choice of the energy density transverse profile. We can see that the region $2-3~{\rm GeV}$ slightly favors the eWN profile but above $3~{\rm GeV}$ the eBC profile is closer to the data. Finally, in the panel (d) we study the sensitivity of the hadronic spectrum to the fit parameter $K=K(\tau_0)$. We see how the default $(K=1.54,\tau_0=0.12~{\rm fm})$ case reproduces the LHC multiplicity best while the case $(K=1.05,\tau=0.14~{\rm fm})$ gives too small a multiplicity and the case $(K=2.22,\tau=0.11~{\rm fm})$ too large a multiplicity.

\vspace{0.3cm}

From Fig.\ \ref{fig:hadpTsys} we observe that with our updated EKRT and ideal hydrodynamics framework we are restricted to a quite narrow window of the parameters $K, \tau_0$ and $T_f$. On the other hand, the uncertainty in the initial transverse profile also plays a role, which is signified if one studies also $v_2$. As discussed below, the next steps were to bring the EKRT model consistently to NLO \cite{PHD2}, localize the model, and exploit viscous hydrodynamics to study centrality dependence of bulk observables \cite{PHD3}.

\section{NLO-improved pQCD + averaged $E_T$ saturation and ideal hydrodynamics framework}

Next, I discuss the results obtained in \cite{PHD2}, where the charged-particle multiplicity and $p_T$ spectra for identified charged particles $(\pi^+,K^+,p,\bar{p})$ in most central \textrm{Au+Au} collisions at RHIC and \textrm{Pb+Pb} collisions at the LHC were computed using the improved NLO pQCD + averaged $E_T$ saturation and ideal hydrodynamics framework. 

\vspace{0.3cm}

\begin{figure}[h!]
\center
\includegraphics[scale=.65]{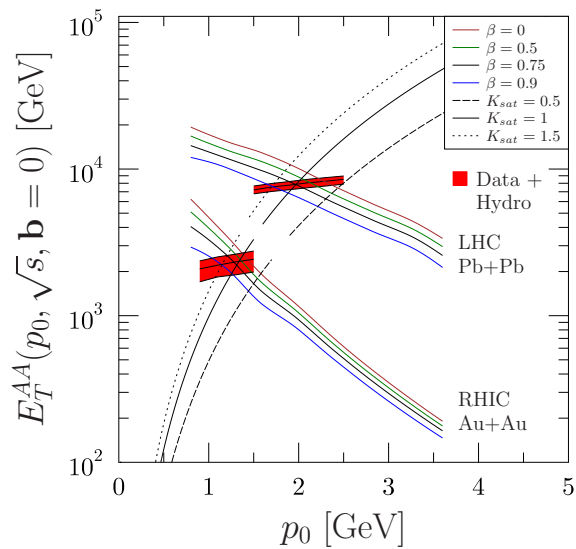}
\caption{The computed NLO minijet $E_T$, as a function of the $p_T$ cut-off $p_0$. The rising curves are the r.h.s of Eq.\ \eqref{eq:newavesat}. Figure from \cite{PHD2}.}
\label{fig:ETLASKU}
\end{figure}

First, in Fig.\ \ref{fig:ETLASKU} we show the averaged NLO minijet $E_T$ computation in the mid-rapidity window $\Delta y = 1$ with several different $(K_{\textrm{sat}},\beta)$ pairs as a function of the $p_0$ scale. The theoretical computations are done for the most central \textrm{Pb+Pb} collisions at the LHC energy $\sqrt{s_{NN}}=2.76~{\rm TeV}$ and \textrm{Au+Au} collisions at the RHIC energy $\sqrt{s_{NN}}=200~{\rm GeV}$. The implementation of the effective centrality selection is explained in detail in \cite{ETEKRT1,PHD2}. 
%%%%%%%%%%%%%%FIGURE%%%%%%%%%%%%%%%%%%%%%%%%%%%%%%%%%%%%
%%%%%%%%%%%%%%%%%%%%%%%%%%%%%%%%%%%%%%%%%%%%%%%%%%%%%%%%
In Fig.\ \ref{fig:ETLASKU} the rising curves are the right-hand side of the averaged saturation criterion in Eq.\ \eqref{eq:newavesat}, with three different values of the proportionality constant $K_{\textrm{sat}}$. Finally, the red bands show the range of values for $E_T(p_0)$ and $p_0$ that reproduce the measured charged-particle multiplicities \cite{Aamodt:2010pb,CMSPHD2} (LHC) and \cite{PHENIXPHD2,STARPHD2,BHRAMSPHD2} (RHIC) after an ideal boost invariant hydrodynamic evolution with $\tau_0 = 1/p_{0}$ mapping and the eBC energy density profile. The EoS was s95-PCE-v1 with a chemical freeze-out at $T_{\textrm{chem}}=150~{\rm MeV}$. As seen from Fig.\ \ref{fig:ETLASKU} there are several different correlated parameter pairs $(K_{\textrm{sat}},\beta)$ that reproduce the measured LHC and RHIC charged-particle multiplicities simultaneously. For example, if we choose $K_{\textrm{sat}}=1$ and $\beta =0.75$ we describe the average LHC multiplicity perfectly and agree nicely also with the RHIC multiplicity without any fine-tuning of the model. Also, it's clear from the figure above that the large values of the parameter $\beta$ can reduce the amount of the produced $E_T$ almost by a factor of 2. 

\vspace{0.3cm}

\begin{figure}[h!]
\center
\includegraphics[scale=.68]{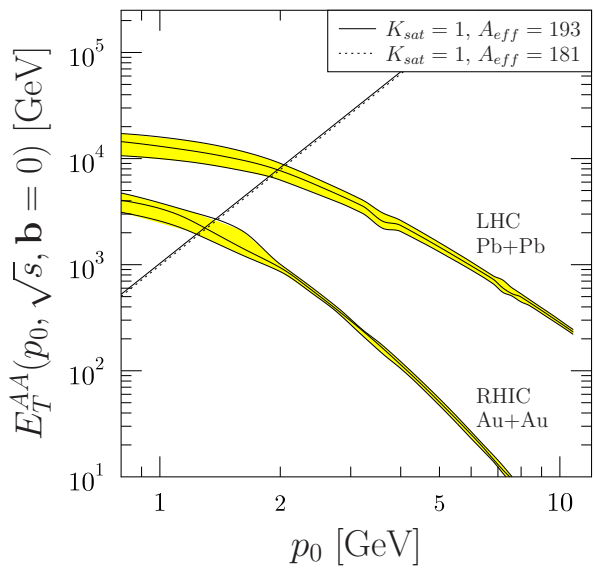}
\vspace{-0.8cm}
\caption{The computed NLO minijet $E_T$ in $\Delta y=1$ with fixed $\beta =0.75$ as a function of $p_0$, including the nPDF uncertainties. Figure from \cite{PHD2}.}
\label{fig:ETER}
\end{figure}

Next we show how the nPDF uncertainties propagate into the computed NLO minijet $E_T$ in mid-rapidity $\Delta y=1$. This calculation is shown in Fig. \ref{fig:ETER} for a fixed $\beta = 0.75$ at RHIC (lower bands) and at the LHC (upper bands), where the rising straight lines are the right-hand side of the averaged saturation equation in Eq.\ \eqref{eq:newavesat} with a fixed $K_{\textrm{sat}}=1$. 
\vspace{-1.4cm}
%%%%%%%%%%%%%%FIGURE%%%%%%%%%%%%%%%%%%%%%%%%%%%%%%%%%%%%
%%%%%%%%%%%%%%%%%%%%%%%%%%%%%%%%%%%%%%%%%%%%%%%%%%%%%%%%
The yellow error bands show the nPDF uncertainties which are computed using the 30 error sets in EPS09, and the black lines inside the yellow error bands are the $E_T$ computed using the EPS09 best fit (see details of the error analysis in \cite{PHD2}). We can conclude that the nPDF-originating uncertainties in the NLO $E_T$ calculation remain rather small even at the low perturbative saturation scales $(p_0 \sim 1,\dots,2~{\rm GeV})$ both at the LHC and RHIC. Saturation makes this uncertainty even smaller, as is seen in the figure.

\vspace{0.3cm}

\begin{figure}[h!]
\center
\includegraphics[scale=1.0]{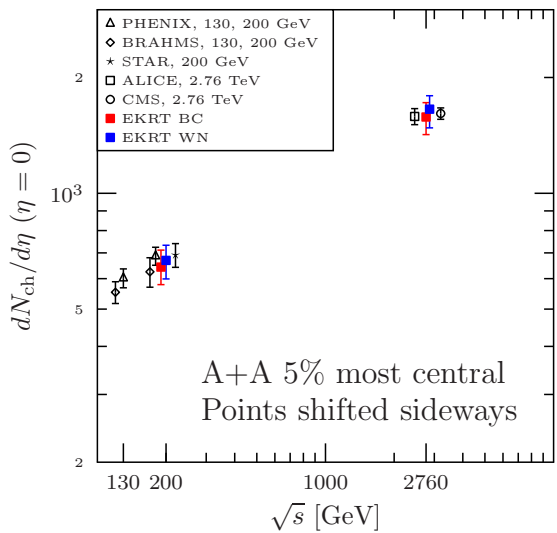}
\caption{Computed charged-particle multiplicity (filled blue and red squares), compared with RHIC and LHC data. Figure from \cite{PHD2}.}
\label{fig:mult}
\end{figure}

Using then one possible parameter combination $\beta=0.75$ and $K_{\textrm{sat}}=1$ with the BC and WN transverse profiles for the initial energy density, we show in Fig.\ \ref{fig:mult} the computed charged-particle multiplicity ${\rm d}N_{\textrm{ch}}/{\rm d}\eta$ in most central \textrm{Au+Au} collisions at RHIC and \textrm{Pb+Pb} collisions at the LHC compared with the measured data \cite{PHENIXPHD2,STARPHD2,BHRAMSPHD2} (RHIC) and \cite{Aamodt:2010pb,CMSPHD2} (LHC).
%%%%%%%%%%%%%%FIGURE%%%%%%%%%%%%%%%%%%%%%%%%%%%%%%%%%%%%
%%%%%%%%%%%%%%%%%%%%%%%%%%%%%%%%%%%%%%%%%%%%%%%%%%%%%%%%
Also in Fig. \ref{fig:mult} the propagation of the nPDF errors into our computed charged-particle multiplicity  (see details in \cite{PHD2}) is shown. First, from Fig.\ \ref{fig:mult} we can conclude that for the fixed parameters $\beta$ and $K_{\textrm{sat}}$ the uncertainties related to the computed charged-particle multiplicity are rather small, both at RHIC and the LHC. The nPDF uncertainty is about $\pm 10$\% and the transverse profile uncertainty is only a few percent (see also Table 1 in \cite{PHD2}). Second, we can conclude that in our new NLO-improved EKRT model with ideal hydrodynamics the $\sqrt{s_{NN}}$ scaling seems to work very well and the measured RHIC and LHC multiplicities are reproduced nicely. Third, and perhaps most importantly, since it is possible to fix the values of $\beta$ and $K_{\textrm{sat}}$ at one given cms-energy $\sqrt{s_{NN}}$ and then genuinely make the computation at another cms-energy without retuning the model further, we can conclude that our NLO-improved EKRT framework has some definite predictive power.

\vspace{0.3cm}

\begin{figure}[h!]
\center
\includegraphics[scale=.765]{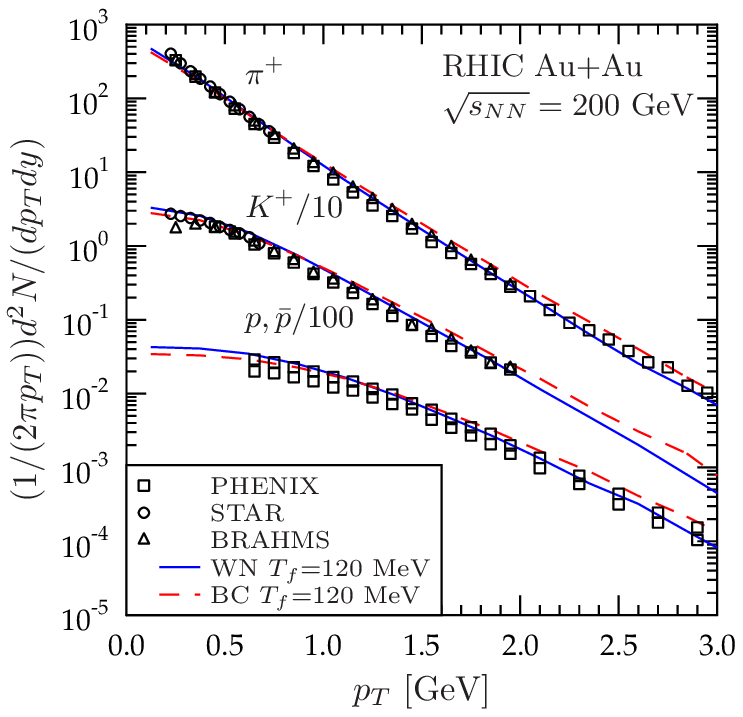}
\includegraphics[scale=.765]{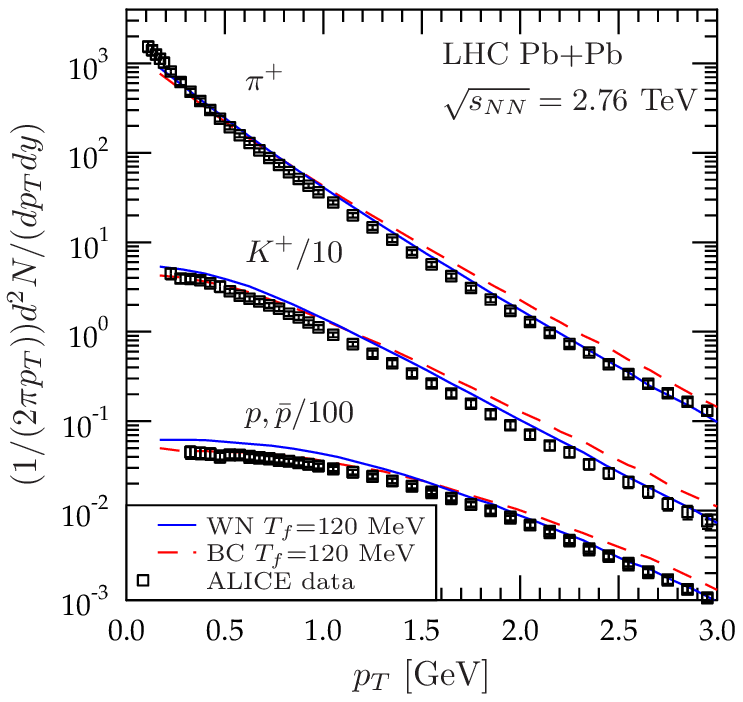}
\caption{The computed $p_T$ spectra of $\pi^{+},K^{+},p$ and $\bar{p}$ in most central $A+A$ collisions at RHIC (Left) and at the LHC (Right), compared with the data. Figure from \cite{PHD2}.}
\label{fig:spectra}
\end{figure}

Finally, in Fig.\ \ref{fig:spectra} we present our results for the computed  $p_{T}$ spectra of $\pi^{+}, K^{+}, p$ and $\bar{p}$ in most central \textrm{Au+Au} collisions at RHIC and in \textrm{Pb+Pb} collisions at the LHC. Also shown is the comparison with the data measured at RHIC \cite{RHICPT1,STARPHD2,RHICPT2} and at the LHC \cite{LHCPT}.
Here, the theory curves are computed as follows: First, we fixed the parameter setup $(\beta = 0.75, K_{\textrm{sat}}=1)$ and BC transverse profile based on the red band shown in Fig.\ \ref{fig:ETLASKU}. Note that this parameter setup also simultaneously reproduces the RHIC charged-particle multiplicity. Furthermore, the kinetic decoupling temperature $T_{\rm f}=120~{\rm MeV}$ is fixed on the basis of the RHIC spectra (left panel in Fig.\ \ref{fig:spectra}). Then, the LHC spectra (right panel in Fig.\ \ref{fig:spectra}) is computed without retuning $T_{\rm f}$. From Fig.\ \ref{fig:spectra} we can conclude that the WN profile would seem to reproduce the identified particle $p_T$ spectra better than the steeper BC profile, and overall the particle $p_T$ spectra are very nicely reproduced simultaneously both at RHIC and the LHC. Finally, it is worth noting that once $\beta, K_{\textrm{sat}}$ and $T_{\rm f}$ were fixed as explained above, the computed identified hadron $p_T$ spectra for the LHC is a genuine prediction from the model, and not a fit to the data.

\section{NLO-improved pQCD + local $E_T$ saturation and viscous hydrodynamics framework}

Finally, I come to the results obtained in \cite{PHD3}, where the centrality dependence of multiplicity, $p_T$-spectra and $v_2$ in \textrm{Au+Au} collisions at RHIC and \textrm{Pb+Pb} collisions at the LHC was computed using the NLO-improved pQCD + local saturation and viscous hydrodynamics framework. In this study we used the computed energy density profiles, see section \ref{locaini}, Figs.\ \ref{fig:initialstate} and \ref{fig:ecce}, and in the boost-invariant viscous hydrodynamic simulations (see chapter \ref{HYDRO}) the s95-PCE-v1 EoS with the chemical freeze-out at $T_{\rm chem} = 175~{\rm MeV}$ and kinetic freeze-out at $T_{\rm f} = 100~{\rm MeV}$.

\vspace{0.3cm}

\begin{figure*}[!]
\epsfxsize 5.9cm \epsfbox{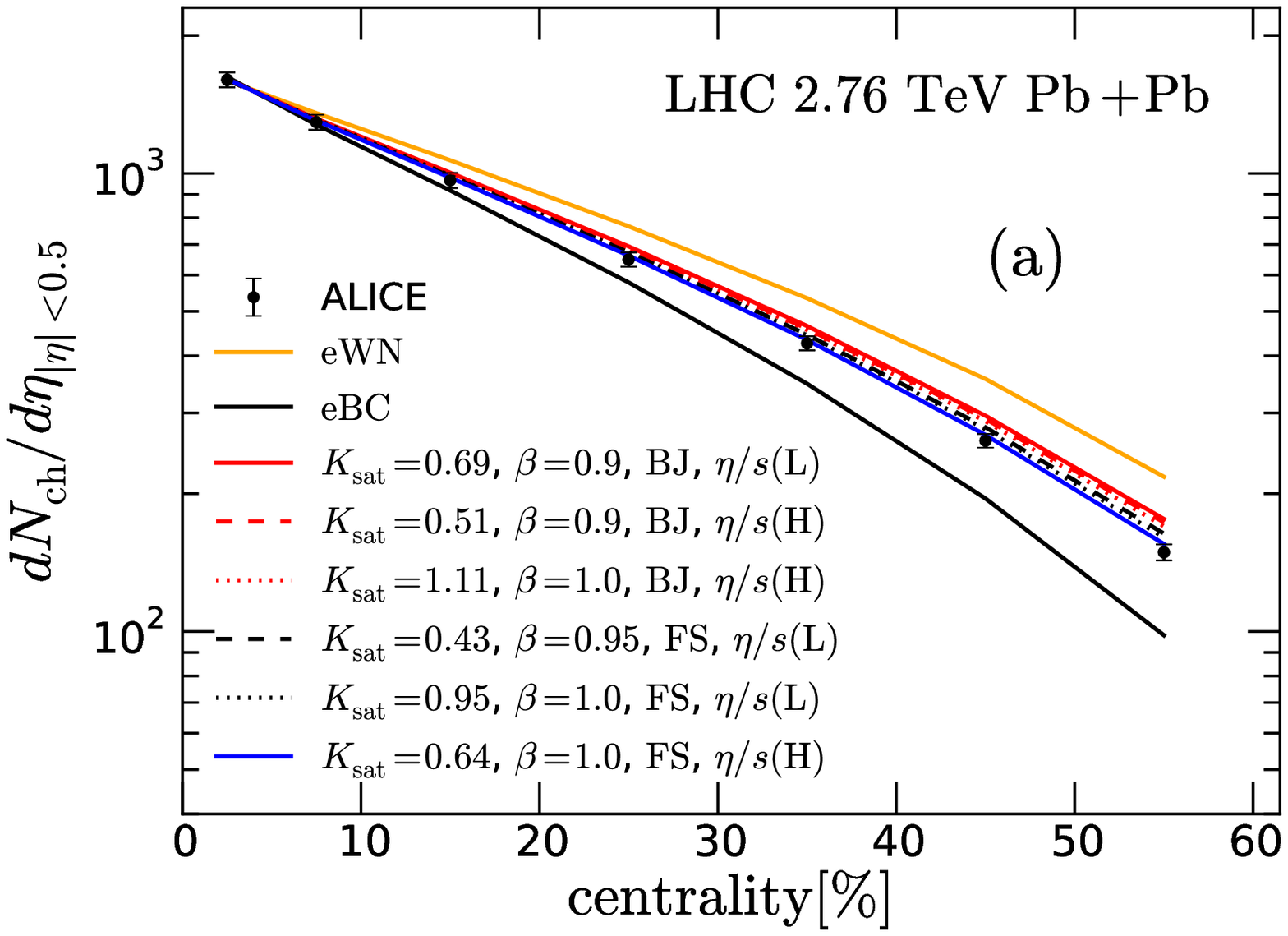} 
\epsfxsize 5.9cm \epsfbox{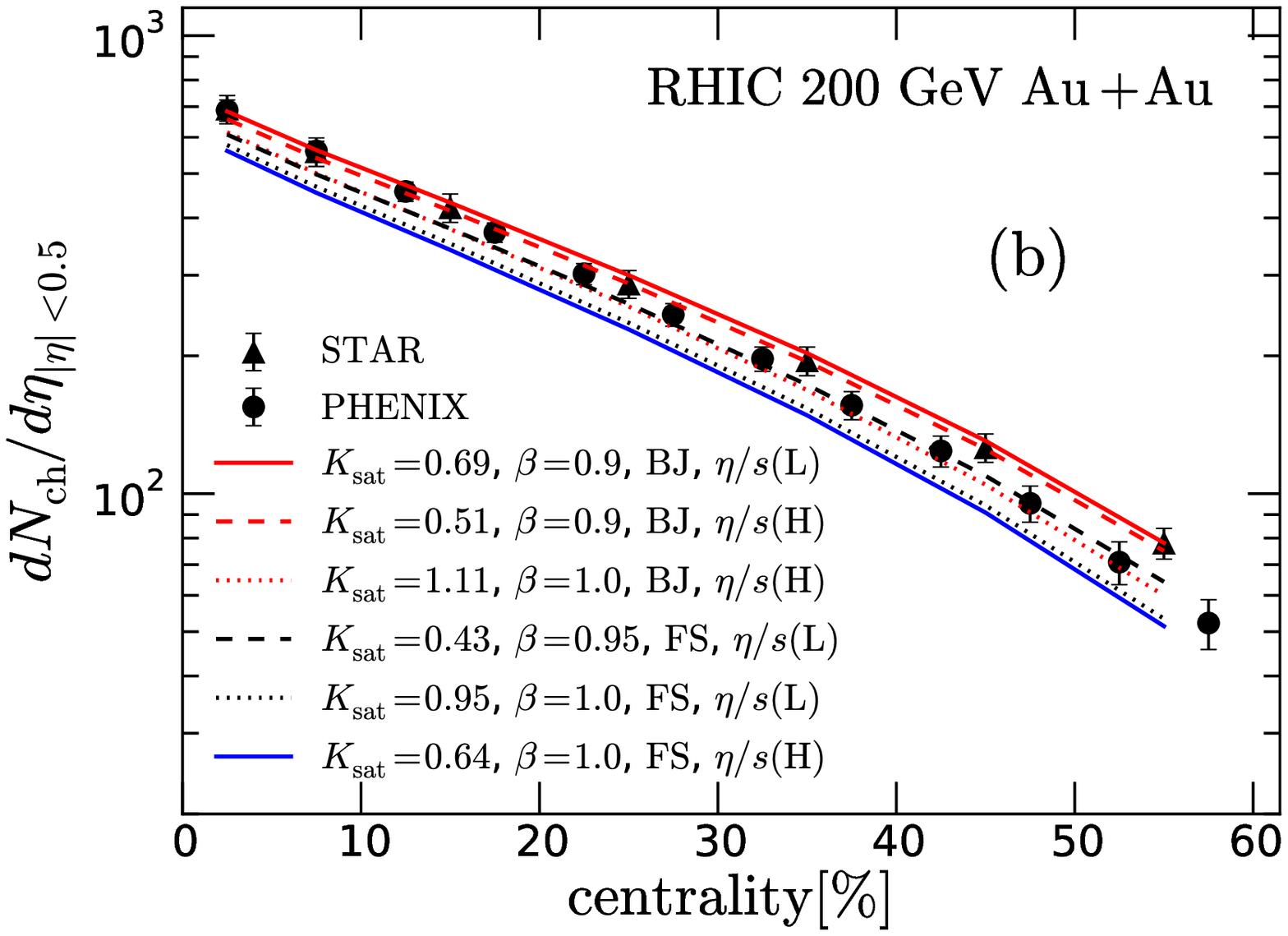} 
\epsfxsize 5.9cm \epsfbox{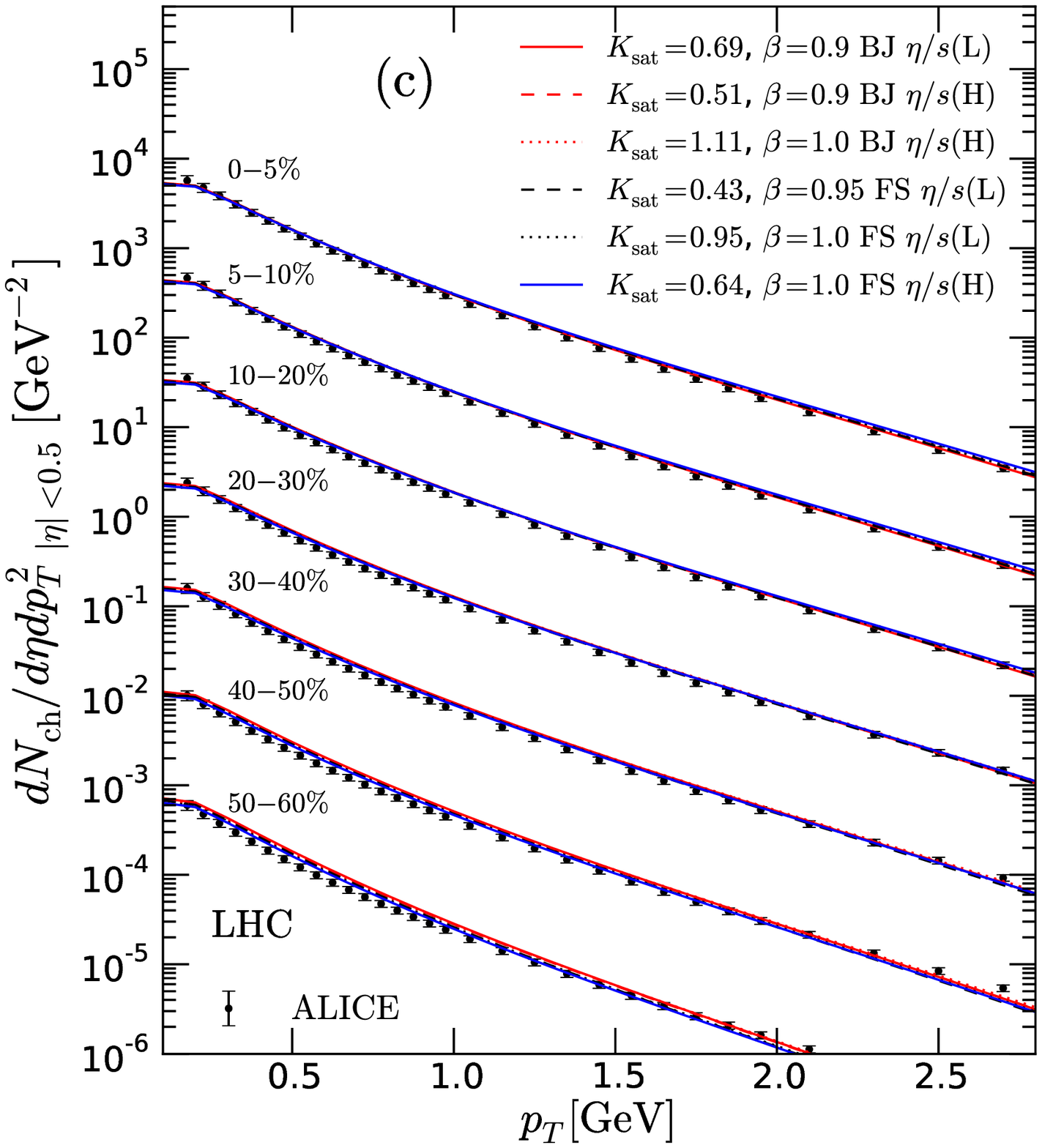} 
\epsfxsize 5.9cm \epsfbox{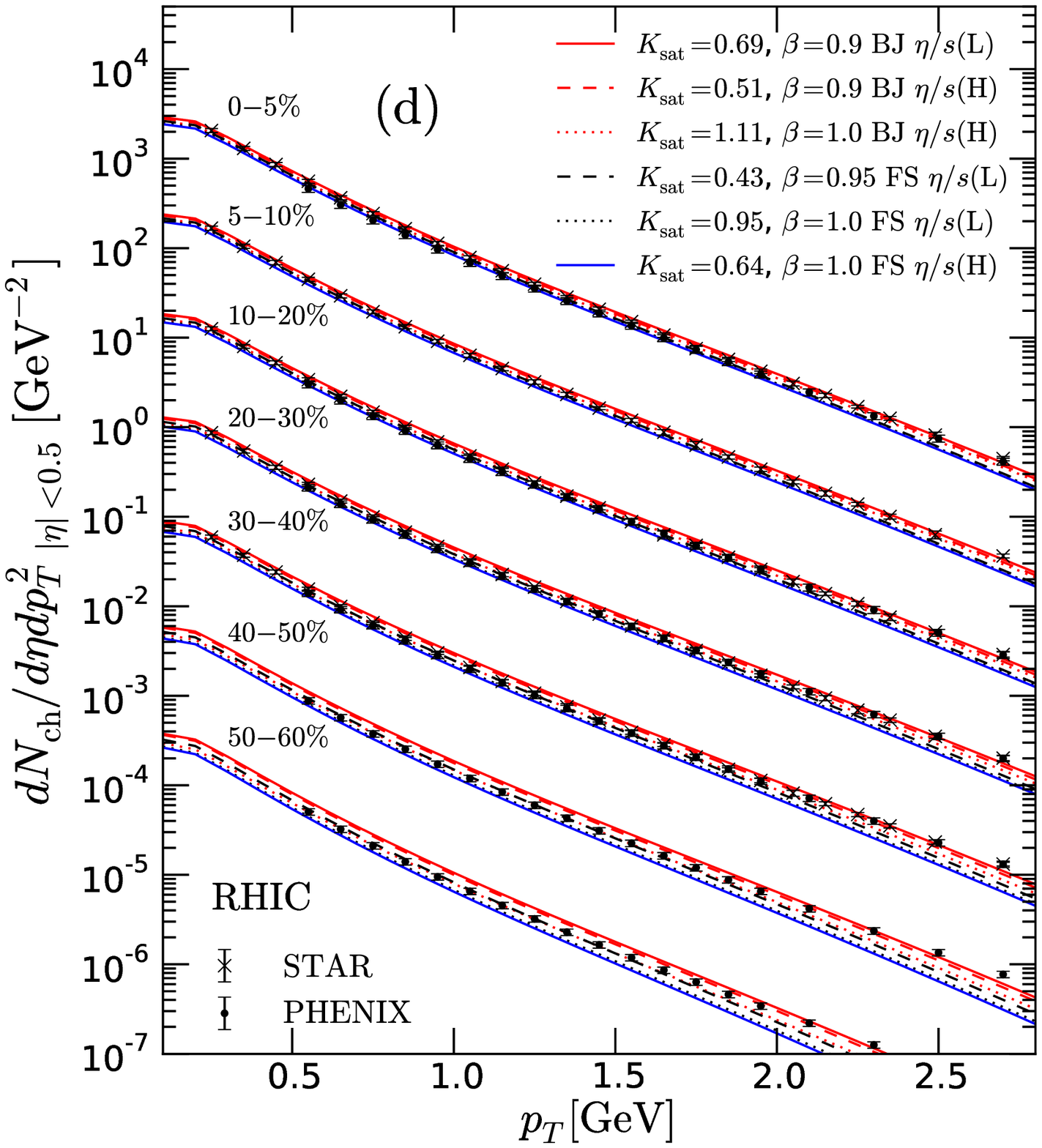} 
\epsfxsize 6.3cm \epsfbox{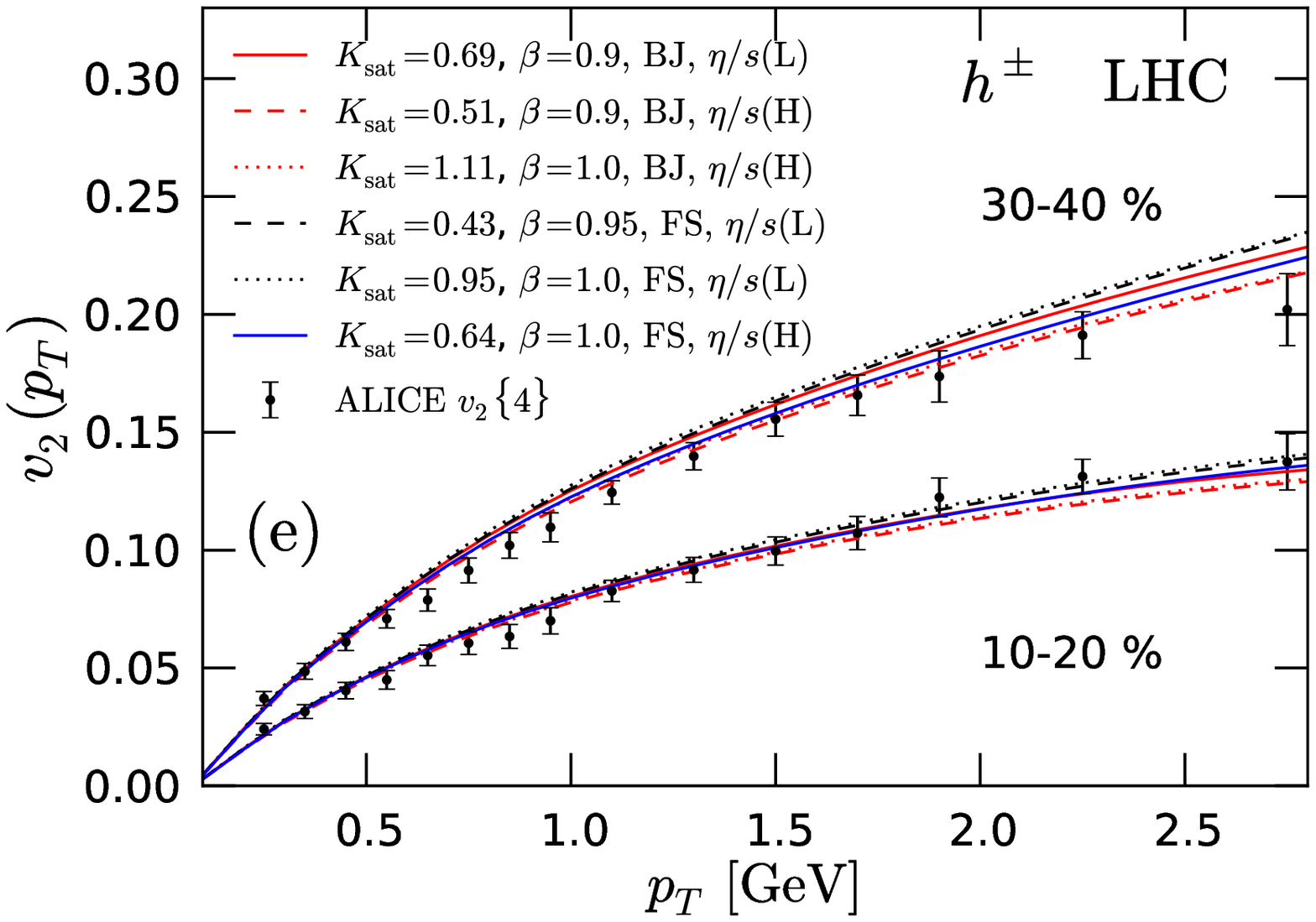} 
\epsfxsize 6.3cm \epsfbox{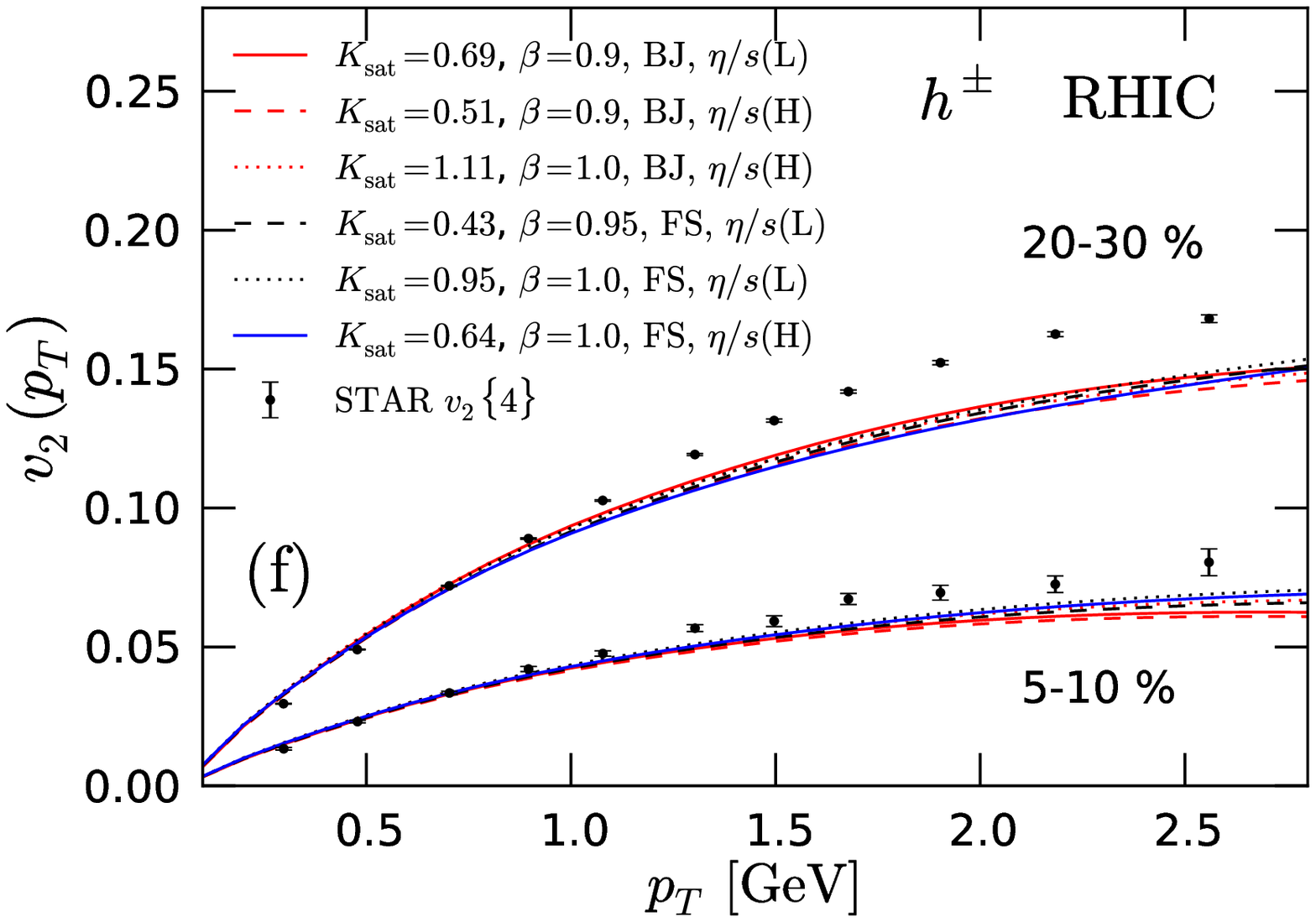} 
\caption{\protect 
Charged-hadron multiplicity as a function of centrality at the LHC (a) and RHIC (b). 
Charged-hadrons $p_T$ spectra at the LHC (c) and RHIC (d), in the centrality classes of the LHC data in panel (a), scaled down by increasing powers of 10. Elliptic flow $v_2(p_T)$ at the LHC (e) and RHIC (f), compared with the measured LHC data. Figure from \cite{PHD3}.
}
\label{fig: results4}
\end{figure*}

First, in Fig.\ \ref{fig: results4}a and \ref{fig: results4}b we show the computed centrality dependence of the charged hadron multiplicity in Pb+Pb collisions at $\sqrt{s_{NN}}=2.76$ TeV and in Au+Au collisions at $\sqrt{s_{NN}}=200$ GeV, and the comparison with the data measured at the LHC \cite{Aamodt:2010pb} and at RHIC 
\cite{PHENIXPHD2,STARPHD2}. Also, the results obtained with the eBC and eWN parametrizations based on the simple use of the Glauber model \cite{BCANDWN} are shown. For a given set of parameters $\{\beta, {\rm BJ}/{\rm FS}, \eta/s(T)\}$, the computations at RHIC and LHC are performed by tuning the remaining parameter $K_{\rm sat}$ in such a way that the multiplicity in the 5\%  most central collisions at the LHC is reproduced. As can be seen in the panels \ref{fig: results4}a and \ref{fig: results4}b, we can simultaneously describe the measured centrality dependence of the multiplicity at the LHC and RHIC with several sets of parameters $\{\beta, {\rm BJ}/{\rm FS}, \eta/s(T)\}$. These figures also demonstrate that the data seem to favor larger values of $\beta$ and the FS scenario over the BJ. Furthermore, it should be emphasized again that once the most central multiplicity at the LHC is fixed, the rest (other centralities and RHIC) is a prediction, and especially that no re-tuning has been performed from the LHC to RHIC.

\vspace{0.3cm}

Next, in Figs.\ \ref{fig: results4}c and \ref{fig: results4}d we show the centrality dependence of the computed $p_T$ spectra of charged hadrons at the LHC and RHIC. The data are from \cite{Abelev:2012hxa} (LHC) and \cite{Adams:2003kv,Adler:2003au} (RHIC). As these figures show, the $p_T$ spectra are not very sensitive to the parameters $\{\beta, {\rm BJ}/{\rm FS}, \eta/s(T)\}$ once the centrality dependence of the multiplicities is under control. Here we should also emphasize that the parameters $(T_{\rm f},T_{\rm dec})$ are kept unchanged from RHIC to LHC.

\vspace{0.3cm}

Finally, in Figs.\ \ref{fig: results4}e and \ref{fig: results4}f we show the computed elliptic flow coefficients $v_2(p_T)$ at the LHC and RHIC, respectively. Also the comparison with the data measured at the LHC \cite{Aamodt:2010pa} and  RHIC \cite{Bai} is shown. We observe that the $v_n$ coefficients depend strongly on the $\eta/s(T)$ parametrization: the "L" and "H" parametrizations shown in Fig.\ \ref{fig:etapers} both give a nice agreement with the data, while an ideal hydrodynamics would fail to reproduce the $v_2(p_T)$ data. Here a simultaneous LHC and RHIC analysis should be emphasized, and especially that as  $\eta/s(T)$ is a property of QCD matter, it must not be retuned when moving from LHC to RHIC.

%%%%%%%%%%%%%%%%%%%%% FIGURE %%%%%%%%%%%%%%%%%%%%%%%%%%%%%%%%
%%%%%%%%%%%%%%%%%%%%% FIGURE %%%%%%%%%%%%%%%%%%%%%%%%%%%%%%%%

\chapter{Conclusions and Outlook}
\label{conclusions}

In this thesis I have presented in detail the theoretical pQCD basis, implementation and the main numerical results of the new NLO-improved EKRT model framework.

\vspace{0.3cm}

As an important part of this thesis, I have documented the calculation techniques and details of an example process $qq'\rightarrow qq'$ to NLO pQCD. From this we can understand in detail how the NLO computation and the UV renormalization proceed. I have also discussed how the subtraction method is exploited in the minijet $E_T$ context.

\vspace{0.3cm}

In \cite{PHD1} the original but updated leading-order EKRT model with ideal hydrodynamics and parton-medium interaction modeling was used to describe simultaneously the low- and high-$p_T$ spectra of charged hadrons in central Pb+Pb collisions at the LHC. It was found that a good agreement with the low-$p_T$ $(\lesssim 4-5~{\rm GeV})$ data can be obtained by hydrodynamics and with the high-$p_T$ by pQCD + jet quenching simulations.

\vspace{0.3cm}

In \cite{PHD2} the original EKRT model was carried to NLO in pQCD as rigorously as possible. In particular, a new set of infrared and collinear safe measurement functions for the NLO minijet transverse energy calculation were constructed, and a new dynamical infrared and collinear safe saturation criterion for central $A+A$ collision was introduced. The framework was updated with the EPS09 NLO nuclear parton distributions and the propagation of the nPDF uncertainties into the computed minijet transverse energy, saturation scales and the final-state multiplicities were studied. 

\vspace{0.3cm}

The saturated minijet transverse energy calculation was converted into the QCD matter initial conditions for boost-invariant hydrodynamics. The new NLO-improved EKRT framework with ideal hydrodynamics was shown \cite{PHD2} to give a good description of the charged-particle multiplicity and identified bulk hadron $p_T$ spectra at $\sqrt{s_{NN}}=200~{\rm GeV}$ Au+Au collisions at RHIC and at $\sqrt{s_{NN}}=2.76~{\rm TeV}$ Pb+Pb collisions at the LHC. The article \cite{PHD2} demosntrated that the NLO-improved EKRT model indeed has a definite predictive power.

\vspace{0.3cm}

In \cite{PHD3} the new NLO-improved EKRT model was extended to non-central $A+A$ collisions, generalizing the new infrared and collinear safe saturation criterion in \cite{PHD2} to non-zero impact parameters by making it local in the transverse plane. In this study we also used the new EPS09s impact-parameter dependent nPDFs. 

\vspace{0.3cm}

Based on this model, the initial energy density profiles and formation times of the produced QGP at the LHC and RHIC were computed. Using the computed initial conditions for (2+1)-dimensional viscous hydrodynamics a good simultaneous agreement with the measured centrality dependence of the low-$p_T$ bulk observables at the LHC and RHIC was found. In particular, some constraints for the temperature dependence of the QCD matter shear viscosity were found, although we were not able to set error bars on the result.

\vspace{0.3cm}

In general, based on the results shown here we can conclude that the new NLO-improved EKRT framework as presented in this thesis gives a viable way to treat the initial parton production and to compute the initial conditions for hydrodynamics. We see the pQCD + saturation as a conveniently complementary description of the particle production mechanism in comparison to \textrm{e.g.} the Color Glass Condensate  based models (see for example \cite{CGCCOMP}), where the soft gluon fields are assumed to dominate the initial energy production.

\vspace{0.3cm}

In the future, our aim is to extend this very promising new NLO EKRT framework to include geometrical event-by-event (EbyE) fluctuations due to random nucleon configurations in the colliding nuclei as well as the dynamical fluctuations in the number of minijet collisions. This work is already in good progress and the first preliminary results have been presented in the Quark Matter 2014 conference \cite{QM2014}. Eventually, for a more global description of the heavy-ion bulk observables, including also studies of $p+A$ collisions and rapidity dependent phenomena, we aim at a Monte Carlo framework, where the different EbyE fluctuations could be coherently built in.

\begin{appendices}
\let\cleardoublepage\clearpage
\chapter{Basics of QCD}

\section{QCD Lagrangian}
\label{QCDL}

The Lagrangian density of Quantum Chromodynamics (QCD) can be written as \cite{MUTA,HBOOK,PESKIN}, 
\begin{equation}
\label{eq:QCDLag}
\mathcal{L}^{\text{eff}}_{\text{QCD}}  = \mathcal{L}_{\text{clas}} + \mathcal{L}_{\text{gauge}} + \mathcal{L}_{\text{ghost}},
\end{equation}
where $\mathcal{L}_{\text{clas}}$ is the classical Lagrange density, $\mathcal{L}_{\text{gauge}}$ is the gauge-fixing term and $\mathcal{L}_{\text{ghost}}$ is the ghost term. The classical Lagrangian density is invariant under local $\textrm{SU}(N_c=3)$ gauge transformations and it takes the form
\begin{equation}
\mathcal{L}_{\text{clas}}  = -\frac{1}{4}F^{a}_{\mu\nu}F^{a \mu\nu} + \sum_{q} \bar\Psi^{(q)}_{i,\alpha}\biggl \{i(\gamma^{\mu})_{\alpha\beta}(D_{\mu})_{ij} - m_{q}\delta_{\alpha\beta}\delta_{ij}\biggr \}\Psi^{(q)}_{\beta,j}.
\end{equation}
The nonabelian field-strength tensor $F^{a}_{\mu\nu}$ in terms of the gluon vector field $A^{a}_{\mu}$ is given by
\begin{equation}
F^{a}_{\mu\nu} = \partial_{\mu}A_{\nu} - \partial_{\nu}A_{\mu} + gf^{abc}A^{b}_{\mu}A^{c}_{\nu},
\end{equation}
where the index $a$ is running from 1 to $N_{c}^2 -1 = 8$, $g$ is the QCD coupling and $f^{abc}$ are the fully antisymmetric structure constants of the $\textrm{SU}(N_c)$ color group. The covariant derivative $(D_{\mu})_{ij}$ is given by
\begin{equation}
\label{eq:covder}
(D_{\mu})_{ij} = \delta_{ij}\partial_{\mu} - ig(T^a_{(F)})_{ij}A^{a}_{\mu},
\end{equation}
where $T^a_{(F)}$ are matrices representing the gauge group generators in the fundamental (F) representation of $\textrm{SU}(N_c)$. The covariant derivative acts on the  4-component spinor quark fields $\bar\Psi^{(q)}_{i,\alpha}$ and $\Psi^{(q)}_{j,\beta}$, where $i,j$ are  color indices running from 1 to $N_c$. There are $n_q$ independent quark fields, labeled by the flavor index $q$. 

\vspace{0.3cm}

The choice of a specific covariant gauge condition is most conveniently implemented already at the Lagrangian level by introducing the gauge fixing term
\begin{equation}
\mathcal{L}_{\text{gauge}} = -\frac{1}{2\xi}(\partial_{\mu}A^{a \mu})^2.
\end{equation}
Throughout the present thesis, we use the Feynman gauge, \textrm{i.e.} choose the gauge parameter as $\xi=1$. In the covariant gauges there are unphysical degrees of freedom in the classical fields which must be absorbed in non-physical fermion-like scalar fields, ghosts. For this, in the effective, physical $\mathcal{L}_{\text{ghost}}$, one introduces the ghost term with interactions between the gluons and ghost fields,
\begin{equation}
\mathcal{L}_{\text{ghost}} = (\partial_{\mu}\bar\omega^{a})\biggl \{\partial^{\mu}\delta^{ca} - gf^{abc}A^{b \mu} \biggr \}\omega^{c},
\end{equation}
where $\omega^a$ and $\bar\omega^a$ are the ghost fields.

\section{Feynman rules for QCD}
\label{QCDFR}

In the following we list the Feynman rules of QCD necessary for the computation of the partonic matrix elements discussed in chapter \ref{VIRCOR}.

\subsection{External lines}

For external quarks $(q)$ or anti-quarks ($\bar{q}$) of momentum $p$, spin $s$ and color index $i=1,2,3$, entering or leaving a diagram, the Feynman rules are:
\begin{figure}[h!]
\center
\includegraphics[scale=.50]{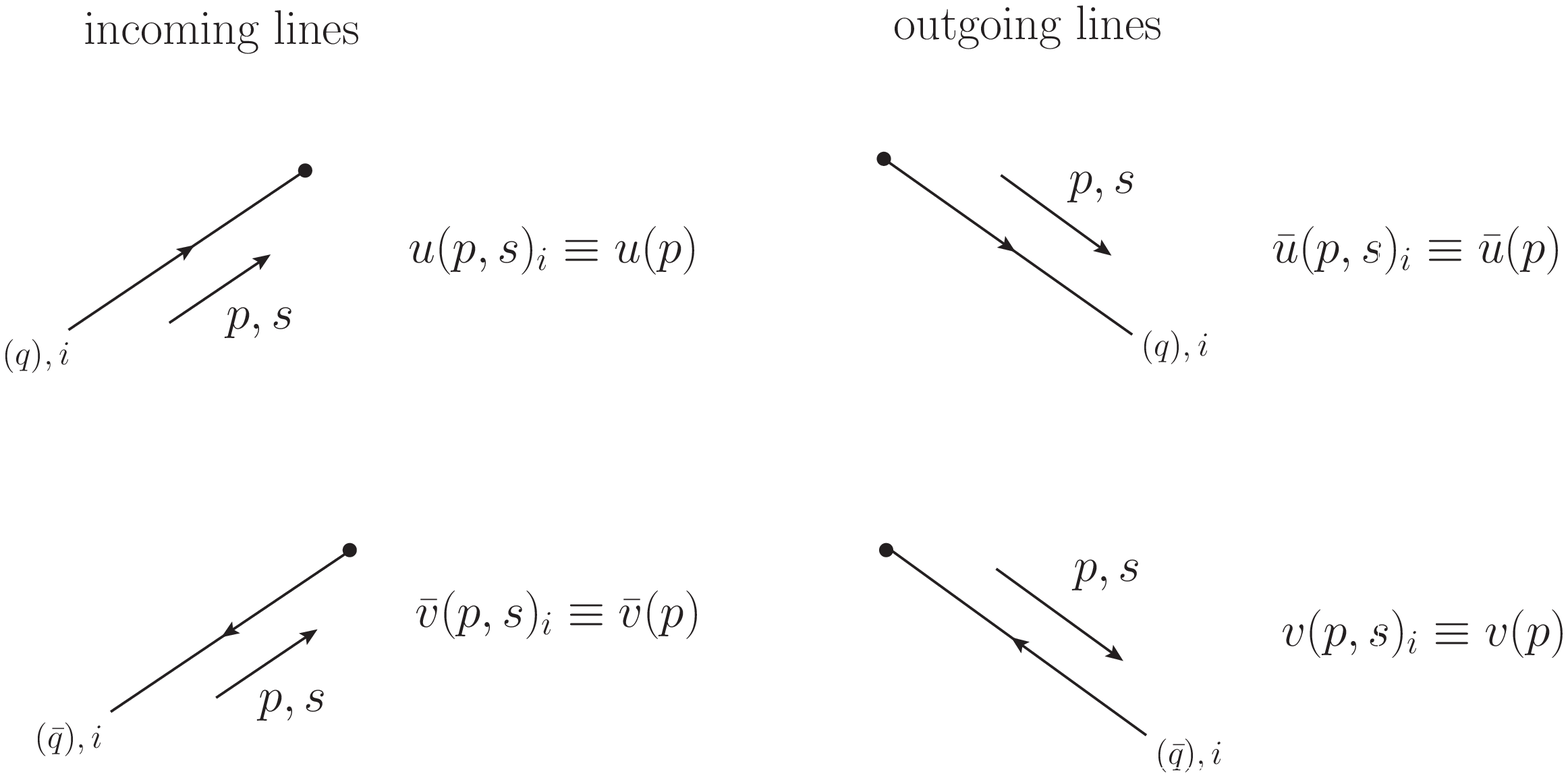}
%\caption{}
\label{fig:legs}
\end{figure}
\newline
where $u, \bar{u}$ and $v, \bar{v}$ are the quark spinors in the momentum space. For the massless quarks we consider here they satisfy the Dirac equation in momentum space
\begin{equation}
\label{eq:diraceq}
\cancel{p}u(p) = \bar{u}(p)\cancel p = 0, \quad \cancel{p}v(p) = \bar{v}(p)\cancel p = 0.
\end{equation}
For an incoming external gluon ($g$) of momentum $k$, polarization $\lambda$ and color index $a=1,\dots,8$, the Feynman rule is
\begin{equation}
\epsilon_{\mu}(k,\lambda)_a \equiv \epsilon_{\mu}(k),
\end{equation}
and for an outgoing gluon
\begin{equation}
\epsilon_{\mu}^{\ast}(k,\lambda)_a \equiv \epsilon_{\mu}(k)^{\ast},
\end{equation}
where $\epsilon_{\mu}$ is the Lorentz four-vector for the gluon's polarization.

\subsection{QCD vertices}
\label{vertices}
The quark-gluon, ghost-gluon and gluon-gluon interaction vertices are the following:

\begin{figure}[h!]
\center
\includegraphics[scale=.55]{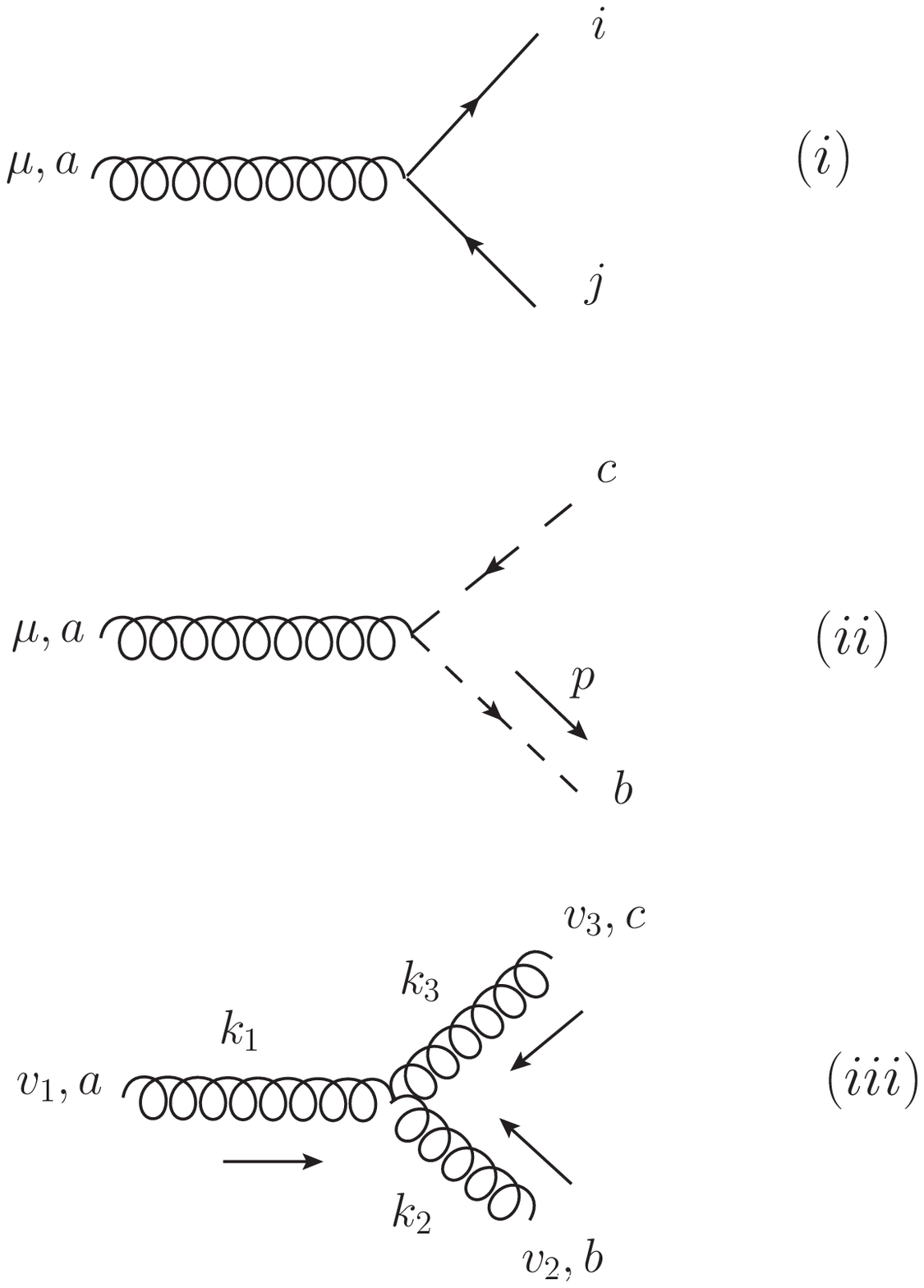}
%\caption{}
\label{fig:legs}
\end{figure}
and the corresponding Feynman rules for these are 
\begin{itemize}
\item[(i)]  \quad $ig(T^a)_{ij}\gamma^{\mu}$,
\item[(ii)] \quad $gf^{abc}p^{\mu}$, with $p^{\mu}$ being the momentum of the outgoing ghost,
\item[(iii)] \quad $gf^{abc}\mathcal{C}^{v_1v_2v_3}(k_1,k_2,k_3)$,
\end{itemize} 
where, with all momenta pointing into the vertex, 
\[
\mathcal{C}^{v_1v_2v_3}(k_1,k_2,k_3) = g^{v_1v_2}(k_1-k_2)^{v_3} + g^{v_2v_3}(k_2-k_3)^{v_1} + g^{v_3v_1}(k_3-k_1)^{v_2}.
\]

\subsection{QCD propagators}

The (massless) quark, gluon and ghost propagators are, correspondingly:
\begin{figure}[h!]
\center
\includegraphics[scale=.50]{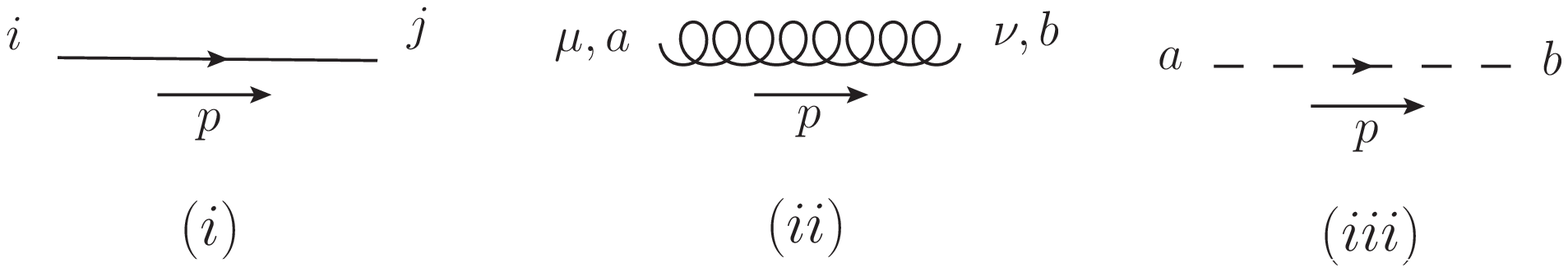}
%\caption{}
\label{fig:props}
\end{figure}

and their Feynman rules read
\begin{itemize}
\item[(i)]  \quad $\frac{i\cancel p\delta_{ij}}{p^2+i\delta}$,
\item[(ii)] \quad $\frac{-i\delta^{ab}}{p^2 + i\delta}\biggl [g^{\mu\nu} - (1-\xi)\frac{p^{\mu}p^{\nu}}{p^2}\biggr ]$,
\item[(iii)] \quad $\frac{i\delta^{ab}}{p^2 + i\delta}$,
\end{itemize} 
where $p$ is the momentum of the particle, $\delta \rightarrow 0_+$, and the notation $p_{\mu}\gamma^{\mu} \equiv \cancel p$ is used. The parameter $\xi$ in the gluon propagator fixes the gauge. 

\subsection{Other relevant Feynman rules}

In addition to the rules given above one has to:
\begin{itemize}
\item For every closed loop, integrate in $d$ dimensions over the loop momentum $k$. We adopt the notation
\begin{equation}
\label{eq:shothand}
\int \frac{{\rm d}^d k}{(2\pi)^d} \equiv \int_k.
\end{equation}
\item Multiply every quark or ghost loop by a factor $(-1)$ .
\item Include, if necessary, a symmetry factor which takes care of the possible permutations of fields.
\end{itemize}

\section{QCD color algebra }
\label{QCDCOLOUR}

In this section we provide a list of useful relations for the Lie algebra $\textrm{SU}(N_c)$ \cite{FIELDS, PESKIN}, which are actively used in chapter \ref{VIRCOR}.

\vspace{0.3cm}

The Lie algebra of $\textrm{SU}(N_c)$ is defined by the commutation relation 
\begin{equation}
[T^a_{(F)}, T^{b}_{(F)}] = if^{abc}T^{c}_{(F)},
\end{equation}
where the generators $T^{a}_{(F)} \equiv T^a$, in the $N_c$-dimensional fundamental representation, are a complete set of $N_c\times N_c$ traceless hermitean matrices. The standard generators of $\textrm{SU}(N_c=3)$ in QCD are $T^a = \lambda^a/2$ where $\lambda^a$ are the Gell-Mann matrices \cite{FIELDS}.

\subsection{Useful relations}

The fully symmetric structure constants $d_{abc}$ are defined according to 
\begin{equation}
\label{eq:anticolor}
\{T^a, T^b \} = \frac{1}{N_c}\delta^{ab} + d_{abc}T^c.
\end{equation}
Some useful relations:
\begin{equation}
\label{eq:COLORR1}
\begin{split}
\mathrm{Tr}(T^aT^b) & = \frac{1}{2}\delta^{ab},\\
\mathrm{Tr}(T^aT^bT^c) & = \frac{1}{4}\left (d^{abc} + if^{abc} \right ),\\
\mathrm{Tr}(T^aT^bT^aT^c) & = -\frac{1}{12}\delta^{bc}.
\end{split}
\end{equation}
The Fierz identity for $SU(N_c)$:
\begin{equation}
\label{eq:FIDE}
(T^a)_{ij}(T^a)_{k\ell} = \frac{1}{2}\left (\delta_{i\ell}\delta_{jk} - \frac{1}{N_c}\delta_{ij}\delta_{k\ell}\right ).
\end{equation}
In addition, some useful relations for the structure constants:
\begin{equation}
\label{eq:COLORR2}
\begin{split}
f_{acd}f_{bcd} &= N_c\delta_{ab}, \quad f_{abb} = 0,\\
d_{acd}d_{bcd} & = \frac{5}{3}\delta_{ab}, \quad d_{abb} = 0,\\
f_{acd}d_{bcd} & = 0.
\end{split}
\end{equation}

\chapter{Dirac $\gamma$ matrices and traces in $d$ dimensions}

In $d$ dimensions, we need to modify the standard 4-dimensional rules of the contractions and traces of Dirac $\gamma$ matrices. In $d$ dimensions, the metric tensor is
\begin{equation}
g^{\mu\nu} = \textrm{diag}(1,\underbrace{-1,-1,\cdots ,-1}_{d-1}),
\end{equation}
which  satisfies the relation $g^{\mu\nu}g_{\mu\nu} = d$. For the Dirac $\gamma$ algebra we can assume that there are $d$ Dirac matrices $\gamma^{\mu}$, $\mu=0,\dots,d-1$, which satisfy the anticommutation relation
\begin{equation}
\{\gamma^{\mu},\gamma^{\nu}\} = \gamma^{\mu}\gamma^{\nu} + \gamma^{\nu}\gamma^{\mu} = 2g^{\mu\nu}.
\end{equation}
Thus, the contraction of Dirac $\gamma$ matrices in $d$ dimensions satisfies the following properties \cite{PESKIN}:
\begin{equation}
\label{eq:gammarel1}
\begin{split}
\gamma^{\mu}\gamma_{\mu} & = d,\\
\gamma^{\mu}\gamma^{\nu}\gamma_{\mu} & = (2-d)\gamma^{\nu},\\
\gamma^{\mu}\gamma^{\nu}\gamma^{\rho}\gamma_{\mu} & = 4g^{\nu\rho} - (4-d)\gamma^{\nu}\gamma^{\rho},\\
\gamma^{\mu}\gamma^{\nu}\gamma^{\rho}\gamma^{\sigma}\gamma_{\mu} & = -2\gamma^{\sigma}\gamma^{\rho}\gamma^{\nu} + (4-d)\gamma^{\nu}\gamma^{\rho}\gamma^{\sigma}.
\end{split}
\end{equation}
Some useful traces over the Dirac $\gamma$-matrices are \cite{PESKIN}:
\begin{equation}
\label{eq:gammatra}
\begin{split}
\mathrm{Tr}[\gamma^{\mu}\gamma^{\nu}] & = 4g^{\mu\nu},\\
\mathrm{Tr}[\gamma^{\mu}\gamma^{\nu}\gamma^{\rho}\gamma^{\sigma}] & = 4\left (g^{\mu\nu}g^{\rho\sigma} - g^{\mu\rho}g^{\nu\sigma} + g^{\mu\sigma}g^{\nu\rho}\right ),\\
\mathrm{Tr}[\textrm{any odd} \# \gamma's] & = 0,\\
\mathrm{Tr}[\gamma^{\mu}\gamma^{\nu}\gamma^{\rho}\gamma^{\sigma}\dots] & = \mathrm{Tr}[\dots\gamma^{\sigma}\gamma^{\rho}\gamma^{\nu}\gamma^{\mu}].
\end{split}
\end{equation}

\chapter{One-loop tensor integrals}
\label{SINT}

In this Appendix \ref{SINT} part, I describe how to calculate the dimensionally regulated one-loop scalar and tensor integrals, which I actively use in chapter \ref{VIRCOR}. Some of the techniques that I use here are quite standard and details can be found in many basic quantum field theory books, see for example \cite{PESKIN,MUTA}. However, many of the more detailed techniques are usually described on quite a general level in the literature, and thus lots of work is needed if we would like to apply these to actual calculations. Hence, I believe it is useful to show here in detail how to use (and derive) these tools for the practical one-loop computations.

\section{One-loop $n$-point tensor integrals}

The one-loop  $n$-point tensor integrals have the general form
\begin{equation}
\label{eq:TIdef}
I_{(n)}^{\mu_1\ldots \mu_p} \equiv \int_k\frac{k^{\mu_1}\ldots k^{\mu_p}}{d_0\ldots d_{n-1}}, \quad \text{where} \quad \int_{k} \equiv \int\frac{{\rm d}^dk}{(2\pi)^{d}}, 
\end{equation}
$k$ is the loop momentum, $n$ is the number of propagators and $p$ the rank of the tensor integral. The scalar propagator factors $d_i$ are defined as
\begin{equation}
d_{j} = (k+r_j)^2  + i\delta,
\end{equation}
where the four-momenta $r_j$ are related with the external momenta $p_i$ through the relations
\begin{equation}
 r_j = \sum_{i=1}^{j} p_i, \quad j = 1,\ldots,n-1, \quad\quad r_0 = 0,
\end{equation} 
and the small imaginary part, $+i\delta$, where $\delta \rightarrow 0_+$, fixes the analytic continuations. We define the four independent massless scalar integrals as:
\begin{equation}
\label{eq:scalarnpoint}
\begin{split}
A_{0} & = \int_k \frac{1}{d_0},\\
B_{0}(p^2) & = \int_k\frac{1}{d_0d_1},\\
C_{0}(p_1,p_2;s_{12}) & = \int_k \frac{1}{d_0d_1d_2},\\
D_{0}(p_1,p_2,p_3,p_4;s_{12},s_{23}) & = \int_k \frac{1}{d_0d_1d_2d_3}, 
\end{split}
\end{equation}
where $p^2\neq 0$, and the invariants 
\begin{equation}
s_{12} = (p_1+p_2)^2,\quad s_{23}=(p_2+p_3)^2, \quad  s_{13}=(p_1+p_3)^2,
\end{equation}
fulfilling $s_{12}+s_{23}+s_{13}=0$. Here, we should note that all external momenta are taken to be incoming.

\section{Calculation of massless scalar integrals in QCD}

The calculation of basic massive and massless $(n \leqslant 4)$ scalar one-loop integrals has required a great amount of work, see for example \cite{VHSCALAR,LOOPS1,LOOPS2,LOOPS3} (and references therein). In the original article by t'Hooft and Veltman \cite{VHSCALAR}, the general scalar one-loop integrals with massive internal lines (\textrm{i.e.} integrals do not contain IR and CL singularities) were computed by using the dimensional regularization approach, and the final formulae were expressed in terms of the logarithm and dilogarithm functions. However, these general results are not so practical in QCD since the gluon propagators are massless and often in the computations where the high-energy limit is considered also the quarks are treated as massless particles. For QCD the $\epsilon$-expanded (in $ 4-2\epsilon$ dimensions) final results for  dimensionally regulated scalar integrals, which also contain the IR and CL singularities, are nicely collected in \cite{LOOPS3}.

\vspace{0.3cm}

In this section I show in detail how the massless $A_0, B_0, C_0$ and $D_0$ scalar one-loop integrals, which are used frequently in chapter \ref{VIRCOR}, can be calculated using the dimensional regularization and the Feynman parameter method. Throughout this section I will apply techniques that are presented in \cite{VHSCALAR,LOOPS1,LOOPS2}.

\subsection{Mathematical toolbox for loops}

To combine the propagator denominators in Eq. \eqref{eq:scalarnpoint}, we introduce integrals over the Feynman parameters $x_i$ \cite{PESKIN},
\begin{equation}
\label{eq:Feypar}
\frac{1}{d_0d_1\dots d_n} = (n-1)!\int_{0}^{1}{\rm d}x_1\dots {\rm d}x_n\delta(1-\sum_{i=1}^{n} x_i)\frac{1}{\mathcal{D}^n},
\end{equation}   
where $\mathcal{D} = x_1d_0 + x_2d_1 + \dots x_nd_n$. Using the identity above, we can reduce the general scalar $n$-point function into a linear combination of $n$-point integrals
\begin{equation}
I_{n} = (n-1)!\int_{0}^{1}{\rm d}x_1\dots {\rm d}x_n\delta(1-\sum_{i=1}^{n} x_i)\int_k\frac{1}{\mathcal{D}^n}.
\end{equation}
All formulae needed in the calculation of the $n$-point functions may be derived from only one general integral \cite{PESKIN},
\begin{equation}
\label{eq:masterformula}
\int_k\frac{1}{(k^2-\Delta + i\delta)^n} = (-1)^n\frac{i(4\pi)^{2-\frac{d}{2}}}{(4\pi)^2}\frac{\Gamma(n-\frac{d}{2})}{\Gamma(n)}\left (\frac{1}{\Delta -i\delta}\right )^{n-\frac{d}{2}},
\end{equation}
where $\Delta = \Delta(p_i,x_i)$, which now is a function of external momentum and Feynman parameters $x_i$. If $\Delta < 0$, the $+i\delta$ convention makes sure that we can perform the integral in Eq.\ \eqref{eq:masterformula} with the correct logarithm branch.  

\subsection{Special functions}

The gamma function $\Gamma(\beta)$ appearing in Eq.\ \eqref{eq:masterformula} is defined by the integral
\begin{equation}
\Gamma(\beta) = \int_{0}^{\infty}t^{\beta-1}e^{-t}{\rm d}t,  
\end{equation}
where the parameter $\beta$ is real and positive. The gamma function satisfies the relations
\begin{equation}
\begin{split}
\Gamma(1+\beta) & = \beta\Gamma(\beta),\\
\Gamma(1-\beta) & = -\beta\Gamma(-\beta).
\end{split}
\end{equation}
The expansion of $\Gamma(x)$ near its pole is
\begin{equation}
\Gamma(x) = \frac{1}{x}-\gamma_E + \mathcal{O}(x),
\end{equation} 
where $\gamma_E$ is the Euler-Mascheroni constant.

\vspace{0.3cm}

The beta function $B(r,s)$ is defined by the integral
\begin{equation}
\label{eq:betafunction}
B(r,s) = \int_{0}^{1}{\rm d} u (1-u)^{s-1} u^{r-1},
\end{equation}
for all $\mathbb{R}e(r)>0$ and $\mathbb{R}e(s)>0$. The relation between the gamma function and beta function is given by
\begin{equation}
B(r,s) = \frac{\Gamma(r)\Gamma(s)}{\Gamma(r+s)}.
\end{equation}

\vspace{0.3cm}

The integral representation of the hypergeometric function $ {}_{2}F_{1}$ is defined by the integral
\begin{equation}
\label{eq:hypegeometric}
{}_{2}F_{1}(a,b,c;z) = \frac{\Gamma(c)}{\Gamma(b)\Gamma(c-b)}\int_{0}^{1}{\rm d}x~x^{b-1}(1-x)^{c-b-1}(1-xz)^{-a}
\end{equation}
for all $\mathbb{R}e(c) > 0$ and $\mathbb{R}e(b) >0$ with $\vert \arg(1-z)\vert < \pi$. 

\vspace{0.3cm}

A useful linear transformation formula for the hypergeometric function $ {}_{2}F_{1}$ reads,
\begin{equation}
\label{eq:hflin1}
\begin{split}
&{}_{2}F_{1}(a,b,c;z) =  \frac{\Gamma(c)\Gamma(c-a-b)}{\Gamma(c-a)\Gamma(c-b)} {}_{2}F_{1}(a,b,1+a+b-c;1-z) \\
& + (1-z)^{c-a-b}\frac{\Gamma(c)\Gamma(a+b-c)}{\Gamma(a)\Gamma(b)} {}_{2}F_{1}(c-a,c-b,1+c-a-b;1-z)
\end{split}
\end{equation}
for $\vert \arg(1-z) \vert < \pi$. In addition, another useful relation is
\begin{equation}
\label{eq:hfarg1}
{}_{2}F_{1}(a,b,c;1) = \frac{\Gamma(c)\Gamma(c-a-b)}{\Gamma(c-a)\Gamma(c-b)}.
\end{equation}

\subsection{Scalar one-point $A_0$ function}

In the dimensional regularization approach the massless (scaleless) integral 
\begin{equation}
\label{eq:A0int}
A_0 = \int_k \frac{1}{k^2+i\delta}
\end{equation} 
is zero, and thus we set $A_0 = 0$ \cite{MUTA}.

\subsection{Scalar two-point $B_0$ function}

Using the Feynman parametrization in Eq. \eqref{eq:Feypar}, we obtain
\begin{equation}
\label{eq:B0int}
B_0(p^2) = \int_{0}^{1}{\rm d}x_1\int_k \frac{1}{\mathcal{D}^2},
\end{equation} 
where
\begin{equation}
\begin{split}
\mathcal{D} & = x_1d_0 + (1-x_1)d_1= (k + x_1p)^2 - \Delta + i\delta 
\end{split}
\end{equation}
and  $\Delta = -p^2x_1(1-x_1)$. Shifting the loop momentum $k$ to $\ell = k + x_1p$, and performing the $d$-dimensional loop momentum integration using Eq.\ \eqref{eq:masterformula} for $d=4-2\epsilon$, we find
\begin{equation}
B_0(p^2) = \frac{i(4\pi)^{\epsilon}}{(4\pi)^2}\Gamma(\epsilon)\int_{0}^{1}{\rm d}x_1\left (\frac{1}{\Delta -i\delta} \right )^{\epsilon}.
\end{equation}
Expanding the integrand in powers of $\epsilon$, we get
\begin{equation}
B_0(p^2) = \frac{(4\pi)^{\epsilon}}{(4\pi)^2}R_{\Gamma}\int_{0}^{1}{\rm d}x_1\biggl \{\frac{1}{\epsilon} -\ln\biggl [-p^2x_1(1-x_1) - i\delta \biggr ] + \mathcal{O}(\epsilon) \biggr \},
\end{equation}
where we have introduced the overall constant $R_{\Gamma}$ which occurs in $d$-dimensional integrals,
\begin{equation}
\label{eq:RGAMMA}
R_{\Gamma} = \frac{\Gamma^2(1-\epsilon)\Gamma(1+\epsilon)}{\Gamma(1-2\epsilon)} = 1 - \epsilon\gamma_E + \epsilon^2\left ( \frac{\gamma_E^2}{2}-\frac{\pi^2}{12}\right ) + \mathcal{O}(\epsilon^3).
\end{equation}
Performing the integral over the Feynman parameter $x_1$, we get our final result for $B_0$,
\begin{equation}
\label{eq:B0final}
B_0(p^2) = i\frac{(4\pi)^{\epsilon}}{(4\pi)^2}R_{\Gamma}\left ( \frac{1}{-p^2-i\delta}\right )^{\epsilon}\biggl \{\frac{1}{\epsilon} + 2 \bigg \} + \mathcal{O}(\epsilon).
\end{equation}

\subsection{Scalar three-point $C_0$ function}

Combining the denominators in the function $C_0$ with the Feynman parametrization formula given in Eq. \eqref{eq:Feypar} and eliminating the delta function by performing the $x_3$ integration, we find
\begin{equation}
C_0(p_1,p_2,s_{12}) = 2\int [{\rm d} \mathbf{x}]\int_k\frac{1}{\mathcal{D}^3},
\end{equation}
where 
\begin{equation}
\label{eq:c0feyint}
\int [{\rm d} \mathbf{x}]  = \int_{0}^{1}{\rm d}x_1\int_{0}^{1-x_1}{\rm d}x_2,
\end{equation}
and
\begin{equation} 
\begin{split}
\mathcal{D} & = x_1d_2 + x_2d_0 + (1-x_1-x_2)d_1 = (k + \xi)^2 -\Delta + i\delta,\\
\xi & = p_1(1-x_1-x_2) +x_1(p_1+p_2),
\end{split}
\end{equation} 
and $\Delta = -s_{12}x_1x_2$. Note that there are also other possible parametrizations of the function $\mathcal{D}$ but the above form is the most convenient for the calculations presented here. The $d$-dimensional loop momentum integration can be done by shifting the loop momentum $k$ to $\ell = k + \xi$ and applying Eq.\ \eqref{eq:masterformula} for $d=4-2\epsilon$,
\begin{equation}
\label{eq:c0int2}
C_0(p_1,p_2,s_{12}) = -\frac{i(4\pi)^{\epsilon}}{(4\pi)^2}\Gamma(1+\epsilon)\int [{\rm d} \mathbf{x}]\left ( \frac{1}{\Delta -i\delta}\right )^{1+\epsilon}.
\end{equation}
Simplifying the double integral in Eq.\ \eqref{eq:c0int2} further by introducing a new integration variable $x_2 = x(1-x_1)$,
\begin{equation}
\label{eq:c0int2v2}
\begin{split}
C_0(p_1,p_2,s_{12}) = \frac{i(4\pi)^{\epsilon}}{(4\pi)^2}&\frac{1}{s_{12}}\left (\frac{1}{-s_{12}-i\delta}\right )^{\epsilon}\Gamma(1+\epsilon)\\
&\times\int_{0}^{1} {\rm d}x_1(1-x_1)^{-\epsilon}x_1^{-\epsilon-1} \int_{0}^{1} {\rm d}x x^{-\epsilon-1}
\end{split}
\end{equation}
and applying Eq.\ \eqref{eq:betafunction}, gives
\begin{equation}
C_0(p_1,p_2,s_{12}) = \frac{i(4\pi)^{\epsilon}}{(4\pi)^2}\frac{1}{s_{12}}\left (\frac{1}{-s_{12}-i\delta}\right )^{\epsilon}\frac{\Gamma(-\epsilon)^2\Gamma(1+\epsilon)}{\Gamma(1-2\epsilon)}.
\end{equation}
Expanding the gamma functions in powers of $\epsilon$, we obtain our final result for $C_0$,
\begin{equation}
\label{eq:C0final}
\begin{split}
C_0(p_1,p_2,s_{12}) & = i\frac{R_{\Gamma}}{(4\pi)^2}\frac{1}{s_{12}}\left (\frac{4\pi}{-s_{12}-i\delta}\right )^{\epsilon}\frac{1}{\epsilon^2}.
\end{split}
\end{equation}
%\\
%& = \frac{i(4\pi)^{\epsilon}}{(4\pi)^2}\frac{R_{\Gamma}}{Q^2}\biggl \{\frac{1}{\epsilon^2} -\frac{1}{\epsilon}\ln(-Q^2-i\delta) + \frac{1}{2}\ln^2(-Q^2-i\delta) \biggr \} + \mathcal{O}(\epsilon).

\subsection{Scalar four-point $D_0$ function}

Following the same Feynman parametrization procedure as before, we obtain
%The one-loop scalar $D_0$ function in $d$-dimension space-time is given by
%\begin{equation}
%D_0(p_1,p_2,p_3,p_4,s_{12},s_{23}) = \int_k\frac{1}{d_0d_1d_2d_3},
%\end{equation}
%where all external momenta are taken to be incoming, so that the massless propagators have the form:
%\begin{equation}
%\begin{split}
%d_0 & = k^2 +i\delta,\\
%d_1 & = (k+p_1)^2 + i\delta,\\
%d_2 & = (k+p_1+p_2)^2 + i\delta,\\
%d_3 & = (k+p_1+p_2+p_3)^2 + i\delta. 
%\end{split}
%\end{equation}
%Here $p_i^2=0$ for all $i=1,2,3,4$. Combining the denominators with the Feynman parametrization formula given in Eq. \eqref{eq:Feypar} and eliminating the delta function by performing the $x_4$ integration we get
\begin{equation}
D_0(p_1,p_2,p_3,p_4;s_{12},s_{23}) = 3!\int [{\rm d} \mathbf{x}]\int_k\frac{1}{\mathcal{D}^4},
\end{equation}
where 
\begin{equation}
\int [{\rm d} \mathbf{x}]  = \int_{0}^{1}{\rm d}x_1\int_{0}^{1-x_1}{\rm d}x_2\int_{0}^{1-x_1-x_2}{\rm d}x_3 
\end{equation}
and
\begin{equation} 
\mathcal{D}  = x_1d_0 + x_2d_1 + x_3d_2 + (1-x_1-x_2-x_3)d_3. 
\end{equation}
Completing the square in the denominator $\mathcal{D}$ by shifting the loop momentum to $\ell = k - \xi$ with $\xi = p_4(1-x_1-x_2) - p_1x_2 + p_3x_3$, we have
\begin{equation}
\mathcal{D} =  \ell^2 - \Delta + i\delta, 
\end{equation}
where
\begin{equation}
\Delta  = -s_{23}(1-x_1-x_2-x_3)x_2 - s_{12}x_1x_3.
\end{equation}
Performing the $d$-dimensional loop momentum integration using Eq.\ \eqref{eq:masterformula}, and taking $d=4-2\epsilon$  we obtain
\begin{equation}
D_0(s_{12},s_{23}) = \frac{i(4\pi)^{\epsilon}}{(4\pi)^2}\Gamma(2+\epsilon)\int [{\rm d} \mathbf{x}]\left (\frac{1}{\Delta - i\delta} \right )^{2+\epsilon}.
\end{equation}
To proceed with the integration $D_0$, we apply the following linear transformation of the integration variables 
\begin{equation}
\begin{split}
x_1 & = (1-x)(1-y),\\
x_2 & = x(1-y),\\
x_3 & = yz,
\end{split}
\end{equation}
where the Jacobian corresponding to this transformation is $y(1-y)$. Thus, the integral above takes the form
\begin{equation}
D_0(s_{12},s_{23}) = \frac{i(4\pi)^{\epsilon}}{(4\pi)^2}\Gamma(2+\epsilon) \int_{0}^{1}{\rm d}z\int_{0}^{1}{\rm d}y\int_{0}^{1}{\rm d}x ~\mathcal{I}(x,z,y;s_{12},s_{23}),
\end{equation}
where
\begin{equation}
\mathcal{I}(x,z,y;s_{12},s_{23}) = \frac{y(1-y)}{\biggl \{-y(1-y)\biggl [x(1-z)s_{23} + z(1-x)s_{12} \biggr ] -i\delta \biggr \}^{2 + \epsilon}}.
\end{equation}
Next, the integration over $x$ is easily performed and we get the result 
\begin{equation}
\begin{split}
D_0(s_{12},s_{23}) = \frac{i(4\pi)^{\epsilon}}{(4\pi)^2}\Gamma(1+\epsilon) \int_{0}^{1}&\frac{{\rm d}z}{zs_{12}-(1-z)s_{23}}\\
& \times \int_{0}^{1}{\rm d}y~\mathcal{R}(z,y;s_{12},s_{23}),
\end{split}
\end{equation}
where 
\begin{equation}
\begin{split}
\mathcal{R}(z,y;s_{12},s_{23}) = &\biggl \{-y(1-y)zs_{12} - i\delta \biggr \}^{-\epsilon-1}\\
& - \biggl \{ -y(1-y)(1-z)s_{23}-i\delta\biggr \}^{-\epsilon-1}.
\end{split}
\end{equation}
Simplifying  both terms in the curly brackets we find
\begin{equation}
\begin{split}
\mathcal{R}(z,y;s_{12},s_{23}) = y^{-\epsilon-1}&\biggl \{ \biggl [-zs_{12}  -i\delta\biggr ]^{-\epsilon-1} (1-y\xi_1)^{-\epsilon-1} \\
& - \biggl [-(1-z)s_{23}-i\delta \biggr ]^{-\epsilon-1}(1-y\xi_2)^{-\epsilon-1}\biggr \},
\end{split}
\end{equation}
where we denote
\begin{equation}
\begin{split}
\xi_1 & = 1 - \frac{i\delta}{zs_{12} + i\delta},\\
\xi_2 & = 1 - \frac{i\delta}{(1-z)s_{23} + i\delta}.
\end{split}
\end{equation}
The integral over $y$  can be performed by applying Eq.\ \eqref{eq:hypegeometric}, and thus we obtain the result
\begin{equation}
\label{eq:D0intstep}
D_0(s_{12},s_{23}) =\frac{i(4\pi)^{\epsilon}}{(4\pi)^2}\frac{\Gamma(1+\epsilon)}{(-\epsilon)} \int_{0}^{1}\frac{{\rm d}z}{zs_{12}-(1-z)s_{23}}\mathcal{T}(z;s_{12},s_{23}),
\end{equation}
where 
\begin{equation}
\label{eq:funktio}
\begin{split}
\mathcal{T}(z;s_{12},s_{23}) = & \biggl [-zs_{12}  -i\delta\biggr ]^{-\epsilon-1}{}_{2}{F}_{1}(1+\epsilon,-\epsilon,1-\epsilon;\xi_1)\\
& - \biggl [-(1-z)s_{23}-i\delta \biggr ]^{-\epsilon-1}{}_{2}{F}_{1}(1+\epsilon,-\epsilon,1-\epsilon;\xi_2).
\end{split}
\end{equation}
Next, we rewrite
\begin{equation}
\label{eq:aputulos}
\frac{1}{zs_{12}-(1-z)s_{23}} = \left (\frac{1}{s_{12}+s_{23}}\right ) \frac{1}{z-z_0} = \frac{1}{\Delta s} \frac{1}{z-z_0},
\end{equation}
where $\Delta s = s_{12} + s_{23}$ and $z_0 = s_{23}/(s_{12}+s_{23})$. Then we express the product between Eq.\ \eqref{eq:aputulos} and Eq.\ \eqref{eq:funktio} as
\begin{equation}
\label{eq:apu1}
\begin{split}
& \frac{1}{\Delta s}\frac{1}{z-z_0}\biggl [-zs_{12}  -i\delta\biggr ]^{-\epsilon-1}{}_{2}{F}_{1}(\dots;\xi_1) =\\
& \frac{-1}{s_{12}s_{23}}\biggl \{\frac{1}{z-z_0}\biggl [-zs_{12}  -i\delta\biggr ]^{-\epsilon} + s_{12}\biggl [-zs_{12}  -i\delta\biggr ]^{-\epsilon-1}\biggr \}{}_{2}{F}_{1}(\dots;\xi_1)
\end{split}
\end{equation}
and
\begin{equation}
\label{eq:apu2}
\begin{split}
& \frac{1}{\Delta s}\frac{1}{z-z_0}\biggl [-(1-z)s_{23}  -i\delta\biggr ]^{-\epsilon-1}{}_{2}{F}_{1}(\dots;\xi_2) =\\
& \frac{-1}{s_{12}s_{23}}\biggl \{\frac{1}{z-z_0}\biggl [-(1-z)s_{23}  -i\delta\biggr ]^{-\epsilon} - s_{23}\biggl [-zs_{12}  -i\delta\biggr ]^{-\epsilon-1}\biggr \}{}_{2}{F}_{1}(\dots;\xi_2).
\end{split}
\end{equation}
Using Eqs.\ \eqref{eq:apu1} and \eqref{eq:apu2}, we can write Eq.\ \eqref{eq:D0intstep} in the following form:
\begin{equation}
\label{eq:D0lastzint}
D_0(s_{12},s_{23}) =\frac{i(4\pi)^{\epsilon}}{(4\pi)^2}\frac{R_{\Gamma}}{s_{12}s_{23}}\frac{2}{\epsilon^2}\frac{\Gamma(-2\epsilon)}{\Gamma^2(-\epsilon)}\biggl \{ \mathcal{J}_{(1)}(s_{12},s_{23}) +  \mathcal{J}_{(2)}(s_{12},s_{23})  \biggr \},
\end{equation}
where we have defined the integrals $\mathcal{J}_{(1,2)}$ as:
\begin{equation}
\label{eq:D0last1}
\begin{split}
\mathcal{J}_{(1)}(s_{12},s_{23}) = \int_{0}^{1} & {\rm d}z\biggl \{ -s_{23}\biggl [-(1-z)s_{23}-i\delta \biggr ]^{-\epsilon-1}{}_{2}{F}_{1}(1+\epsilon,-\epsilon,1-\epsilon;\xi_2)\\
&  - s_{12}\biggl [-zs_{12}  -i\delta\biggr ]^{-\epsilon-1}{}_{2}{F}_{1}(1+\epsilon,-\epsilon,1-\epsilon;\xi_1) \biggr \}.
\end{split}
\end{equation}
and
\begin{equation}
\label{eq:D0last2}
\begin{split}
\mathcal{J}_{(2)}(s_{12},s_{23}) = \int_{0}^{1} & \frac{{\rm d}z}{z-z_0}\biggl \{ \biggl [-(1-z)s_{23}-i\delta \biggr ]^{-\epsilon}{}_{2}{F}_{1}(1+\epsilon,-\epsilon,1-\epsilon;\xi_2)\\
&  - \biggl [-zs_{12}  -i\delta\biggr ]^{-\epsilon}{}_{2}{F}_{1}(1+\epsilon,-\epsilon,1-\epsilon;\xi_1) \biggr \}.
\end{split}
\end{equation}
Let us first evaluate the integral in Eq.\ \eqref{eq:D0last1}. Making the change $z = (1-z)$ in the second term on the right-hand side in Eq.\ \eqref{eq:D0last1} and simplifying the hypergeometric functions with the help of the result in Eq.\ \eqref{eq:hfarg1}, we obtain
\begin{equation}
\begin{split}
\mathcal{J}_{(1)}(s_{12},&s_{23}) =  \frac{-\epsilon\Gamma^2(-\epsilon)}{\Gamma(-2\epsilon)}\biggl \{(-s_{23}-i\delta)^{-\epsilon}\int_{0}^{1}{\rm d}z\biggl [1-z\left (\frac{s_{23}}{s_{23}+i\delta}\right )\biggr ]^{-(1+\epsilon)}\\
&\quad\quad + (-s_{12}-i\delta)^{-\epsilon}\int_{0}^{1}{\rm d}z\biggl [1-z\left (\frac{s_{12}}{s_{12}+i\delta}\right )\biggr ]^{-(1+\epsilon)}\biggr \} .
\end{split}
\end{equation}
Performing the last trivial $z$ integration, we find
\begin{equation}
\label{eq:D0last1f}
\mathcal{J}_{(1)}(s_{12},s_{23}) = \frac{\Gamma^2(-\epsilon)}{\Gamma(-2\epsilon)}\biggl \{(-s_{23}-i\delta)^{-\epsilon} + (-s_{12}-i\delta)^{-\epsilon}\biggr \}.
\end{equation} 
To carry out the remaining integration $\mathcal{J}_{(2)}$, we apply Eq.\ \eqref{eq:hfarg1}. As a result, Eq.\ \eqref{eq:D0last2} reduces to
\begin{equation}
\begin{split}
\mathcal{J}_{(2)}(s_{12},s_{23}) = \frac{-\epsilon\Gamma^2(-\epsilon)}{\Gamma(-2\epsilon)}\int_{0}^{1}  \frac{{\rm d}z}{z-z_0} \biggl \{ \biggl  [& -(1-z)s_{23}-i\delta \biggr ]^{-\epsilon}\\
&  - \biggl [-zs_{12}  -i\delta\biggr ]^{-\epsilon} \biggr \}.
\end{split}
\end{equation} 
Expanding this expression into a power series in $\epsilon$, we obtain 
\begin{equation}
\label{eq:D0last2fin}
\mathcal{J}_{(2)}(s_{12},s_{23}) = -2\int_{0}^{1}  \frac{{\rm d}z}{z-z_0}\ln\left (\frac{-(1-z)s_{23}-i\delta}{-zs_{12}-i\delta} \right )\epsilon + \mathcal{O}(\epsilon^2).
\end{equation}
Performing the last $z$ integration, we find that the integral in Eq.\ \eqref{eq:D0last2fin} can be written in the following form
\begin{equation}
\label{eq:D0last2fin2}
\mathcal{J}_{(2)}(s_{12},s_{23}) = \biggr\{ \ln^2\left ( \frac{s_{12}+i\delta}{s_{23}+i\delta} \right ) + \pi^2 \biggl \}\epsilon + \mathcal{O}(\epsilon^2).
\end{equation}
Finally, on the basis of Eqs.\ \eqref{eq:D0lastzint}, \eqref{eq:D0last1f} and \eqref{eq:D0last2fin2}, we find that the massless scalar box integral is given by
\begin{equation}
\label{eq:D0final}
\begin{split}
D_0(s_{12},s_{23}) = \frac{i(4\pi)^{\epsilon}}{(4\pi)^2} \frac{R_{\Gamma}}{s_{12}s_{23}}\biggl \{&\frac{2}{\epsilon^2}\biggl [(-s_{12}-i\delta)^{-\epsilon} + (-s_{23}-i\delta)^{-\epsilon}\biggr ]\\
& - \ln^2\left (\frac{s_{12} + i\delta}{s_{23}+i\delta} \right ) - \pi^2 \biggr \} + \mathcal{O}(\epsilon).
\end{split}
\end{equation}
Here, of course, the $D_0(s_{12},s_{23})$ function must be understood as 
\begin{equation}
D_0(s_{12},s_{23}) = D_0(p_1,p_2,p_3;s_{12},s_{23}).
\end{equation}

\section{Reduction of tensor integral to scalar integrals}

In this section we discuss one of the key-methods applied in the theory part of this thesis, the reduction of a general loop structure into a set of scalar integrals, needed for the evaluation of the virtual corrections at NLO.

\vspace{0.3cm}

In general, the higher-order one-loop $n$-point $(n\geq 3)$ tensor integrals are quite complicated to calculate. A very useful and straightforward method to evaluate these integrals is the procedure where the tensor integrals are reduced to the scalar integrals. This method was first proposed by Passarino and Veltman \cite{PASVELT}, and it is also conveniently summarized in \cite{PASVELT2}. However, for the specific problem one should do a lot of extra work to get the relevant integral reductions from the general expressions presented in the literature. Thus, in the following section, I show how to reduce the rank one and rank two massless 2, 3 and 4-point tensor integrals,
\begin{equation}
\label{eq:tensintgen}
\begin{split}
\{B^{\mu},B^{\mu\nu}\}(p^2) & = \int_k\frac{\{k^{\mu},k^{\mu}k^{\nu}\}}{d_0d_1},\\
\{C^{\mu},C^{\mu\nu}\}(p_1,p_2;s_{12}) & = \int_k \frac{\{k^{\mu},k^{\mu}k^{\nu}\}}{d_0d_1d_2},\\
\{D^{\mu},D^{\mu\nu}\}(p_1,p_2,p_3;s_{12},s_{23}) & = \int_k \frac{\{k^{\mu},k^{\mu}k^{\nu}\}}{d_0d_1d_2d_3},
\end{split}
\end{equation}
to the scalar integrals $A_0, B_0, C_0$ and $D_0$, which are all the necessary ingredients for computing the QCD virtual corrections at NLO.

\subsection{Tensor decomposition of two-point ($B$) functions}
The rank one and two $B$ integrals can be decomposed as 
\begin{equation}
\label{eq:twoPV}
B^{\mu}(p^2)  = p^{\mu}B_1(p^2), \quad\quad B^{\mu\nu}(p^2)  = g^{\mu\nu}B_{00}(p^2) + p^{\mu}p^{\nu}B_{11}(p^2),
\end{equation}
where the coefficients $B_1, B_{11}$ and $B_{00}$ are often referred to as form factors. These form factors can be algebraically reduced to the scalar two-point integral $B_0$ with the Passarino-Veltman reduction procedure in the following way: First, we note that the scalar product of the integration momentum $k^{\mu}$ with the external momentum $p^{\mu}$ can be expressed in terms of the denominators as
\begin{equation}
\label{eq:trick}
k\cdot p = \frac{1}{2}(d_0-d_1-p^2), \quad \text{where} \quad p^2 \neq 0.
\end{equation}
Using this trick the rank one one-point tensor integral $B^{\mu}$ can be reduced to $B_0$ by contracting with $p_{\mu}$, which yields
\begin{equation}
\begin{split}
p_{\mu}B^{\mu}(p^2) & \overset{\eqref{eq:trick}}{=} \int_k\frac{k\cdot p}{d_0d_1} = \frac{1}{2}\biggl \{\int_k \frac{1}{d_1} - \int_k \frac{1}{d_0} - p^2\int_k \frac{1}{d_0d_1}\biggr \}\\
& \overset{\eqref{eq:A0int}, \eqref{eq:B0int}}{=} -\frac{p^2}{2}B_0(p^2).
\end{split}
\end{equation}
This relation together with Eq. \eqref{eq:twoPV} implies
\begin{equation}
\label{eq:BRANK11}
B_{1}(p^2) = -\frac{1}{2}B_0(p^2).
\end{equation}
The rank two two-point tensor integral $B^{\mu\nu}$ can be reduced to the scalar integral $B_0$ by contracting with a product of the external momenta $p_{\mu}p_{\nu}$ and the metric tensor $g_{\mu\nu}$. Thus, in $d$ dimensions we obtain:
\begin{equation}
\label{eq:twoPVrank2}
\begin{split}
g_{\mu\nu}B^{\mu\nu}(p^2) & = dB_{00}(p^2) + p^2B_{11}(p^2)\\
p_{\mu}p_{\nu}B^{\mu\nu}(p^2) & = p^2B_{00}(p^2) + p^4B_{11}(p^2),\\
\end{split}
\end{equation}
where
\begin{equation}
g_{\mu\nu}B^{\mu\nu}(p^2) = 0, \quad \text{and} \quad p_{\mu}p_{\nu}B^{\mu\nu}(p^2) = \frac{p^4}{4}B_0(p^2).
\end{equation}
From Eq.\ \eqref{eq:twoPVrank2} we obtain two linear equations with two unknowns, which we may solve for $B_{00}$ and $B_{11}$:
\begin{equation}
\label{eq:BRANK2}
\begin{split}
B_{00}(p^2)  = \frac{p^2}{4(1-d)}B_0(p^2), \quad\quad B_{11}(p^2) = \frac{d}{4(d-1)}B_0(p^2). 
\end{split}
\end{equation} 

\subsection{Tensor decomposition of three-point ($C$) functions}
\label{Cfunctions}

For the three-point integrals we can write the expansion:
\begin{equation}
\label{eq:threePV}
C^{\mu}(p_1,p_2;s_{12})  = p_1^{\mu}C_1 + p_2^{\mu}C_2,
\end{equation}
and
\begin{equation}
\label{eq:threeOV2}
C^{\mu\nu}(p_1,p_2;s_{12})  = g^{\mu\nu}C_{00} + p_1^{\mu}p_1^{\nu}C_{11} + p_2^{\mu}p_2^{\nu}C_{22} + (p_1^{\mu}p_2^{\nu} + p_2^{\mu}p_1^{\nu})C_{12},
\end{equation}
where $C_i = C_i(p_1,p_2;s_{12})$ and $C_{ij} = C_{ij}(p_1,p_2;s_{12})$. These two equations can be reduced to scalar integrals $B_0$ and $C_0$ by contracting with the external momenta and $g_{\mu\nu}$.

\vspace{0.3cm}

For the rank one form factors $C_1$ and $C_2$, we obtain the following results:
\begin{equation} 
\label{eq:Crankoneform}
\begin{split}
C_1(p_1,p_2;s_{12}) & = -\frac{1}{s_{12}}\biggl [B_0(s_{12}) + s_{12}C_0(p_1,p_2;s_{12}) \biggr ],\\
C_2(p_1,p_2;s_{12})	& = \frac{1}{s_{12}}B_0(s_{12}),
\end{split}
\end{equation}
and for the rank two form factors $C_{00}, C_{11}, C_{22}$ and $C_{12}$:
\begin{equation}
\label{eq:Cranktwoform}
\begin{split}
C_{00}(p_1,p_2;s_{12}) & = \frac{1}{2(d-2)}B_0(s_{12}),\\
C_{11}(p_1,p_2;s_{12}) & = \frac{1}{s_{12}}\biggl [\frac{3}{2}B_0(s_{12}) + s_{12}C_0(p_1,p_2;s_{12}) \biggr ],\\
C_{22}(p_1,p_2;s_{12}) & = -\frac{1}{s_{12}}B_{0}(s_{12}),\\
C_{12}(p_1,p_2;s_{12}) & = \frac{1}{2s_{12}}\left (\frac{d}{2-d} \right )B_0(s_{12}).
\end{split}
\end{equation}

\subsection{Tensor decomposition of four-point ($D$) functions}
\label{Dfunctions}

The rank one and two $D$ integrals can be decomposed as 
\begin{equation}
\label{eq:fourPV}
\begin{split}
D^{\mu}(p_1,p_2,p_3;s_{12},s_{23})  & = p_1^{\mu}D_1 + p_2^{\mu}D_2 + p_3^{\mu}D_3 ,\\
\end{split}
\end{equation}
and
\begin{equation}
\begin{split}
D^{\mu\nu}(p_1,p_2,p_3;s_{12},s_{23})   = g^{\mu\nu}D_{00} & + p_1^{\mu}p_1^{\nu}D_{11} + p_2^{\mu}p_2^{\nu}D_{22}+ p_3^{\mu}p_3^{\nu}D_{33} \\
& + \left (p_1^{\mu}p_2^{\nu} + p_{2}^{\mu}p_1^{\nu}\right )D_{12}\\
& +\left (p_1^{\mu}p_3^{\nu} + p_{3}^{\mu}p_1^{\nu}\right )D_{13}\\
 & + \left (p_2^{\mu}p_3^{\nu} + p_{3}^{\mu}p_2^{\nu}\right )D_{23},
\end{split}
\end{equation}
where $D_i=D_i(p_1,p_2,p_3;s_{12},s_{23})$ and $D_{ij} = D_{ij}(p_1,p_2,p_3;s_{12},s_{23})$ for $i=\{1,2,3\}$ and $ij =\{11,22,33,12,13,23\}$. These two equations can be reduced to scalar integrals $B_0, C_0$ and $D_0$.

\vspace{0.3cm}

The rank-one form factors $D_i$ satisfy equations of the form
\begin{equation}
\label{eq:Drankone}
\left( \begin{array}{c}
D_{1}  \\
D_{2}  \\
D_{3} 
\end{array} \right) =
\mathbf{G}_{3\times 3}^{-1}
\left( \begin{array}{c}
R_1\\
R_2\\
R_3
\end{array} \right),
\end{equation}
where $\mathbf{G}_{3\times 3}$ is the $3\times 3$ Gram matrix 
\begin{equation}
\mathbf{G}_{3\times 3} = \left( \begin{array}{ccc}
0 & \frac{s_{12}}{2} & \frac{s_{13}}{2} \\
\frac{s_{12}}{2} & 0 & \frac{s_{23}}{2} \\
\frac{s_{13}}{2} & \frac{s_{23}}{2} & 0
\end{array} \right)
\end{equation}
and the definitions of the functions $R_1, R_2$ and $R_3$ are given by
\begin{equation}
\begin{split}
R_1 & = \frac{1}{2}\biggl [C_0(p_3,p_4;s_{12}) - C_0(p_2,p_3;s_{23}) \biggr ] ,\\
R_2 & = \frac{1}{2}\biggl [-s_{12}D_0 + C_0(p_4,p_1;s_{23})- C_0(p_3,p_4;s_{12})\biggr ],\\
R_3 & = \frac{1}{2}\biggl [+s_{12}D_0 + C_0(p_1,p_2;s_{12})-C_0(p_4,p_1;s_{23})\biggr ].
\end{split}
\end{equation}
We can simplify the solution of Eq.\ \eqref{eq:Drankone} by noting that:
\begin{equation}
\begin{split}
C_0(p_1,p_2;s_{12}) &= C_0(p_3,p_4;s_{12}) = C_0(s_{12}),\\
C_0(p_4,p_1;s_{23}) &= C_0(p_2,p_3;s_{23}) = C_0(s_{23}),\\
\end{split}
\end{equation}
and 
\begin{equation}
D_0 = D_0(p_1,p_2,p_3;s_{12},s_{23}) = D_0(s_{12},s_{23}),
\end{equation}
where the scalar integrals $C_0(s_{ij})$ and $D(s_{12},s_{23})$ are given by Eqs.\ \eqref{eq:C0final} and \eqref{eq:D0final}, respectively. Thus, for the rank-one form factors $D_1, D_2$ and $D_3$, we obtain:
\begin{equation}
\label{eq:boxrankoneform}
\begin{split}
D_1 & = \frac{C_0(s_{23})-C_0(s_{12})}{s_{12}+s_{23}} -\frac{1}{2}\left (\frac{2s_{12}+s_{23}}{s_{12}+s_{23}} \right )D_0(s_{12},s_{23}),\\
D_2 & = -\frac{D_0(s_{12},s_{23})}{2},\\
D_3 & = \frac{C_0(s_{23})-C_0(s_{12})}{s_{12}+s_{23}} - \frac{1}{2}\left (\frac{s_{12}}{s_{12}+s_{23}}\right )D_0(s_{12},s_{23}). \\
\end{split}
\end{equation}
The results in Eq.\ \eqref{eq:boxrankoneform} have been analytically confirmed with \cite{PQCDPAP,PASSARINO}.

\vspace{0.3cm}

The rank two form factors $D_{ij}$ with $ij=\{11,22,33,12,13,23\}$  satisfy equations of the form
\begin{equation}
\label{eq:Dranktwo}
\mathbf{D}^{(k)} = \mathbf{G}_{3\times 3}^{-1}\mathbf{R}^{(k)} \quad \text{for} \quad  k=1,2,3,
\end{equation}
where the $\mathbf{D}^{(k)}$'s are given by
\begin{equation}
\mathbf{D}^{(1)} =\left( \begin{array}{c}
D_{11}  \\
D_{12}  \\
D_{13} 
\end{array} \right)
,\quad \mathbf{D}^{(2)} = \left( \begin{array}{c}
D_{12}  \\
D_{22}  \\
D_{23} 
\end{array} \right), 
\quad \mathbf{D}^{(3)} =\left( \begin{array}{c}
D_{13}  \\
D_{23}  \\
D_{33} 
\end{array} \right).
\end{equation} 
The definitions of the functions $\mathbf{R}^{(k)}$ are given by
\begin{equation}
\mathbf{R}^{(i)} = \left( \begin{array}{c}
R_{1}^{(i)}  \\
R_{2}^{(i)}  \\
R_{3}^{(i)} 
\end{array} \right),
\end{equation}
where
\begin{equation}
\begin{split}
R_1^{(1)} & = \frac{1}{2}\biggl [\mathcal{A}^{(1)} + C_0(p_2,p_3;s_{23})-2D_{00}\biggr ],\\
R_2^{(1)} & = \frac{1}{2}\biggl [-s_{12}D_1 + \mathcal{A}^{(2)}-\mathcal{A}^{(1)}\biggr ],\\
R_3^{(1)} & = \frac{1}{2}\biggl [+s_{12}D_1 + \mathcal{A}^{(3)}-\mathcal{A}^{(2)}\biggr ],\\
R_1^{(2)} & = \frac{1}{2}\biggl [\mathcal{A}^{(1)}-\mathcal{A}^{(0)}\biggr ],\\
R_2^{(2)} & = \frac{1}{2}\biggl [-s_{12}D_2 + \mathcal{B}^{(2)}-\mathcal{A}^{(1)}-2D_{00}\biggr ],\\
R_3^{(2)} & = \frac{1}{2}\biggl [+s_{12}D_2 + \mathcal{B}^{(3)}-\mathcal{B}^{(2)} \biggr ],\\
R_1^{(3)} & = \frac{1}{2}\biggl [\mathcal{B}^{(1)}-\mathcal{B}^{(0)}\biggr ],\\
R_2^{(3)} & = \frac{1}{2}\biggl [-s_{12}D_3 +\mathcal{B}^{(2)}-\mathcal{B}^{(1)}\biggr ],\\
R_3^{(3)} & = \frac{1}{2}\biggl [+s_{12}D_3 -\mathcal{B}^{(2)}-2D_{00}\biggr ],\\
\end{split}
\end{equation}
with
\begin{equation}
\begin{split}
\mathcal{A}^{(0)} & = C_1(p_2,p_3;s_{23}), \\
\mathcal{A}^{(1)} & = -C_0(p_3,p_4;s_{12})-C_2(p_3,p_4;s_{12}),\\
\mathcal{A}^{(2)} & = -C_0(p_4,p_1;s_{23})-C_1(p_4,p_1;s_{23})+C_2(p_4,p_1;s_{23}),\\
\mathcal{A}^{(3)} & = C_1(p_1,p_2;s_{12}),\\
\mathcal{B}^{(0)} & = C_2(p_2,p_3;s_{23}), \\
\mathcal{B}^{(1)} & = C_1(p_3,p_4;s_{12})-C_2(p_3,p_4;s_{12}),\\
\mathcal{B}^{(2)} & = -C_0(p_4,p_1;s_{23})-C_1(p_4,p_1;s_{23}),\\
\mathcal{B}^{(3)} & = C_2(p_1,p_2;s_{12}),
\end{split}
\end{equation}
and
\begin{equation}
\begin{split}
D_{00} & = \frac{1}{2(d-3)}\biggl [C_0(p_2,p_3;s_{23}) + s_{12}D_2 - s_{12}D_3\biggr ].\\
\end{split}
\end{equation}
Again, we can simplify the solution of Eqs.\ \eqref{eq:Dranktwo} by noting that:
\begin{equation}
\begin{split}
C_1(p_1,p_2;s_{12}) &= C_1(p_3,p_4;s_{12}) = C_1(s_{12}),\\
C_2(p_1,p_2;s_{12}) &= C_2(p_3,p_4;s_{12}) = C_2(s_{12}),\\
C_1(p_4,p_1;s_{23}) &= C_1(p_2,p_3;s_{23}) = C_1(s_{23}),\\
C_2(p_4,p_1;s_{23}) &= C_2(p_2,p_3;s_{23}) = C_2(s_{23}),\\
\end{split}
\end{equation}
where
\begin{equation}
\begin{split}
C_1(s_{ij}) & = -\frac{1}{s_{ij}}\biggl [B_0(s_{ij})+s_{ij}C_0(s_{ij})\biggr ],\\
C_2(s_{ij}) & = \frac{1}{s_{ij}}B_0(s_{ij}),
\end{split}
\end{equation} 
for $ij=\{12,23\}$. Here, the form factors $D_k$ for $k=1,2,3$ are given by Eq. \eqref{eq:boxrankoneform}. The results in Eqs.\ \eqref{eq:Dranktwo} have been analytically confirmed with \cite{PQCDPAP,PASSARINO}.

\end{appendices}

\end{document}